\documentclass[aps,prd,reprint,groupedaddress,nofootinbib]{revtex4-1}
\usepackage[dvipsnames]{xcolor}
\usepackage{graphicx}
\usepackage{hyperref}
\usepackage{url} 
\usepackage{ulem}
\usepackage{makecell}
\hypersetup{
    pdfnewwindow=true,      
    colorlinks=true,        
    linkcolor=blue,         
    citecolor=blue,        
    filecolor=blue,         
    urlcolor=blue           
}

\usepackage{amsmath}
\usepackage{amssymb}

\usepackage[acronym]{glossaries}
\setacronymstyle{long-short}
\glsdisablehyper

\newacronym{ligo}{LIGO}{Laser Interferometer Gravitational-wave Observatory}
\newacronym{llo}{LLO}{LIGO Livingston Observatory}
\newacronym{er14}{ER14}{LIGO's Engineering Run 14}
\newacronym{o3}{O3}{LIGO's Observing Run 3}
\newacronym{o3b}{O3b}{LIGO's Observing Run 3b}
\newacronym{roc}{ROC}{Receiver Operating Characteristic}
\newacronym{snr}{SNR}{signal-to-noise ratio}
\newacronym{far}{FAR}{false-alarm rate}

\newacronym{hog}{HOG}{histogram of oriented gradients}
\newacronym{cnn}{CNN}{convolutional neural network}
\newacronym{rnn}{RNN}{recurrent neural network}
\newacronym{lstm}{LSTM}{long short-term memory}
\newacronym{relu}{ReLU}{rectified linear unit}

\renewcommand{\mathbf}{\boldsymbol}

\newcommand{\LF}{\texttt{LF}}

\newcommand{\vggoneBN}{\texttt{VGG13-BN}}

\begin{document}

\title{Detecting and Diagnosing Terrestrial Gravitational-Wave Mimics Through Feature Learning}

\author{Robert E. Colgan$^{1,2}$, Zsuzsa M\'arka$^{5}$, Jingkai Yan$^{2,3}$, Imre Bartos$^{4}$, John N. Wright$^{2,3}$, and Szabolcs M\'arka$^{6}$}

\address{$^1$Department of Computer Science, Columbia University in the City of New York, 500 W. 120th St., New York, NY 10027, USA\\
$^2$Data Science Institute, Columbia University in the City of New York, 550 W. 120th St., New York, NY 10027, USA\\
$^3$Department of Electrical Engineering, Columbia University in the City of New York, 500 W. 120th St., New York, NY 10027, USA\\
$^4$Department of Physics, University of Florida, PO Box 118440, Gainesville, FL 32611-8440, USA\\
$^5$Columbia Astrophysics Laboratory, Columbia University in the City of New York, 538 W. 120th St., New York, NY 10027, USA\\
$^6$Department of Physics, Columbia University in the City of New York, 538 W. 120th St., New York, NY 10027, USA}

\begin{abstract}
As engineered systems grow in complexity, there is an increasing need for automatic methods that can detect, diagnose, and even correct transient anomalies that inevitably arise and can be difficult or impossible to diagnose and fix manually.
Among the most sensitive and complex systems of our civilization are the detectors that search for incredibly small variations in distance caused by gravitational waves---phenomena originally predicted by Albert Einstein to emerge and propagate through the universe as the result of collisions between black holes and other massive objects in deep space.
The extreme complexity and precision of such detectors causes them to be subject to transient noise issues that can significantly limit their sensitivity and effectiveness. 
They are also subject to nearly constant development, improvement, commissioning and other invasive actions that change the nature of the data and its artifact contamination.
In this work, we present a demonstration of a method that can detect and characterize emergent transient anomalies of such massively complex systems.
We illustrate the performance, precision, and adaptability of the automated solution via one of the prevalent issues limiting gravitational-wave discoveries: noise artifacts of terrestrial origin that contaminate gravitational wave observatories' highly sensitive measurements and can obscure or even mimic the faint astrophysical signals for which they are listening.
Specifically, we demonstrate how a highly interpretable convolutional classifier can automatically learn to detect transient anomalies from auxiliary detector data without needing to observe the anomalies themselves.
We also illustrate several other useful features of the model, including how it performs automatic variable selection to reduce tens of thousands of auxiliary data channels to only a few relevant ones; how it identifies behavioral signatures predictive of anomalies in those channels; and how it can be used to investigate individual anomalies and the channels associated with them.
The solution outlined is broadly applicable from medical to automotive and space systems, enabling automated anomaly discovery and characterization and human-in-the-loop anomaly elimination.

\end{abstract}

\maketitle


\section{Introduction}\label{sec:sl_introduction}

Gravitational-wave detectors such as KAGRA, Virgo, GEO600, and \gls{ligo} use highly sensitive interferometers 
~\citep{2021PTEP.2021eA101A,2016CQGra..33g5009D,2015CQGra..32g4001L,2015CQGra..32b4001A}
to measure the extraordinarily tiny variations gravitational waves cause as they pass through the detectors. Discoveries~\cite{2016PhRvL.116f1102A,PhysRevLett.119.161101,2017ApJ...848L..12A} of gravitational-wave detectors are documented in extensive catalogs, such as GWTC-1 \citep{Abbott_2019}, GWTC-2~\citep{2020arXiv201014527A}, GWTC-2.1~\citep{2021arXiv210801045T}, GWTC-3~\citep{2021arXiv211103606T}, 1-OGC~\citep{2019ApJ...872..195N}, 2-OGC~\citep{2020ApJ...891..123N}, 3-OGC~\citep{2021arXiv210509151N}, IAS-Princeton~\citep{venumadhav2019new,zackay2019detecting,Zackay_2019}, and eccentric localizations~\citep{gayathri2020gw190521}.
Due to the detectors' precision, they are also subject to various forms of noise that can appear in the gravitational-wave measurement data and interfere with their ability to detect and characterize gravitational waves.
Among the most troublesome forms of noise are short, loud anomalies known within \gls{ligo} as ``glitches,'' which can obscure or even mimic real gravitational waves.
Unlike ever-present background noise~\citep{2019CQGra..36e5011D,2020PhRvD.101d2003V}, glitches cannot simply be subtracted from the desired signal.
They have a wide variety of causes, some of which are well-understood and many of which are not.
Attempts to mitigate various types of glitches have been a significant focus of \gls{ligo}'s engineering efforts for more than a decade, and have included multiple machine learning\textendash based approaches 
\cite{GSpy,2021arXiv210812044M,2021CQGra..38m5014D,2020CQGra..37q5001S,2020CQGra..37n5001D,2020arXiv200512761E,2020arXiv200503745C,2018CQGra..35i5016R,2017PhRvD..95j4059M,2017PhDT........25V,2016PhDT.......149M,2015CQGra..32x5005N,2013PhRvD..88f2003B,2013PhDT.......555M,2012CQGra..29o5002A,2010CQGra..27s4010C,2010JPhCS.243a2005I,2010JPhCS.243a2006M,2008CQGra..25r4004B,2002nmgm.meet.1841S,gurav2020unsupervised,2020PhRvD.101d2003V}.
In addition to the main interferometer data channel sensitive to gravitational waves, known as the ``strain,'' each \gls{ligo} detector continuously records hundreds of thousands of channels of ``auxiliary'' data describing various aspects of the detector's state and internal and external environment.
Although the vast majority of these channels do not record useful information, some have been found to measure behavior predictive of or associated with certain types of glitches.
Understanding which channels may be associated with glitches and how is therefore an important avenue for diagnosing the root causes of glitches and in some cases reducing or eliminating them.

Recent machine learning\textendash based approaches to glitch understanding and mitigation include works that investigate the association between glitches and \gls{ligo}'s auxiliary data channels by casting glitch detection as a classification problem using only non-astrophysically-sensitive auxiliary channel data as input \cite{EMU,learned_features}.
Many other glitch detection methods directly analyze the gravitational-wave strain
~\citep{2016CQGra..33m4001A,galaxies10010012,2021PhRvD.104j2004M,GSpy,2021CQGra..38m5014D,2021CQGra..38n5001N,2020CQGra..37n5001D,2019APS..APRK01063W,2019PhDT.......117D,2017PhDT........25V,2016PhDT.......149M,2015CQGra..32x5005N,2013PhDT.......555M,2012CQGra..29o5002A,2010CQGra..27s4010C,2010JPhCS.243a2005I,2010JPhCS.243a2006M,2008CQGra..25r4004B}.
However, models trained to replicate their results using only auxiliary channels can provide strain-independent corroboration of their results.
Because such models base their predictions only on astrophysically-insensitive auxiliary data, they present a lower risk of inadvertently labeling an astrophysical event as a glitch.
Notably, achieving a sufficient level of accuracy with a strictly auxiliary channel\textendash based model necessarily implies that it has recognized and learned to be sensitive to behavior observed in only the auxiliary channels that is predictive of glitches in the gravitational-wave strain data stream.
These automatically identified associations can then be analyzed by detector domain experts for potentially unexplored insights into the origins and causes of glitches.
This pursuit could ultimately lead to mitigations that reduce glitch frequency, increase the sensitivity of the detector, and enable sufficiently confident detections of astrophysical events that would otherwise have been rejected as too uncertain \cite{EMU}.

In this work, we demonstrate and apply one such model to a real dataset of a certain type of glitch that was particularly prevalent at \gls{llo} at various times during \gls{o3b}.
We demonstrate the model's ability to identify such glitches with a remarkable accuracy greater than 97\%.
We also illustrate clearly how it arrives at its predictions and highlight several other useful aspects of the model, including its high degree of interpretability and the insights it can provide into the glitches and channels analyzed.
The results provide a demonstration of the capabilities of the models described.
We hope they will stimulate further exploration and application to additional glitch types and datasets, eventually leading to breakthroughs that allow glitches to be reduced or eliminated and the detectors' sensitivity to be increased---which could enable sufficiently confident detections of more gravitational waves and further our understanding of our universe.


\section{Feature Learning for Transient Anomalies}\label{sec:sl_methods}
\begin{figure*}[t]
\includegraphics[width=.49\linewidth]{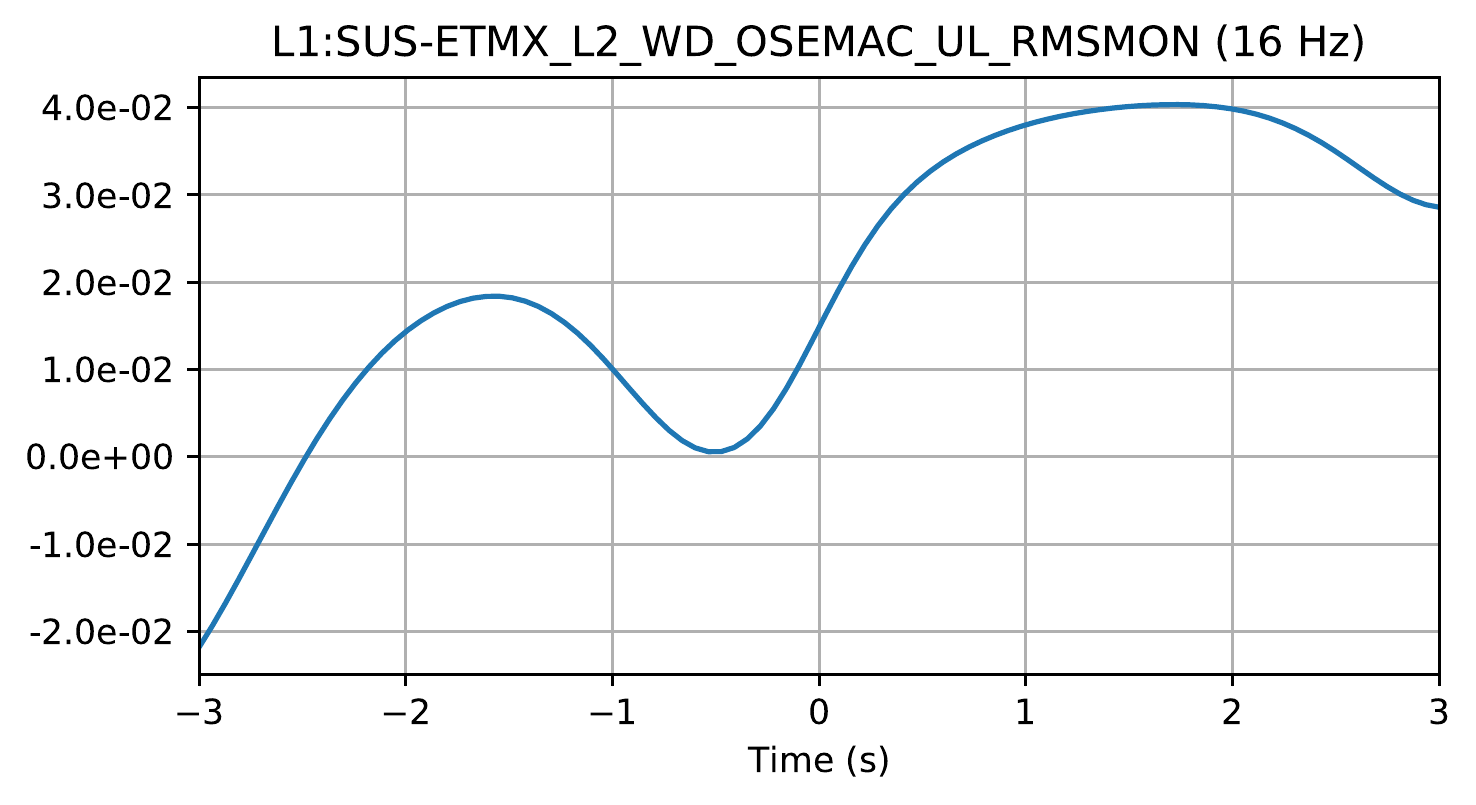}
\includegraphics[width=.49\linewidth]{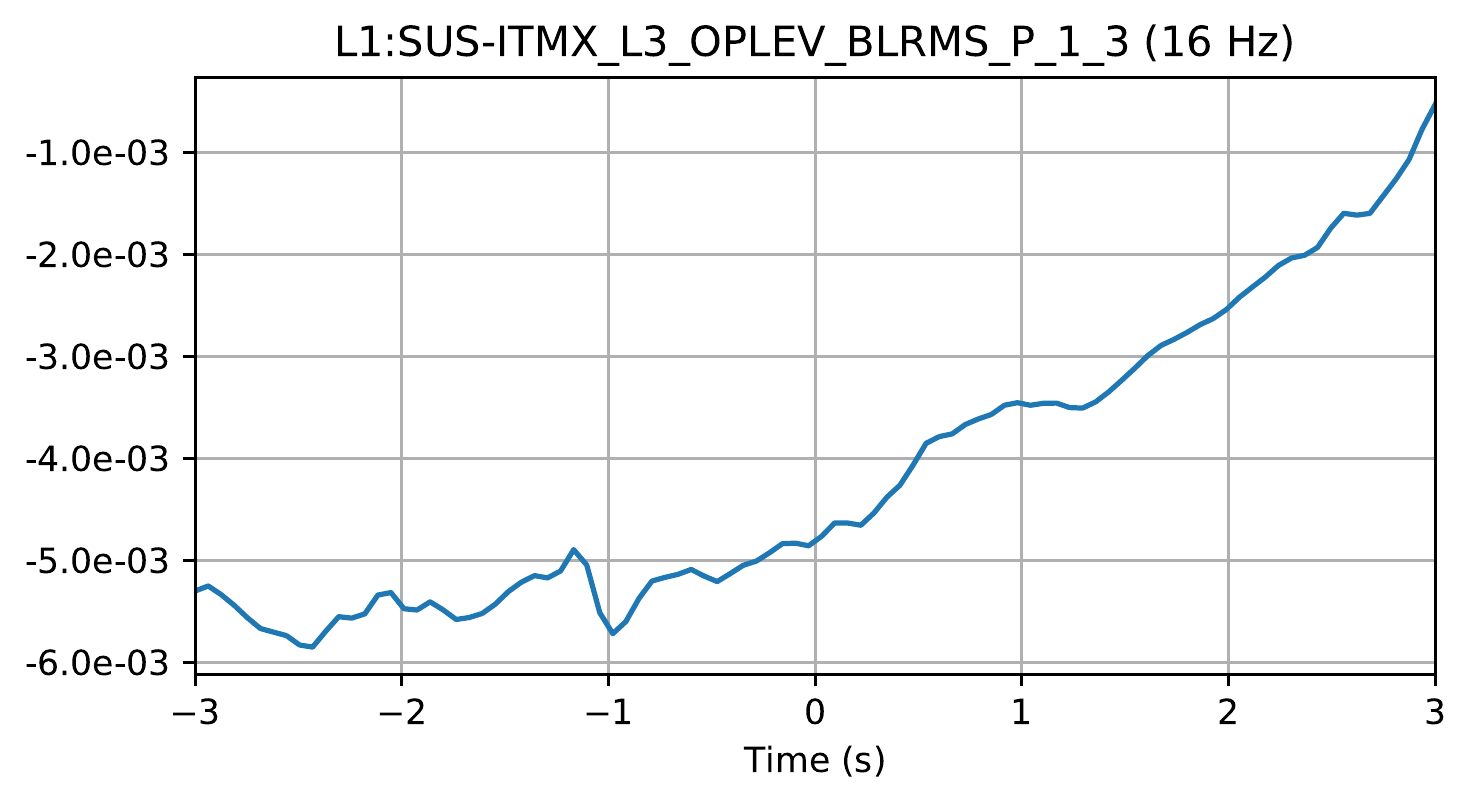}
\caption{
An example of two learned filters in a trained \LF{} model corresponding to two of the channels found by the model to contain behavior associated with scattered light glitches.
See Sec. \ref{sec:sl_results} for more details about what the channels in question measure and what behaviors the model has learned to be sensitive to.
}\label{fig:sl_example_filters}
\end{figure*}

Machine learning models have been demonstrated in recent works \cite{EMU, learned_features} to be able to learn to predict the presence or absence of glitches in high-dimensional gravitational-wave astronomy data by considering only auxiliary channel information, without looking at the gravitational-wave strain data stream in which the glitches themselves appear.
In this section, we describe one such model initially proposed in \cite{learned_features}, which we refer to as \LF{}.
The model is highly interpretable and well-suited to efficient strain-independent glitch detection, offering several useful features including:

\begin{itemize}
    \item (i) {\bf Strain-independent glitch prediction.}
    It is extremely important for potential gravitational-wave detections to be rigorously vetted to ensure they are in fact true astrophysical events.
    Terrestrial-origin glitches can obscure or mimic gravitational waves, so it is also important to identify and distinguish them from astrophysical signals with high confidence.
    Unlike existing glitch-detection methods such as Omicron \cite{Omicron} and GravitySpy \cite{GSpy} that directly analyze the gravitational-wave strain data stream to search for high-power transient noise events, the methods discussed here consider only the detector's auxiliary data channels.
    As such, they provide independent corroboration from an entirely different data source; for example, because auxiliary channels are not astrophysically sensitive, it is unlikely that these methods would inadvertently identify a true gravitational wave as a glitch.
    [Section \ref{sec:sl_results_predictions}]
    \item (ii) {\bf Channel selection.}
    The models are trained on tens of thousands of auxiliary channels.
    Certain channels are known to detector experts to monitor aspects of the detector's behavior associated with glitches, but it would be impossible to comprehensively analyze all channels manually.
    Absent such domain expert knowledge and time, machine learning models can discover connections between auxiliary channels and glitches that might otherwise have gone undetected.
    Notably, the models discussed here are explicitly tuned to learn to ignore input channels that are not useful for making its predictions---the vast majority of them, in this setting.
    Such a practice has previously been demonstrated (e.g., in~\cite{EMU, learned_features}) to improve both interpretability and accuracy.
    These methods achieve excellent performance while ignoring all but a few dozen to few hundred channels, which can then be further investigated manually by detector domain experts.
    [Section \ref{sec:sl_results_channels}]
    \item (iii) {\bf Learned features.}
    In addition to simply telling us which channels to look at, the models discussed here also learn behavioral signatures that correspond to patterns of behavior associated with glitches (or the lack thereof) in individual channels.
    For example, if a certain type of glitch is frequently preceded by a level change in a particular channel, we can expect the feature corresponding to that channel to learn to be sensitive to such a pattern.
    A few of these learned features are illustrated in Figs. \ref{fig:sl_example_filters} and \ref{fig:sl_scattered_light_EN0NoFC_chans}.
    [Section \ref{sec:sl_results_filters}]
    \item (iv) {\bf Channel contributions to individual glitches.}
    For any individual glitch (or set of glitches) of interest, we can feed the model the auxiliary channel data for those glitches and examine the contributions of each channel to the classifier's result for those glitches.
    This could be useful if, for example, there are multiple potential causes of a given glitch and we need to determine which one is most likely at play.
    [Section \ref{sec:sl_results_correlations}]
\end{itemize}

For the experiments presented here, we employ a slightly modified version of the \LF{} model introduced in \cite{learned_features}.
The model is a flat convolutional binary classifier parameterized by a learned filter $\mathbf{f}_p \in \mathbb{R}^T$ of desired length $T$\footnote{
We set the filter length $T$ to 96 samples for all channels based on the finding of \cite{learned_features} that six seconds of data for 16-Hz channels performed well.
As discussed in Sec. \ref{sec:sl_training}, for computational efficiency we used the same number of samples for channels regardless of sample rate, so the model sees six seconds of data for a 16-Hz channel but only 3/8 seconds of data for a 256-Hz channel, for example.}
for each of the $P$ input time series (auxiliary channels) as well as a single learned bias term $b_{s_p}$ for all channels of each sample rate $s_p$.\footnote{
The minor modification we make to the \LF{} model of \cite{learned_features} is the introduction of a separate bias term $b_{s_p}$ for each sample rate $s_p$ rather than a single bias term for all channels, as we found this slightly improved performance. (\cite{learned_features} used only 16-Hz channels in its experiments.)
}

To make a prediction at time $t$, the model takes as input the normalized data (centered at $t$) for all channels $\mathbf{x}_{p}[t] \in \mathbb R^{T}$ of each sample rate $s_p$; cross-correlates the data with the corresponding filter $\mathbf{f}_p$; sums the results for each sample rate $s_p$ and adds the corresponding bias term $b_{s_p}$; and sums the results across all sample rates.
The result is then translated into a probability estimate between 0 and 1 (where 0 indicates the model is certain no glitch is present and 1 indicates it is certain there is a glitch) by passing it through a sigmoid function.
Mathematically,
\begin{equation} \label{eqn:sl_LF}
\hat{y}_t = \sigma \left(\sum_{s_p} \left(\sum_{p} \mathbf{f}_p \star \mathbf{x}_{p}[t]\right) + b_{s_p} \right),
\end{equation}
where $\sigma$ denotes the logistic function $\sigma(x) = (1+\exp(-x))^{-1}$ and $\star$ denotes discrete correlation.

As in \cite{learned_features}, we employ a sparsifying regularizer---the elastic net \cite{elastic}---during training to encourage most of the filters to be 0, i.e. $\lVert{\mathbf{f}_p\rVert} = 0$ for most $p$, because sparsity was found to be essential to good performance as well as beneficial to interpretability \cite{EMU,learned_features}.

After training, the remaining nonzero filters correspond to the channels the model finds useful in making its predictions, indicating channels with behavior that was associated with the appearance of glitches or lack thereof during the training period.
The magnitude of the filters also provides an indication of the relative importance of each channel to the model's predictions (intuitively, when correlated with a higher-magnitude filter, an input data segment will contribute more heavily to the sum and resulting probability estimate than the same segment correlated with a lower-magnitude filter) \cite{learned_features}.

\begin{figure*}[t]
\includegraphics[width=.49\linewidth]{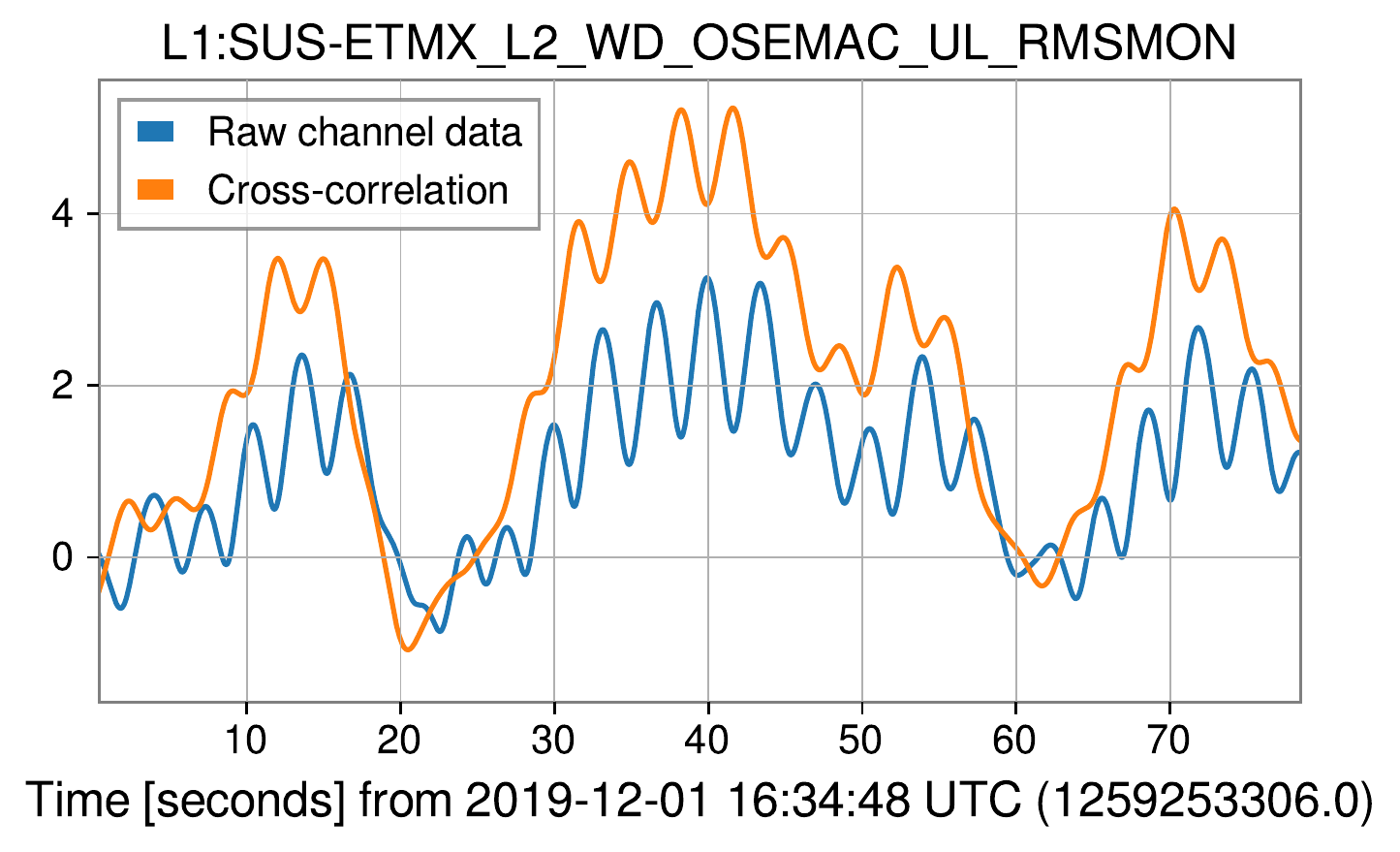}
\includegraphics[width=.49\linewidth]{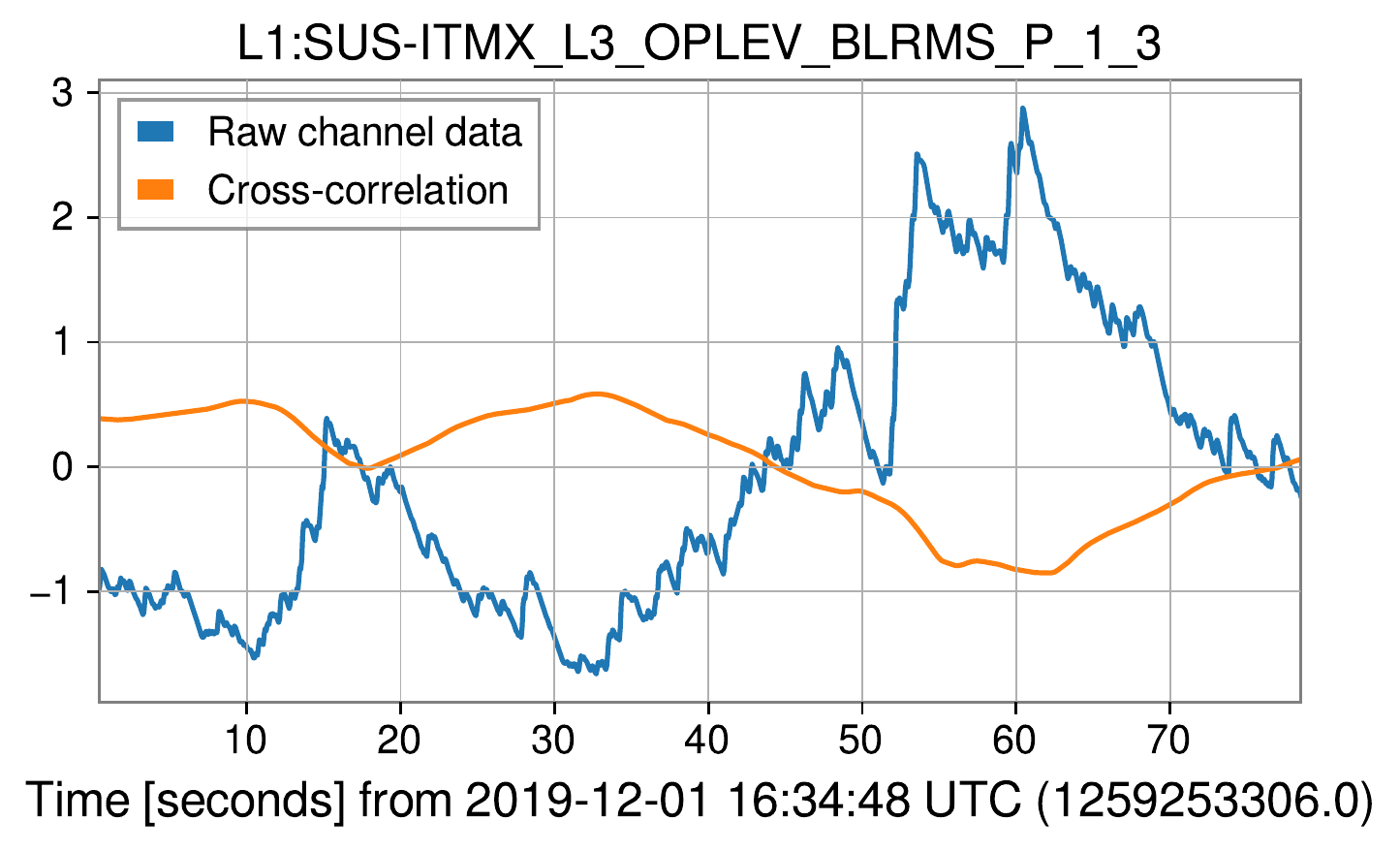}
\caption{
Blue: normalized channel data from the two channels whose corresponding filters are illustrated in Fig. \ref{fig:sl_example_filters} over time during several scattered light glitches, including a 19.75-second-long one centered at GPS time 1,259,253,341.438 (approximately 35.4 seconds on the x-axis), a 9.5-second-long one centered at 1,259,253,318.188 (12.2 seconds), and several shorter ones from 1,259,253,370.438 to 1,259,253,382.719 (64.4 to 76.7 seconds). 
The y-axis corresponds to standard deviations above or below the mean (computed over all samples during the training period).
Orange: the result of the cross-correlation between the filters shown in Fig. \ref{fig:sl_example_filters} and the raw channel data over time.
At each point along the x-axis, the filter is cross-correlated with the corresponding number of samples of raw data centered at that point; the result is plotted on the y-axis.
Note that the filter for the channel on the right is negative, so when the raw value drops below 0 the cross-correlation is positive.
Fig. \ref{fig:sl_omegagram1_with_prob} shows a visualization of the strain data during the same time period illustrated above---in which the glitches are clearly visible---as well as the model's final glitch probability estimate over that period.
}\label{fig:sl_example_convs}
\end{figure*}

In Figs. \ref{fig:sl_example_filters} and \ref{fig:sl_example_convs}, we illustrate visually how the model arrives at its predictions given a potential glitch time $t$.
Fig. \ref{fig:sl_example_filters} shows two examples of learned filters $\mathbf{f}_p$.
Fig. \ref{fig:sl_example_convs} illustrates two channels' data $\mathbf{x}_{p}[t]$ (blue) during several glitches around GPS time $t = 1,259,253,345$ (the center of the x-axis) as well as the cross-correlation (orange) between the channel data centered at each point on the x-axis and the corresponding filter illustrated in Fig. \ref{fig:sl_example_filters}.
To arrive at an estimate that there is a glitch present at that time, the model would sum the cross-correlations at that time over all channels considered.
In Fig. \ref{fig:sl_example_convs}, the cross-correlation values for both channels at that time are positive, so both channels are contributing to the model predicting the presence of a glitch.
The left channel's cross-correlation is significantly greater, however, so it makes a larger contribution to the model's confidence that a glitch is present.


\section{Application to Scattered Light Glitches}\label{sec:sl_data}

\begin{figure*}[t]
\includegraphics[width=\linewidth]{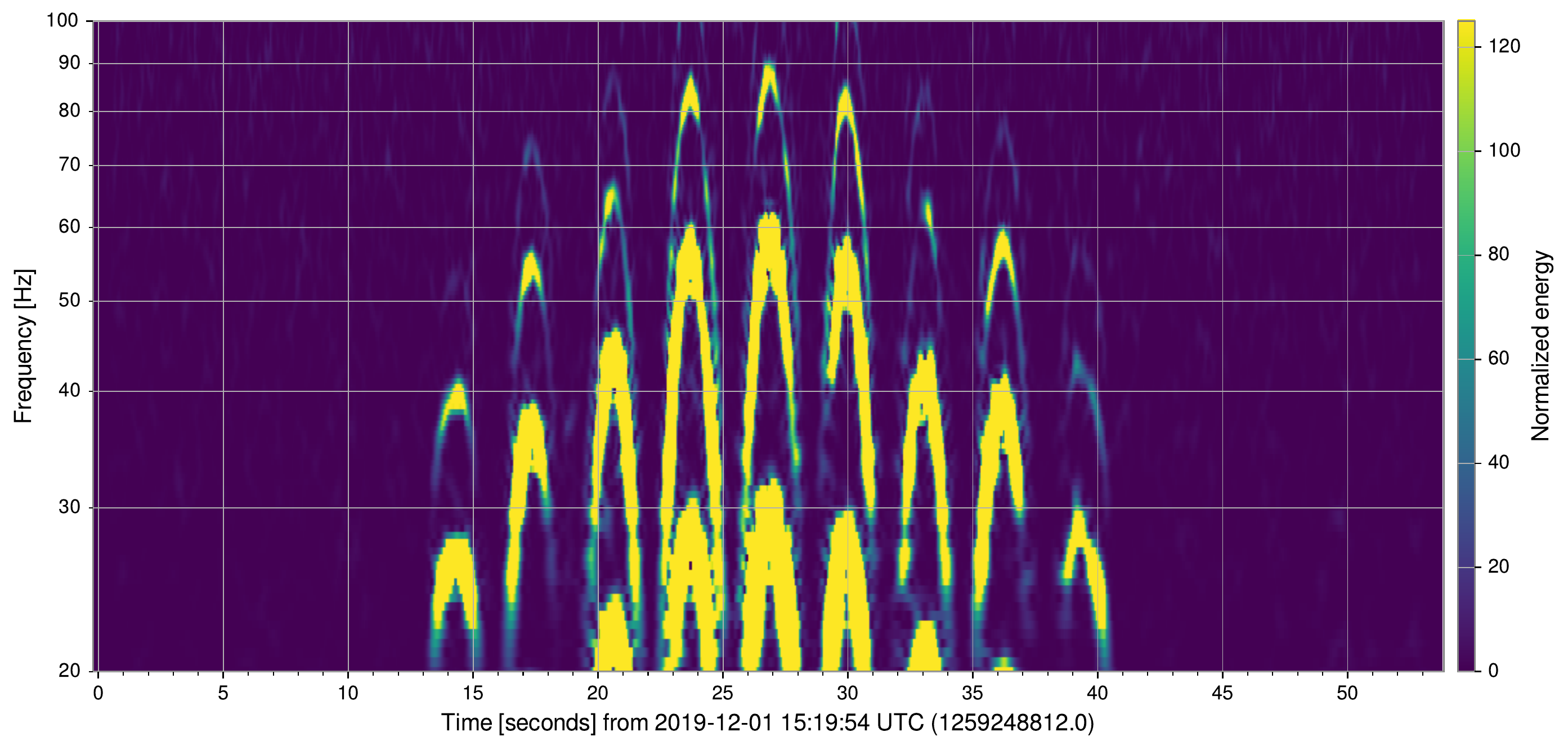}
\caption{Omegagram (a Gabor-wavelet time-frequency representation) of the strain data \cite{rollins_thesis}\cite{2004CQGra..21S1809C}) of a loud scattered light glitch around GPS time 1,259,248,838.813. Note the characteristic arches with peaks spaced approximately 3-4 seconds apart.}
\label{fig:sl_omegagram_only}
\end{figure*}

To illustrate how this method can be employed to interrogate a particular variety of recurring glitch, in this section we demonstrate its application to a particularly troublesome glitch type known as ``scattered light''~\citep{Soni_2021,2021arXiv210707565B,2020PhDT........41A,2012JOSAA..29.1722M,2011arXiv1108.1598T,1997PhRvD..56.6085V,1997PhDT.......130S} during a period when the \gls{llo} detector was experiencing an elevated frequency of this type of glitch.
We chose this particular class of glitch to validate our machine learning and auxiliary channel\textendash based approach because it represents a well-characterized area that has been studied with multiple different approaches \cite{Soni_2021,GSpy}.

\subsection{Scattered Light Glitches}\label{sec:sl_data_scattered_light}

On December 1, 2019, the \gls{llo} detector experienced a high level of scattered light glitches, one of which is illustrated in a time-frequency plot in Fig. \ref{fig:sl_omegagram_only}.
Scattering is observed as wide ``arches" on the Omegagram, a time-frequency representation of the gravitational-wave channel.
These glitches pollute the low-frequency regime (\textless100Hz) of the strain data. This lowest frequency part of the \gls{ligo} sensitivity region is observationally important, as the highest-mass binary black holes observed by \gls{ligo} merge in this regime.
As a result, \gls{ligo} observed an increased trigger rate in the output of compact binary coalescence pipelines, especially for short duration templates \cite{T2000052}.
Recognizing, characterizing, and mitigating scattered light glitches thus can significantly increase the confidence in \gls{ligo} gravitational-wave detections. 

\begin{figure*}[t]
\includegraphics[width=.49\linewidth]{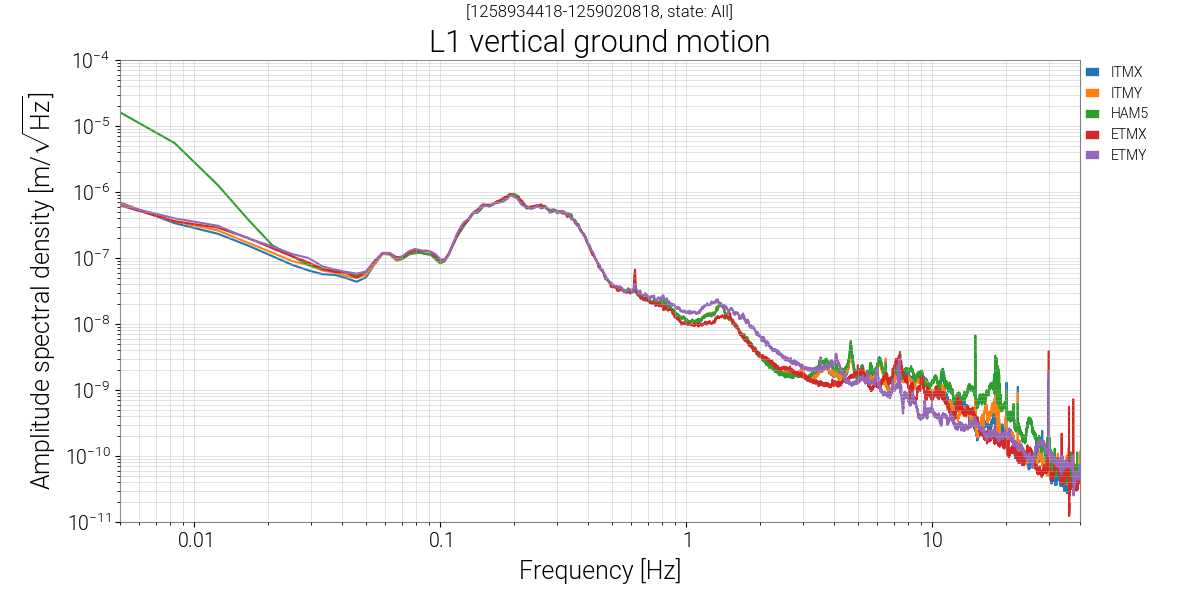}
\includegraphics[width=.49\linewidth]{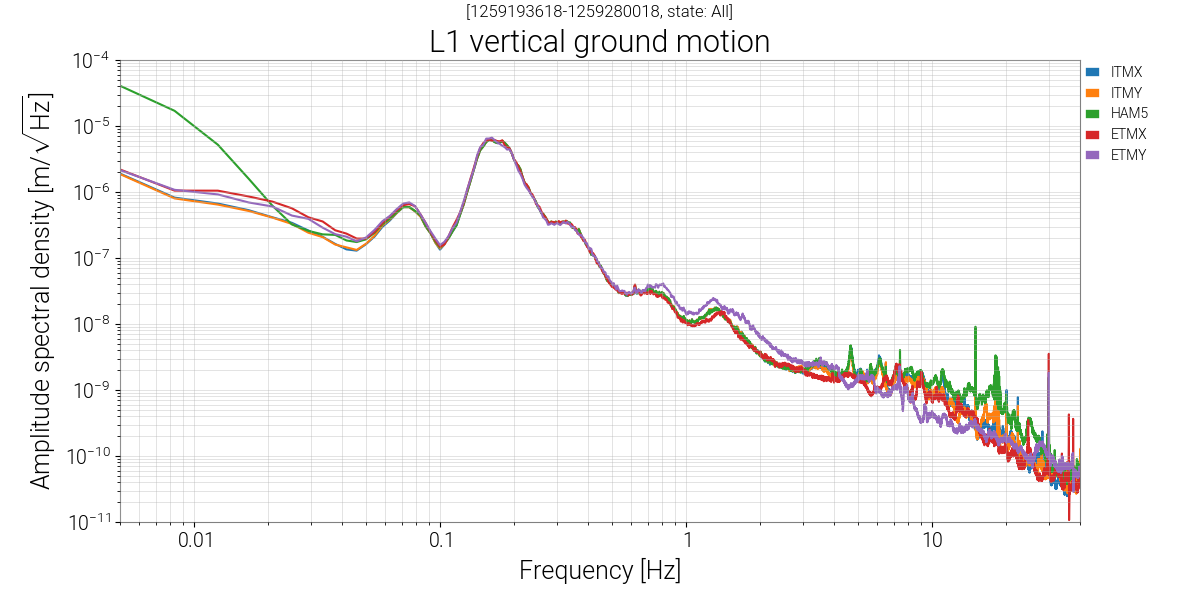}
\caption{Microseismic noise on (left) a typical day at \gls{llo} and (right) December 1, 2019, showing significantly greater activity in the 0.1-0.3 Hz range.}
\label{fig:sl_microseismic}
\end{figure*}

Increased scattered light glitch rates have been observed to correlate with increased microseismic activity at the \gls{ligo} detector sites \cite{T2000052,Soni_2021}.
The ``microseism'' band refers to the 0.03-0.5 Hz region in the amplitude spectral density curve of the observed ground motion produced by ocean waves.
\gls{llo} generally experiences increased microseismic activity due to its close proximity to the Gulf of Mexico.
The microseismic noise at \gls{llo} usually peaks approximately between 0.1-0.3 Hz.
This frequency is below the dominant pendulum frequency of the quadruple suspension (0.45Hz \cite{Soni_2021}); therefore, any force applied below this frequency will move the end test mass suspension chain together and allow spurious optical noise to be fed back to the interferometer's output, giving rise to scattering arches. 

\subsection{Data for Glitch Analysis}
We obtained raw auxiliary channel data from the 70,000-second (17 hour, 40 minute) period from GPS times 1,259,197,952 to 1,259,267,952 (December 1, 2019, 1:12:14 to 20:38:54 UTC).
During that period, the Omicron glitch detector \cite{Omicron}, which monitors the gravitational-wave strain data for events of excess power, recorded a total 48,879 glitches of \gls{snr} 5 or greater.
The GravitySpy project \cite{GSpy} selects a subset of high-\gls{snr} Omicron glitches that are well-suited to morphological classification through a combination of \gls{cnn}-based machine learning and citizen science. 
Of the 6,515 glitches classified in total by GravitySpy during that period, 5,499 were classified as ``scattered light."\footnote{
Of the 6,515 total GravitySpy glitches, all but 297 were also recorded as glitches by Omicron.}
There were several breaks in lock during the above period; we discarded data that fell outside of a lock segment.

We draw training data from the 50,000-second period from GPS times 1,259,197,952 to 1,259,247,952 and validation and test data respectively from the 10,000-second periods from 1,259,247,952 to 1,259,257,952 and 1,259,257,952 to 1,259,267,952.
There were a total of 3,555, 924, and 1,016 scattered light glitches during the training, validation, and test periods respectively.

We follow the same procedure as \cite{EMU} and \cite{learned_features} to reduce the number of auxiliary channels considered from approximately 250,000 to approximately 40,000 by excluding channels that are constant or only vary in a predictable fashion (e.g., counting time cycles).
We also exclude any auxiliary channels known or suspected to be coupled to the gravitational-wave data stream following the procedure of \cite{EMU}.
We include all remaining channels at all sample rates (a total of 39,147).
We normalize each channel by computing the mean and standard deviation of the raw channel data over the entire training data period; then we subtract the training mean and divide by the standard deviation for all data in the training, validation, and test periods.

Our positive samples are drawn from points in time identified by GravitySpy \cite{GSpy} as scattered light glitches.
As in \cite{EMU} and \cite{learned_features}, our negative samples are drawn randomly from periods where no glitch was identified by the Omicron glitch detector \cite{Omicron} within two seconds.
We select the same number of negative samples as there are positive samples in each dataset.


\section{Results}\label{sec:sl_results}
In this section we describe our findings upon applying the \LF{} model to the scattered light data described in Sec. \ref{sec:sl_data}.
We note that that neither the model nor the authors incorporated any specific prior knowledge of the characteristics of scattered light glitches nor the input auxiliary channels; these results were achieved independently and automatically by the method presented here.

The model achieves an overall 97.1\% accuracy distinguishing between glitches and glitch-free points from our test dataset.
It is able to achieve this level of accuracy after being trained on data from all 39,147 potentially informative auxiliary channels and automatically selecting only 25 of them (Table \ref{tab:sl_scattered_light_EN0NoFC_chans}) as relevant to predict this type of glitch.
For each selected channel, the model also provides a visually interpretable feature for each selected channel, suggesting a signature of behavior in that channel associated with the presence or absence of a scattered light glitch.

These results and their potential interpretations may suggest novel directions for investigation of the origins and causes of scattered light glitches deserving of deeper investigations by instrument science domain experts.

\subsection{Predictive Accuracy}\label{sec:sl_results_predictions}

\begin{figure}[t]
\includegraphics[width=\linewidth]{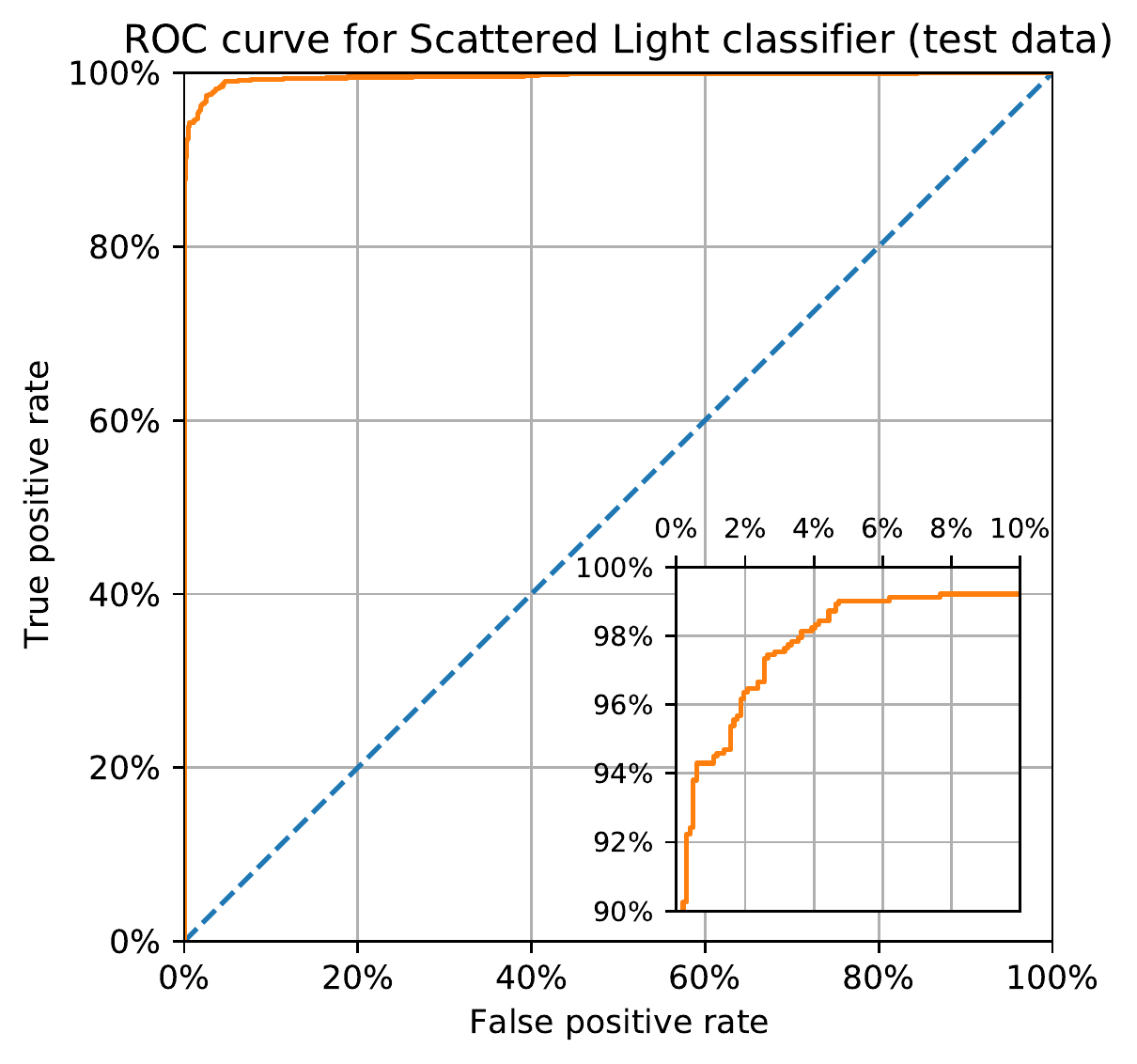}
\caption{ROC curve of \LF{} classifier on scattered light test dataset.}\label{fig:sl_scattered_light_EN0NoFC_ROC_test}
\end{figure}

The best-performing (by loss) model on the scattered light validation dataset achieved a loss of 0.0933, corresponding to an accuracy of 97.0\% (true positive rate 97.4\%, true negative rate 96.6\%).
On the test dataset, it achieved a loss of 0.0866, corresponding to an accuracy of 97.1\% (true positive rate 98.4\%, true negative rate 95.8\%).
An ROC curve for the test dataset is shown in Fig. \ref{fig:sl_scattered_light_EN0NoFC_ROC_test}.

\begin{figure}[t]
\includegraphics[width=\linewidth]{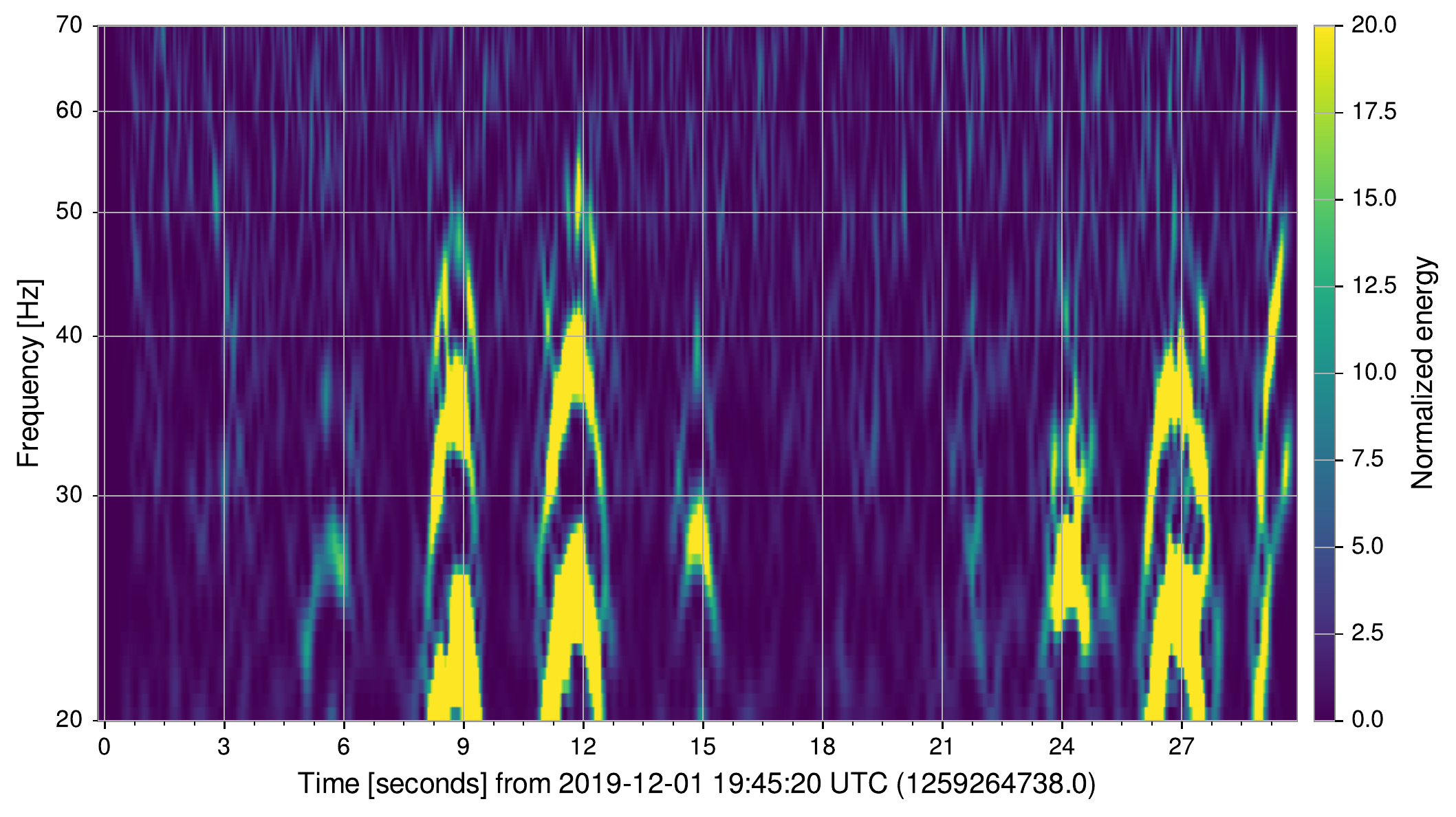}
\caption{Omegagram around GPS time 1259264752.9 (14.9 s on the x-axis), which was classified as a glitch by our model but not by GravitySpy or Omicron.}\label{fig:sl_false_positive}
\end{figure}

Additionally, we manually examined Omegagrams around the 43 instances classified as a glitch by our model but not by GravitySpy or Omicron from our test dataset and found faint but clear signatures of possible scattered light glitches in as many as 30 of them.
If we excluded those 30 from the 1,016 samples labeled ``glitch-free'' in our ground truth, our model's true negative rate would rise from 95.8\% to 98.7\% and overall accuracy would rise from 97.1\% to 98.6\%.
We show an example of one such false positive by our model in Fig. \ref{fig:sl_false_positive}.

For comparison, we also performed an experiment with the same type of model and with data drawn from the same periods as for the scattered light classifier (as described in Sec. \ref{sec:sl_data_scattered_light}) but using all Omicron glitches as the positive class rather than scattered light glitches only.
It achieved a maximum validation accuracy of 82.8\%, indicating that scattered light glitches are significantly easier to classify than general Omicron glitches.

\begin{figure*}[t]
\includegraphics[width=\linewidth]{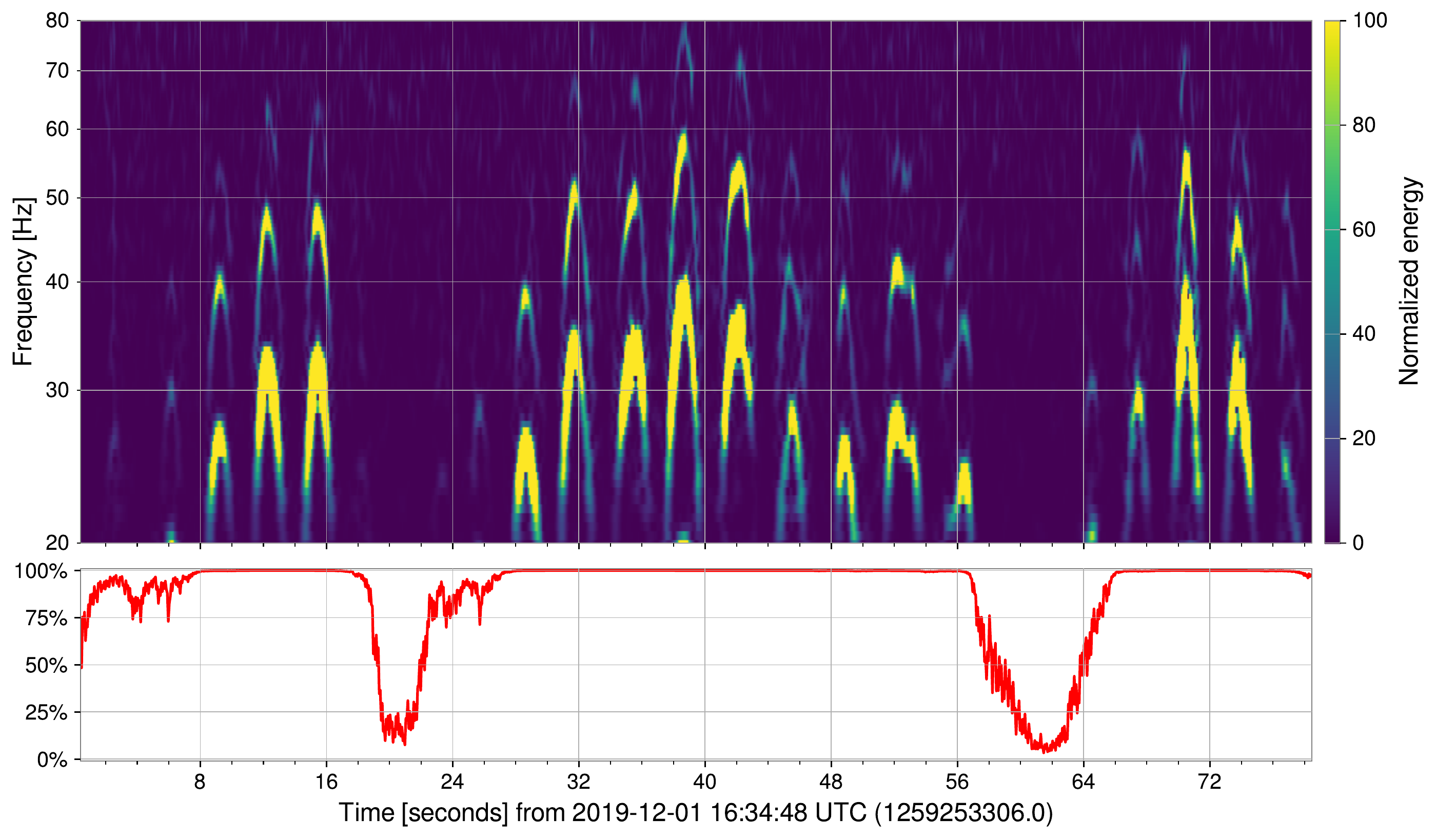}
\caption{Omegagram of several scattered light glitches around GPS time 1,259,253,345 (top) and our model's estimate over time of the probability that there is a glitch (bottom). 
GravitySpy reports multiple scattered light glitches during this period, including a 19.75-second-long one centered at GPS time 1,259,253,341.438 (approximately 35.4 seconds on the x-axis), a 9.5-second-long one centered at 1,259,253,318.188 (12.2 seconds), and several shorter ones from 1,259,253,370.438 to 1,259,253,382.719 (64.4 to 76.7 seconds).}
\label{fig:sl_omegagram1_with_prob}
\end{figure*}

As an example of how the model's output varies over time depending on the presence or absence of a glitch, we fed it a continuous segment of auxiliary channel data around several scattered light glitches that occurred around GPS time 1,259,253,345.
Fig. \ref{fig:sl_omegagram1_with_prob} shows an Omegagram \cite{rollins_thesis, 2004CQGra..21S1809C} of these glitches (top) and our model's estimate of the probability of a glitch at any given time (bottom).
GravitySpy reports a 19.75-second scattered light glitch centered at 1,259,253,341.438 (35 seconds on the x-axis of Fig. \ref{fig:sl_omegagram1_with_prob}), which our model classified as a glitch with very high confidence.
GravitySpy also reports several other shorter scattered light glitches around that time, which are visible in the Omegagram and reflected in our model's output.

\begin{figure}[t]
\includegraphics[width=\linewidth]{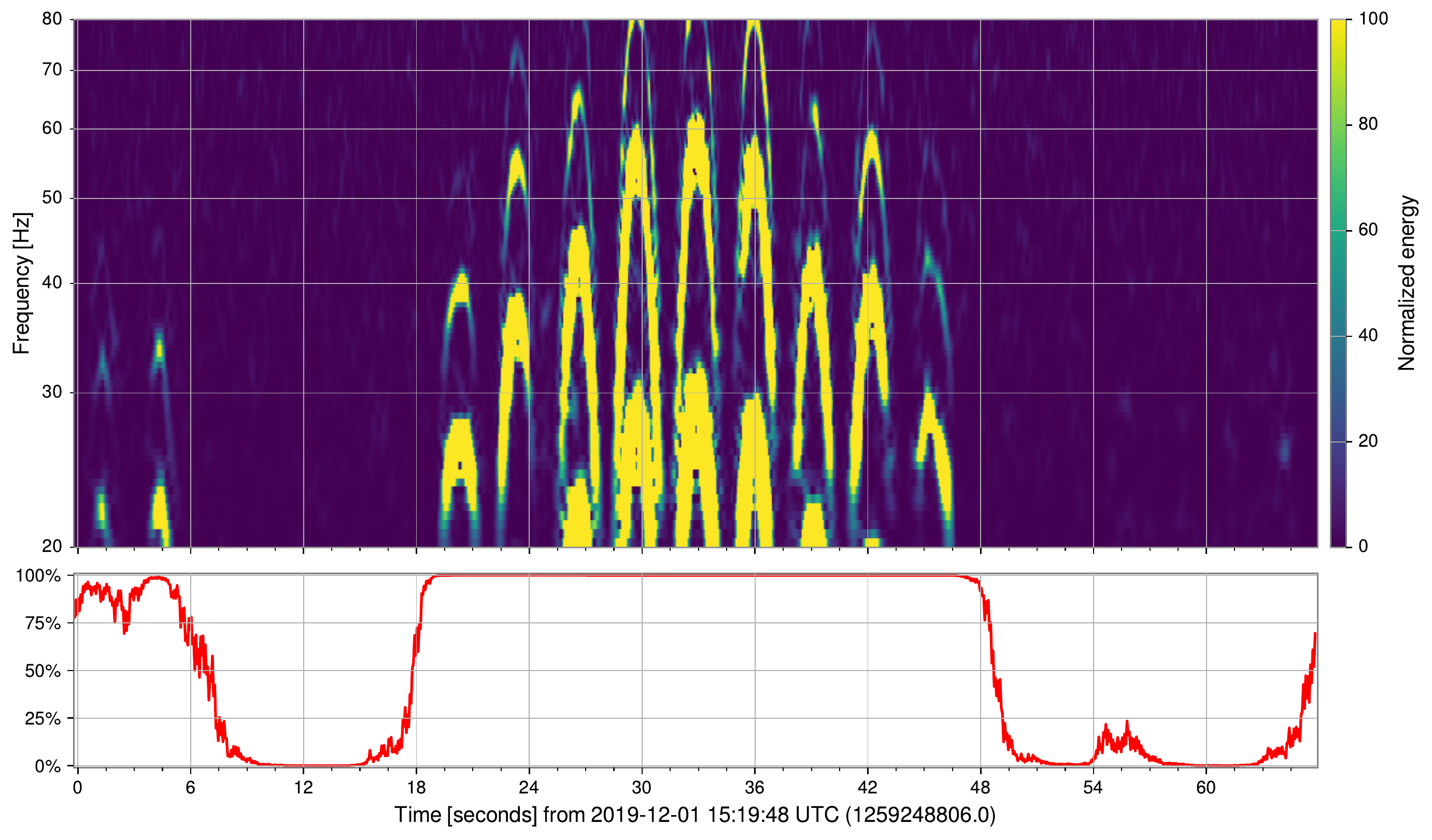}
\caption{Omegagram of several scattered light glitches around GPS time 1,259,248,839 (top) and our model's estimate of the probability that there is a glitch over time (bottom).}\label{fig:sl_omegagram_with_prob_2}
\end{figure}

\begin{figure}[t]
\includegraphics[width=\linewidth]{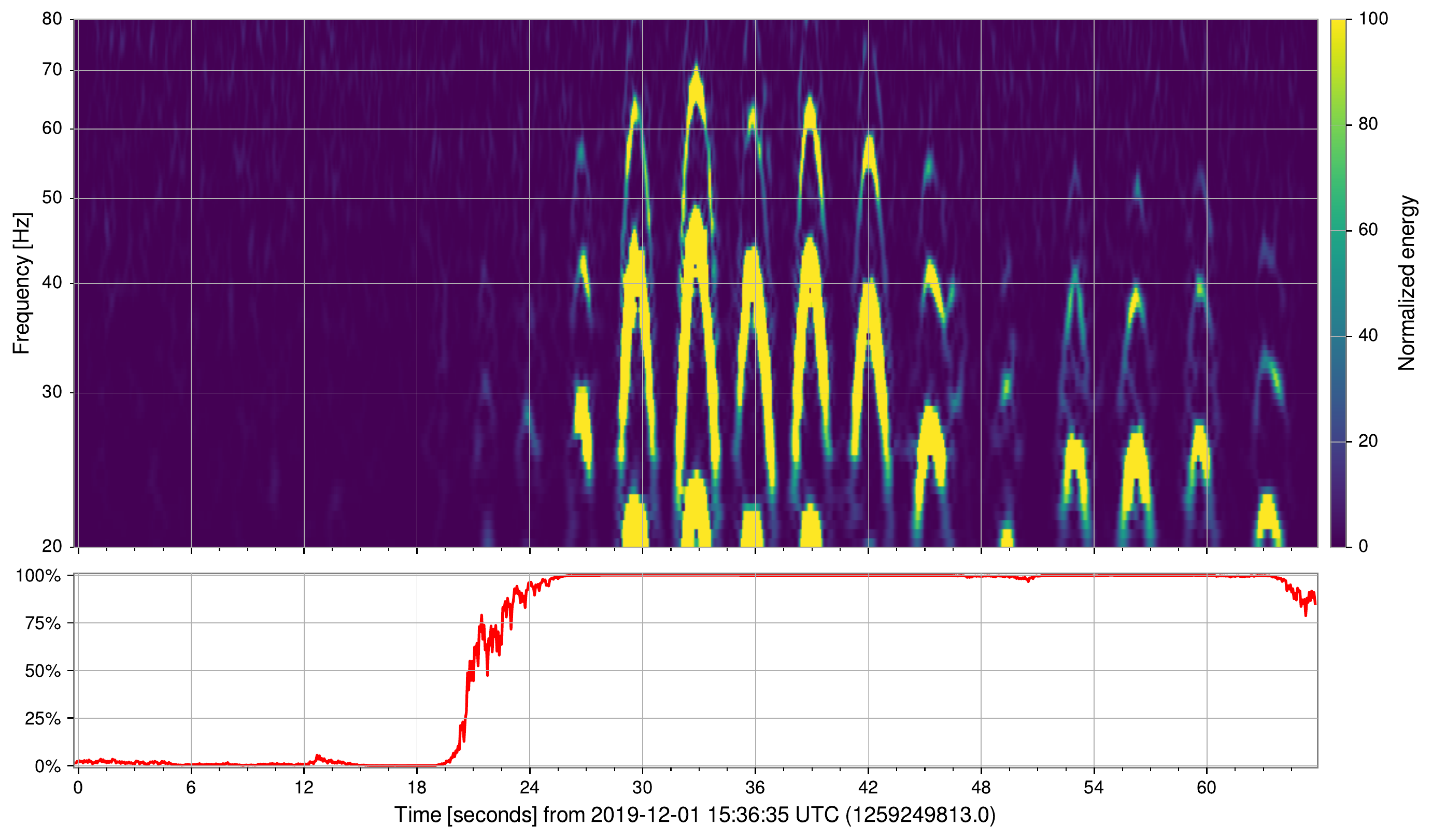}
\caption{Omegagram of several scattered light glitches around GPS time 1,259,249,846 (top) and our model's estimate of the probability that there is a glitch over time (bottom).}\label{fig:sl_omegagram_with_prob_3}
\end{figure}

\begin{figure}[t]
\includegraphics[width=\linewidth]{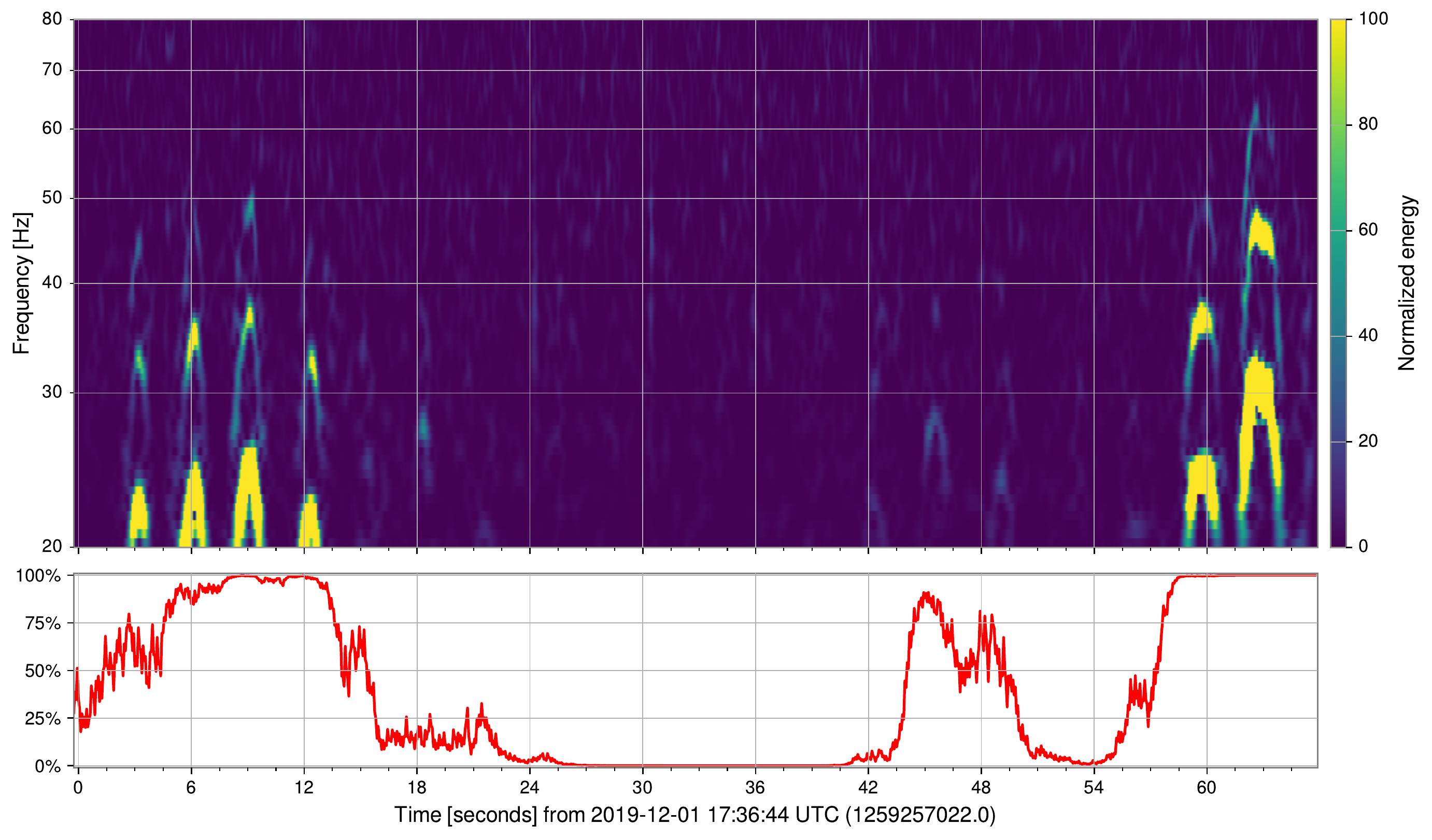}
\caption{Omegagram of a glitch-free period of about 30 seconds around GPS time 1,259,257,055 (top) and our model's estimate of the probability that there is a glitch over time (bottom).}\label{fig:sl_omegagram_with_prob_gf}
\end{figure}

We illustrate the Omegagrams of two more scattered light glitches along with the model's output over the surrounding time period in Figs. \ref{fig:sl_omegagram_with_prob_2} and \ref{fig:sl_omegagram_with_prob_3}.
We illustrate the corresponding raw channel value and correlation for selected channels in Figs. \ref{fig:sl_raw_and_conv_2021-10-13_2} and \ref{fig:sl_raw_and_conv_2021-10-13_3} (see Sec. \ref{sec:sl_results_correlations}).
In Fig. \ref{fig:sl_omegagram_with_prob_gf}, we show the Omegagram of a glitch-free point along with the model's output over the surrounding time period; in Fig. \ref{fig:sl_raw_and_conv_2021-10-13_gf}, we also show the corresponding raw channel value and correlation for the same channels as shown in Fig. \ref{fig:sl_raw_and_conv_2021-10-13}.

\subsection{Selected Channels}\label{sec:sl_results_channels}

\begin{table*}[t]
    \centering
    \begin{tabular}{|c|c|c|}
        \hline
         \bf Filter    & \bf Channel name & \bf Frequency \\
         \bf magnitude & \bf              & \bf           \\
         \hline \hline
            0.258 & L1:SUS-ETMX\_L2\_WD\_OSEMAC\_UL\_RMSMON & 16 Hz \\
            \hline
            0.180 & L1:SUS-ETMX\_L2\_WD\_OSEMAC\_UR\_RMSMON & 16 Hz \\
            \hline
            0.117 & L1:SUS-ETMX\_L2\_WD\_OSEMAC\_LR\_RMSMON & 16 Hz \\
            \hline
            0.112 & L1:SUS-ETMX\_L2\_WD\_OSEMAC\_LL\_RMSMON & 16 Hz \\
            \hline
            0.093 & L1:SUS-SR2\_M1\_WD\_OSEMAC\_T2\_RMSMON & 16 Hz \\
            \hline
            0.066 & L1:SUS-SR2\_M2\_RMSIMON\_LR\_OUT16 & 16 Hz \\
            \hline
            0.056 & L1:SUS-ETMX\_L1\_WD\_OSEMAC\_UL\_RMSMON & 16 Hz \\
            \hline
            0.054 & L1:SUS-ITMY\_L3\_OPLEV\_BLRMS\_S\_100M\_300M & 16 Hz \\
            \hline
            0.050 & L1:SUS-ETMX\_R0\_WD\_OSEMAC\_RT\_RMSMON & 16 Hz \\
            \hline
            0.046 & L1:SUS-ITMX\_M0\_WD\_OSEMAC\_SD\_RMSMON & 16 Hz \\
            \hline
            0.043 & L1:SUS-ITMX\_L3\_OPLEV\_BLRMS\_P\_1\_3 & 16 Hz \\
            \hline
            0.038 & L1:SUS-ITMY\_L3\_ISCINF\_L\_IN1\_DQ & 16384 Hz \\
            \hline
            0.031 & L1:IOP-SUS\_OMC\_WD\_OSEM4\_RMSOUT & 16 Hz \\
            \hline
            0.031 & L1:ISI-BS\_ST2\_BLND\_BLRMS\_X\_300M\_1 & 16 Hz \\
            \hline
            0.029 & L1:SUS-ETMX\_L1\_WD\_OSEMAC\_UR\_RMSMON & 16 Hz \\
            \hline
            0.022 & L1:CAL-PCALY\_IRIGB\_DQ & 16384 Hz \\
            \hline
            0.014 & L0:FMC-CS\_AHU1\_FAN1\_VS & 16 Hz \\
            \hline
            0.011 & L1:SUS-ETMY\_PI\_DOWNCONV\_DC7\_SIG\_INMON & 16 Hz \\
            \hline
            0.011 & L1:SUS-ETMY\_PI\_DOWNCONV\_DC4\_SIG\_INMON & 16 Hz \\
            \hline
            0.008 & L1:HPI-HAM6\_BLRMS\_X\_30\_100 & 16 Hz \\
            \hline
            0.007 & L1:ASC-AS\_B\_RF72\_I\_SUM\_OUT16 & 16 Hz \\
            \hline
            0.006 & L1:SUS-ETMX\_M0\_WD\_OSEMAC\_F2\_RMSMON & 16 Hz \\
            \hline
            0.002 & L1:ASC-AS\_B\_RF72\_I\_SUM\_INMON & 16 Hz \\
            \hline
            0.002 & L1:ASC-AS\_B\_RF72\_I\_SUM\_OUTPUT & 16 Hz \\
            \hline
            0.000 & L1:OAF-STS\_SEN2ACT\_1\_1\_IN1\_DQ & 256 Hz \\
            \hline
    \end{tabular}
    \caption{Channels with nonzero filters and their associated magnitudes from the \LF{} model trained on the scattered light dataset.}
    \label{tab:sl_scattered_light_EN0NoFC_chans}
\end{table*}

The nonzero channels from the best-performing scattered light model and the magnitudes of their associated learned filters are listed in Table \ref{tab:sl_scattered_light_EN0NoFC_chans}.
Only 25 of the 39,147 channels have associated filters with magnitude greater than zero.

The magnitude of the filters also provides an indication of the relative importance of each channel to the model’s predictions (intuitively, when correlated with a higher-magnitude filter, an input data segment will contribute more heavily to the sum and resulting probability estimate than the same segment correlated with a lower-magnitude filter) \cite{learned_features}.
Among those 25, the magnitudes decay rapidly, suggesting the model's predictions are dominated by the contributions of even fewer channels.

Many of the channels in question are related to \gls{llo}'s suspension system (SUS), test masses (ETMX, ITMX), and sensors (OSEM, OPLEV).
The details of these subsystems are described in ~\citep{2015CQGra..32r5003M,2016SPIE.9960E..09S,2016RScI...87i4502R}.

\subsection{Learned Filters}\label{sec:sl_results_filters}
\begin{figure*}[t]
\centerline{
            \includegraphics[width=.31\linewidth]{figures/filters_2021-10-13/ch16-24795.pdf}
            \includegraphics[width=.31\linewidth]{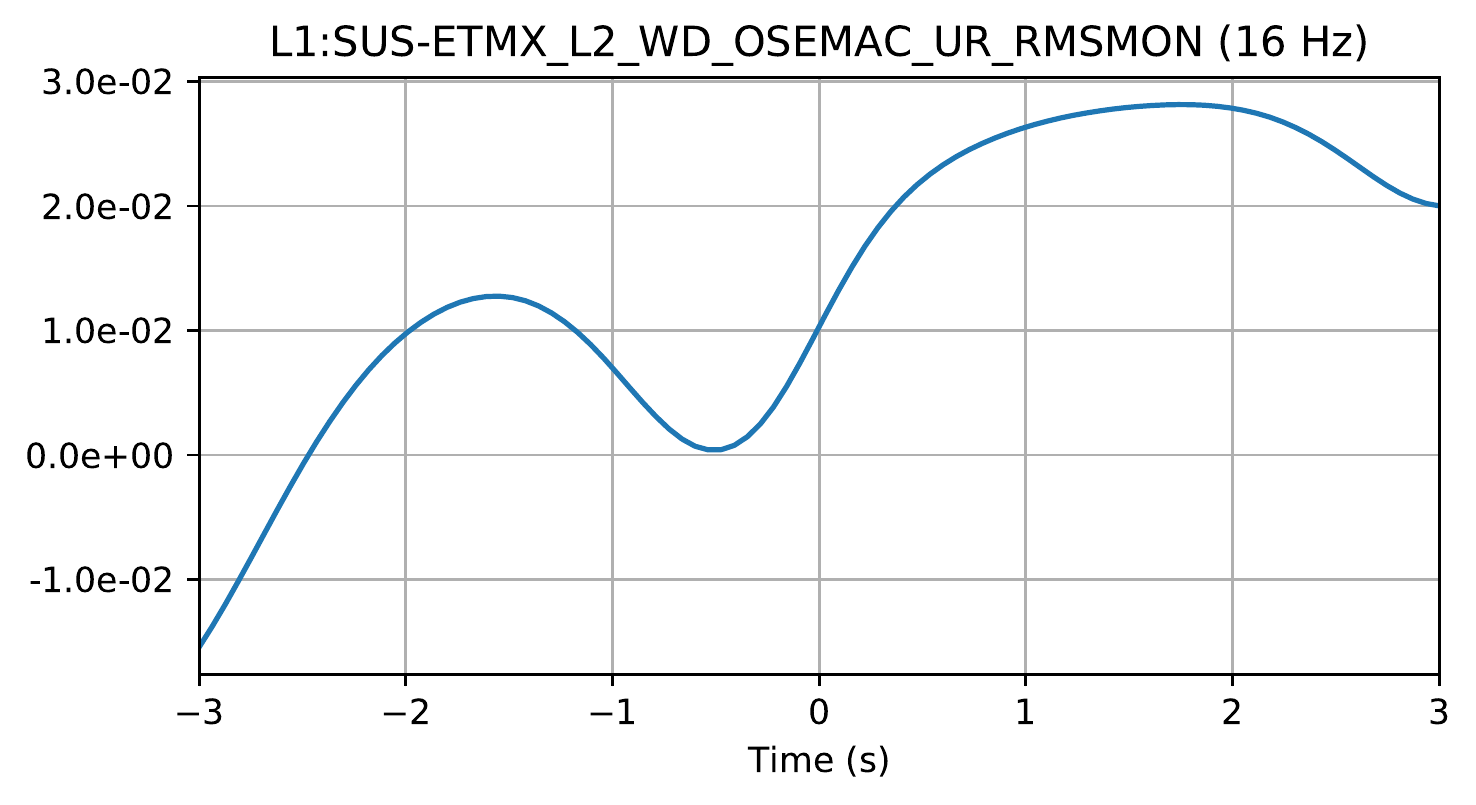}
            \includegraphics[width=.31\linewidth]{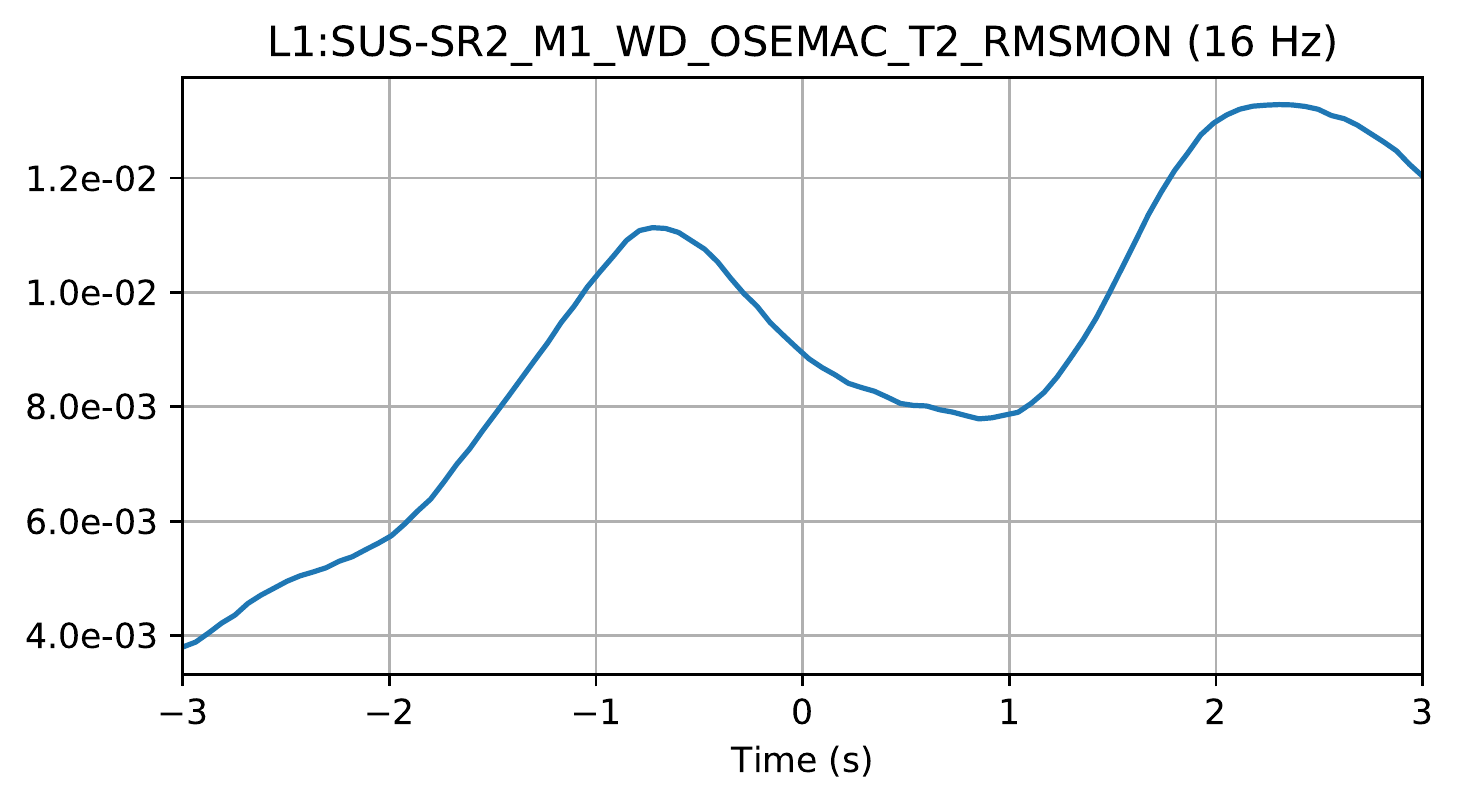}
            }
\centerline{
            \includegraphics[width=.31\linewidth]{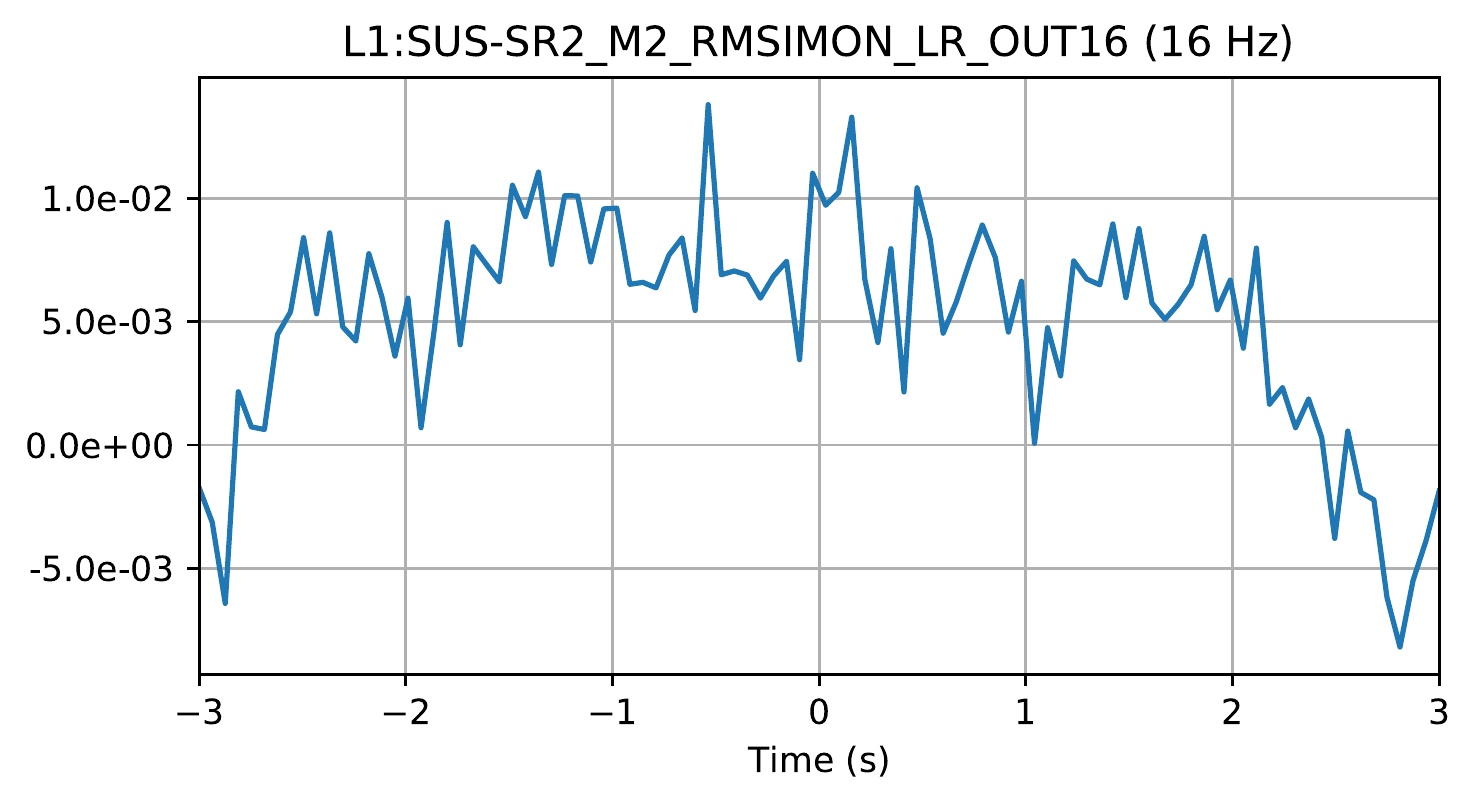}
            \includegraphics[width=.31\linewidth]{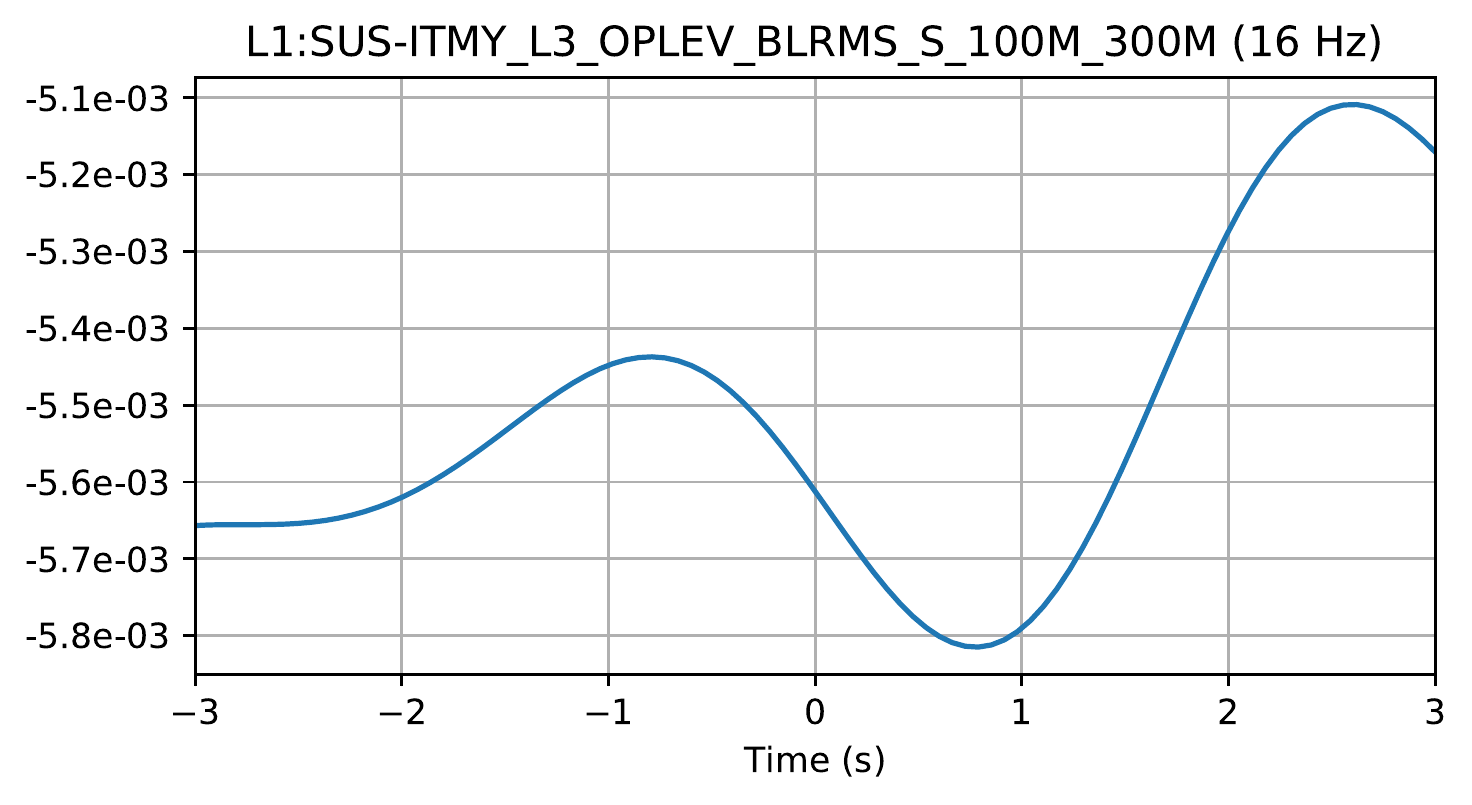}
            \includegraphics[width=.31\linewidth]{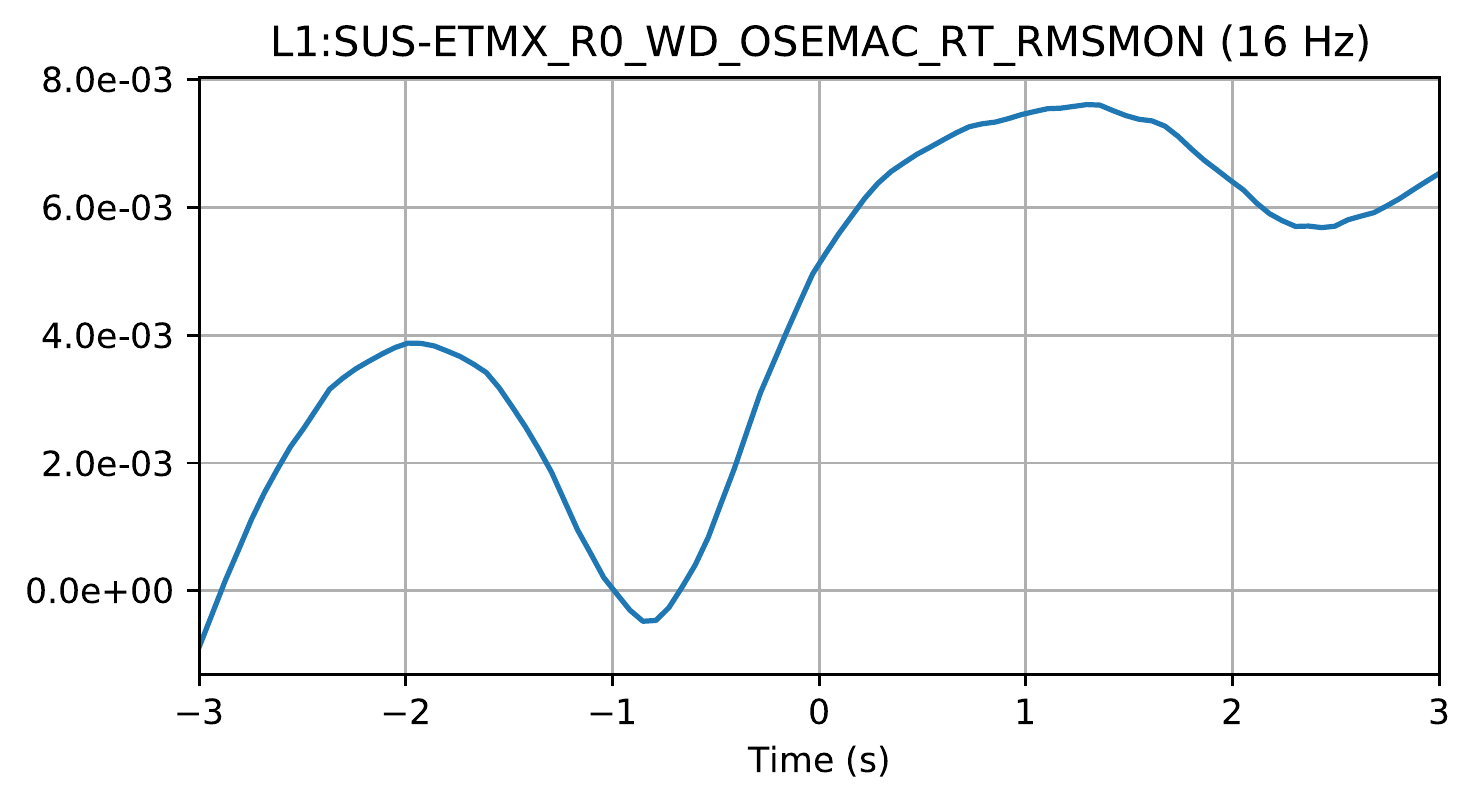}
            }
\centerline{
            \includegraphics[width=.31\linewidth]{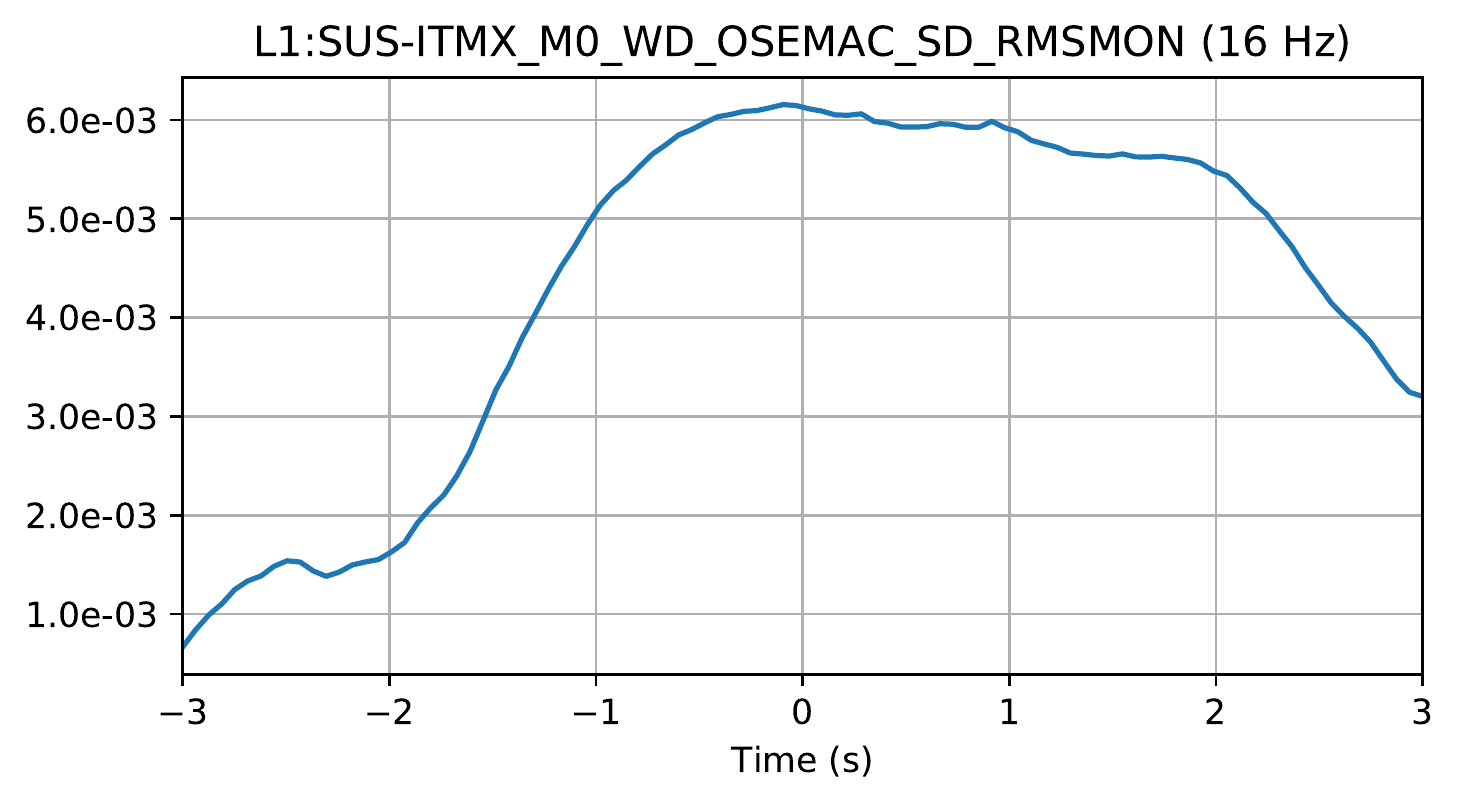}
            \includegraphics[width=.31\linewidth]{figures/filters_2021-10-13/ch16-27345.pdf}
            \includegraphics[width=.31\linewidth]{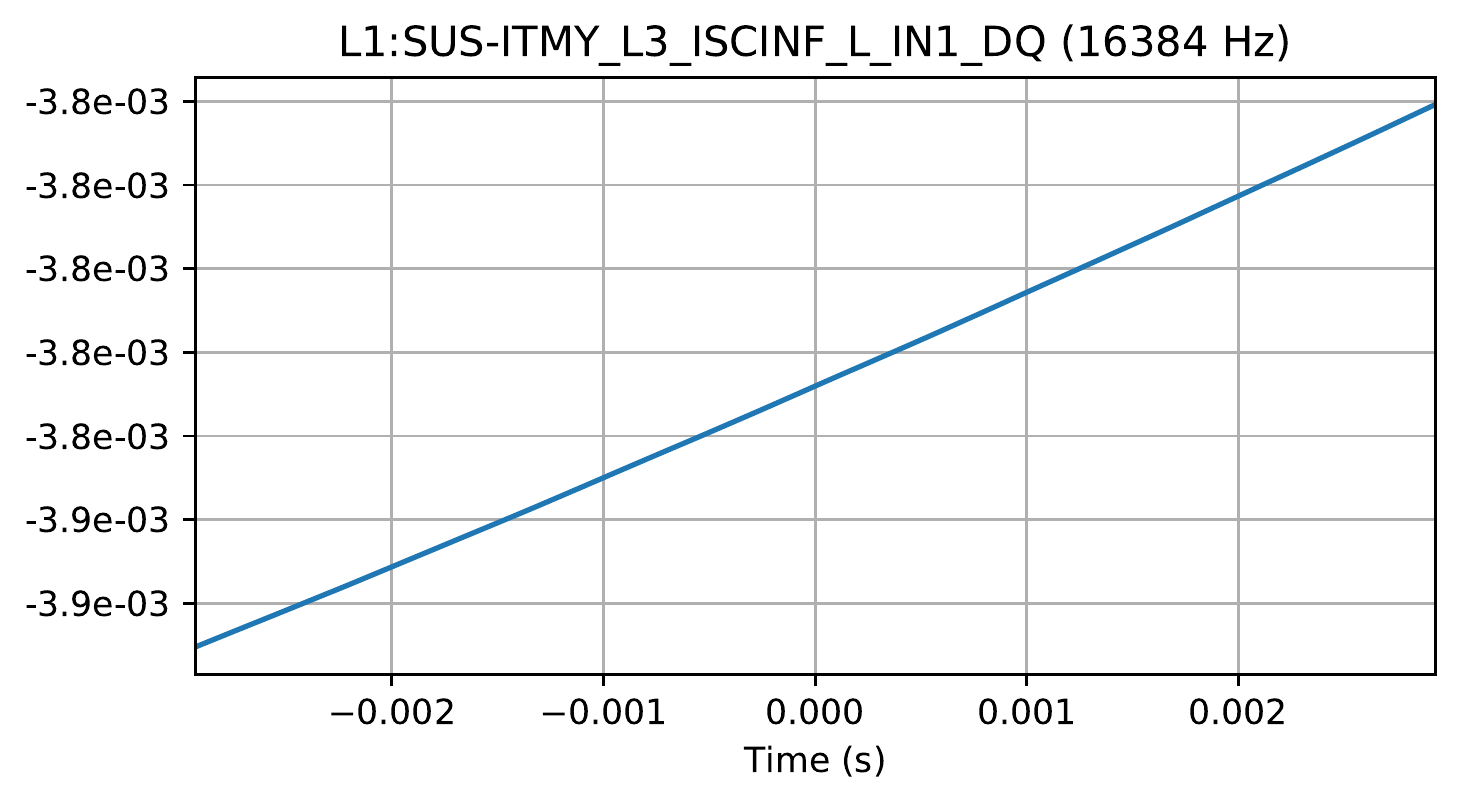}
            }
\centerline{
            \includegraphics[width=.31\linewidth]{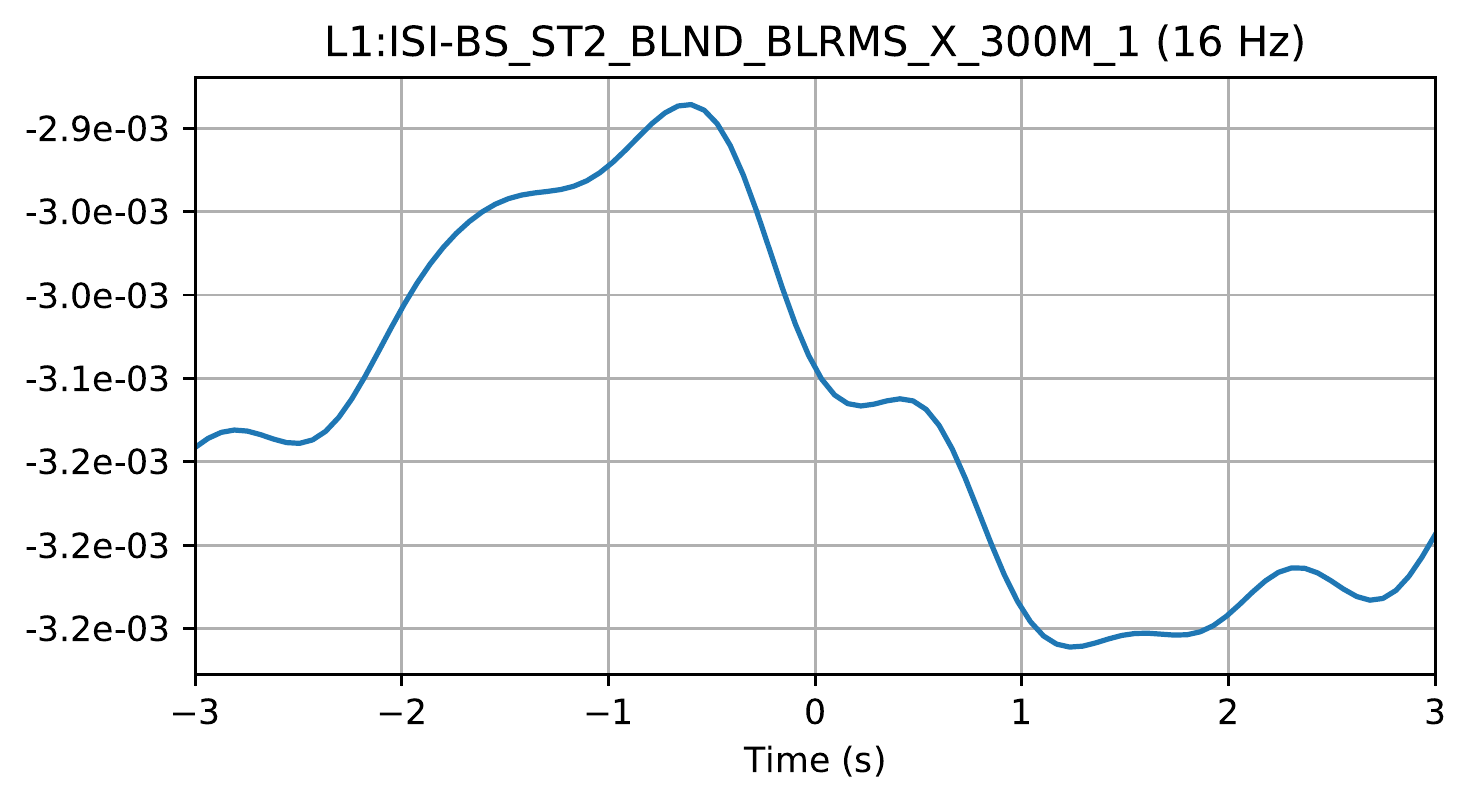}
            \includegraphics[width=.31\linewidth]{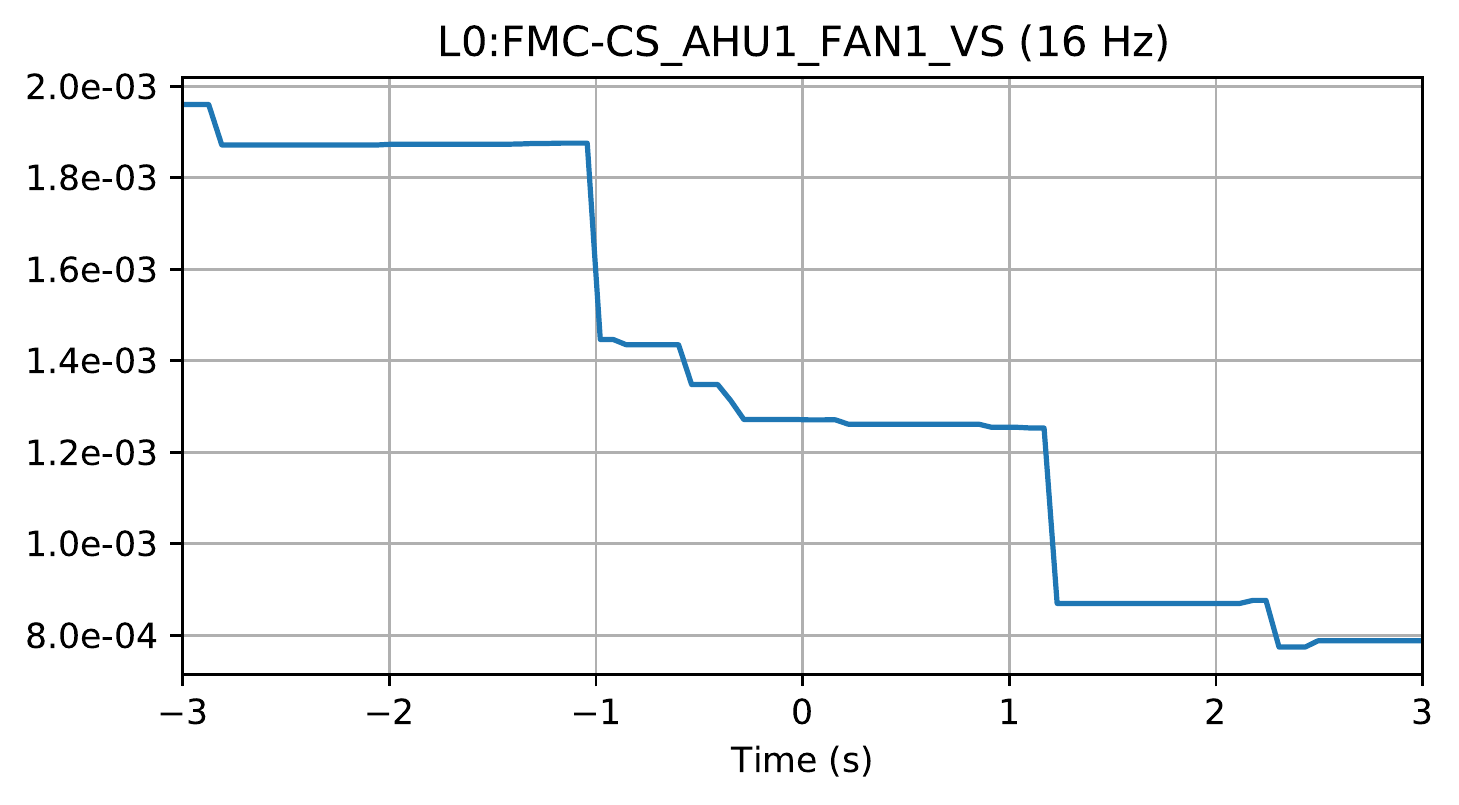}
            \includegraphics[width=.31\linewidth]{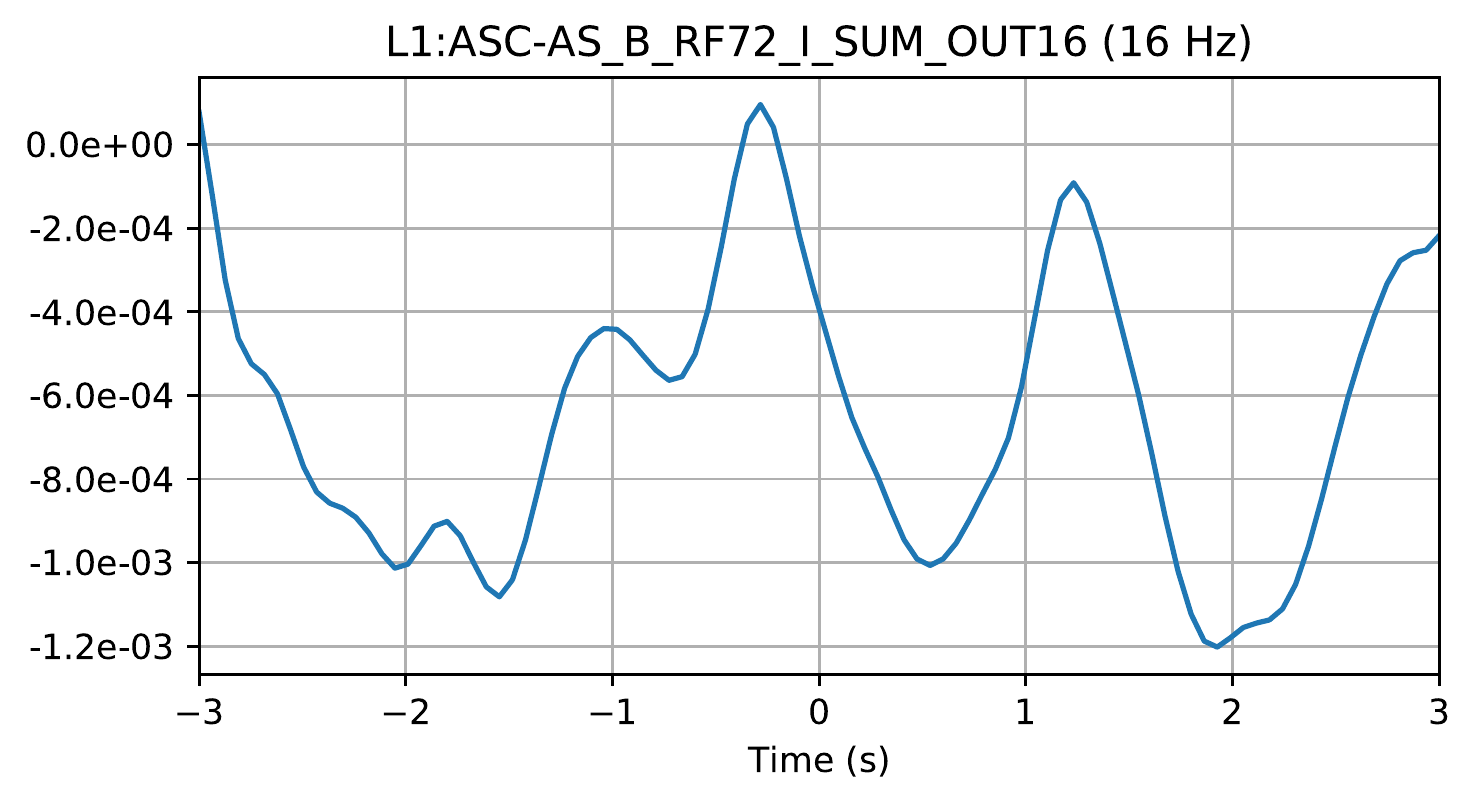}
            }
\caption{Twelve of the features learned by the best-performing \LF{} model on the scattered light dataset.
The filters shown are ordered by magnitude (see Table \ref{tab:sl_scattered_light_EN0NoFC_chans}), although we omit some channels for illustration (for example, all of the first four highest-magnitude filters are nearly identical in shape).
The x-axes correspond to filter length in seconds; the y-axes show the magnitude of the filter over time.
Intuitively, a higher overall magnitude indicates the associated channel is more important to the model's decisions: when cross-correlated with a higher-magnitude filter, an input data segment will contribute more heavily to the sum and resulting probability estimate than the same segment cross-correlated with a lower-magnitude filter.}
\label{fig:sl_scattered_light_EN0NoFC_chans}
\end{figure*}

Several of the learned filters are illustrated in Fig. \ref{fig:sl_scattered_light_EN0NoFC_chans}.
We leave detailed interpretive analysis and instrument hardware\textendash related detective work regarding the channels and features themselves to future work.
However, we qualitatively observe that many of the channels have a similar rising oscillatory shape.
All four of the highest-magnitude channels listed in Table \ref{tab:sl_scattered_light_EN0NoFC_chans}, the first two of which are shown in the top left and top center of Fig. \ref{fig:sl_example_filters}, are nearly identical in shape, likely indicating that they are sensitive to the same or similar phenomena; this is also unsurprising given the similarity of their names.
We also note that the period of oscillation in the features displaying that pattern is of a similar (3-4 second) duration to the individual arches characteristic of scattered light glitches, perhaps suggesting that these channels are sensitive to the same phenomena that result in the appearance of this type of glitch in the strain.

\subsection{Channel Importance for Individual Glitch Classification}\label{sec:sl_results_correlations}
\begin{figure*}[t]
\centerline{
            \includegraphics[width=.31\linewidth]{figures/raw_and_conv_2021-10-13/1259253345.438_ch16-24795.pdf}
            \includegraphics[width=.31\linewidth]{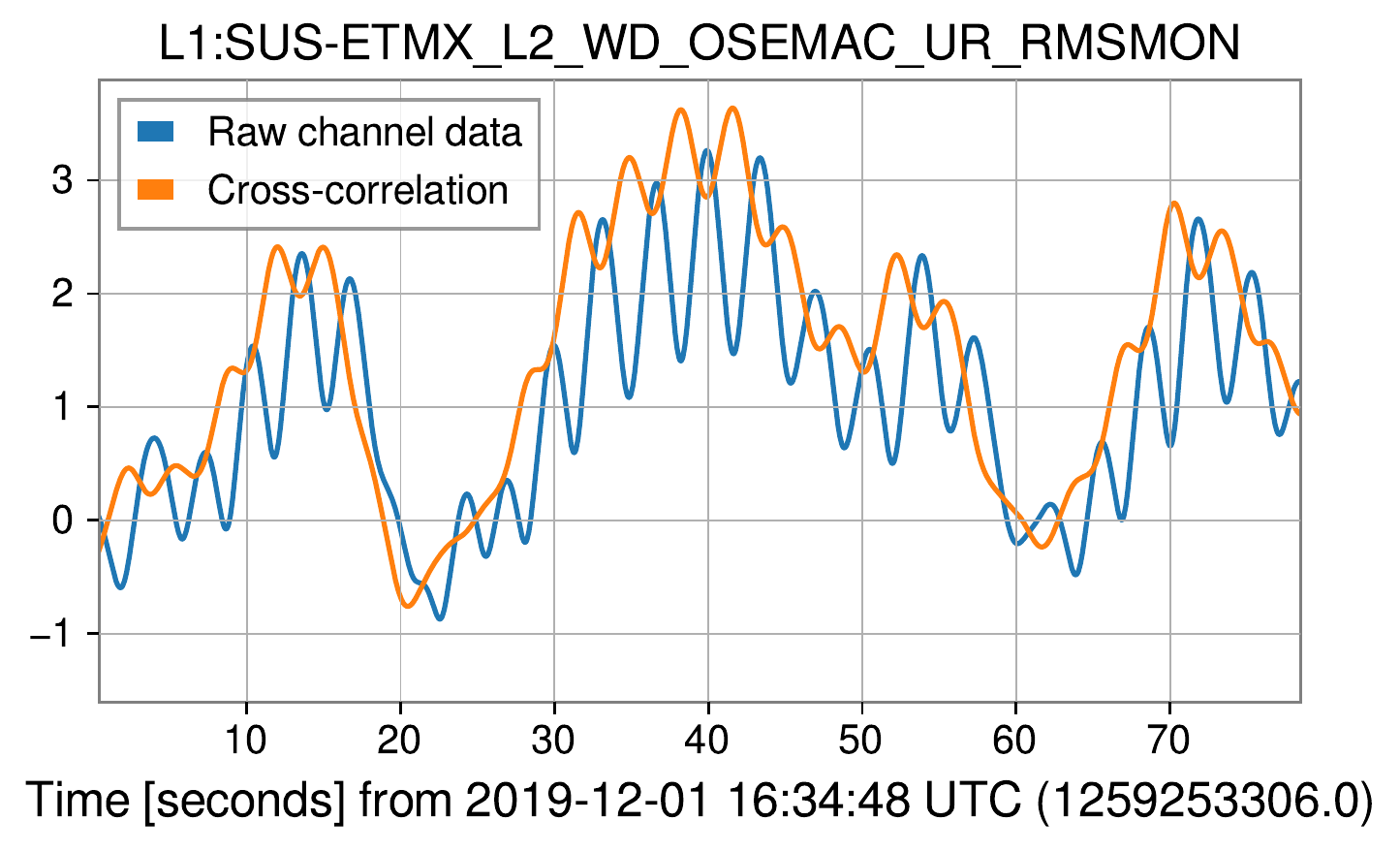}
            \includegraphics[width=.31\linewidth]{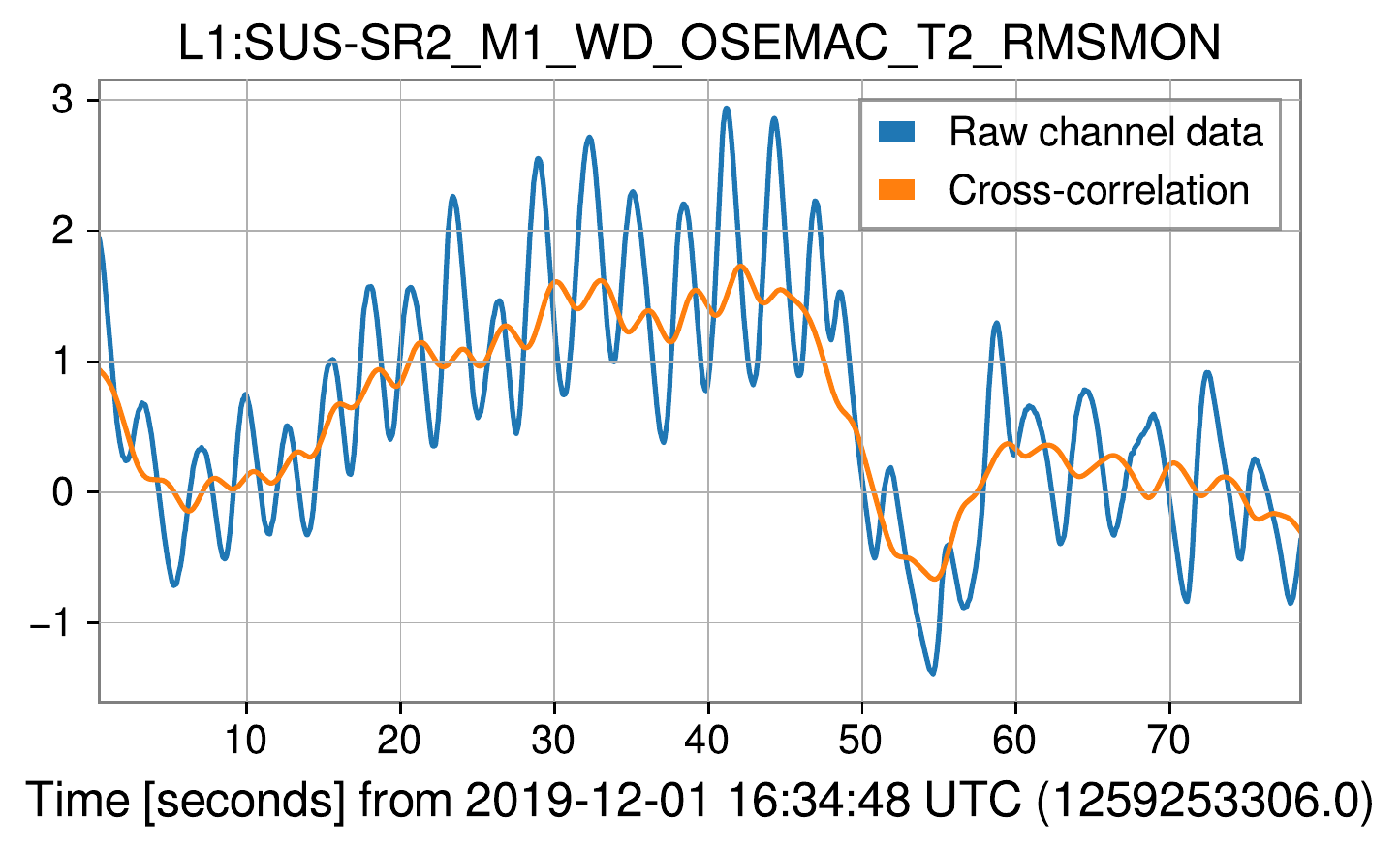}
            }
\centerline{
            \includegraphics[width=.31\linewidth]{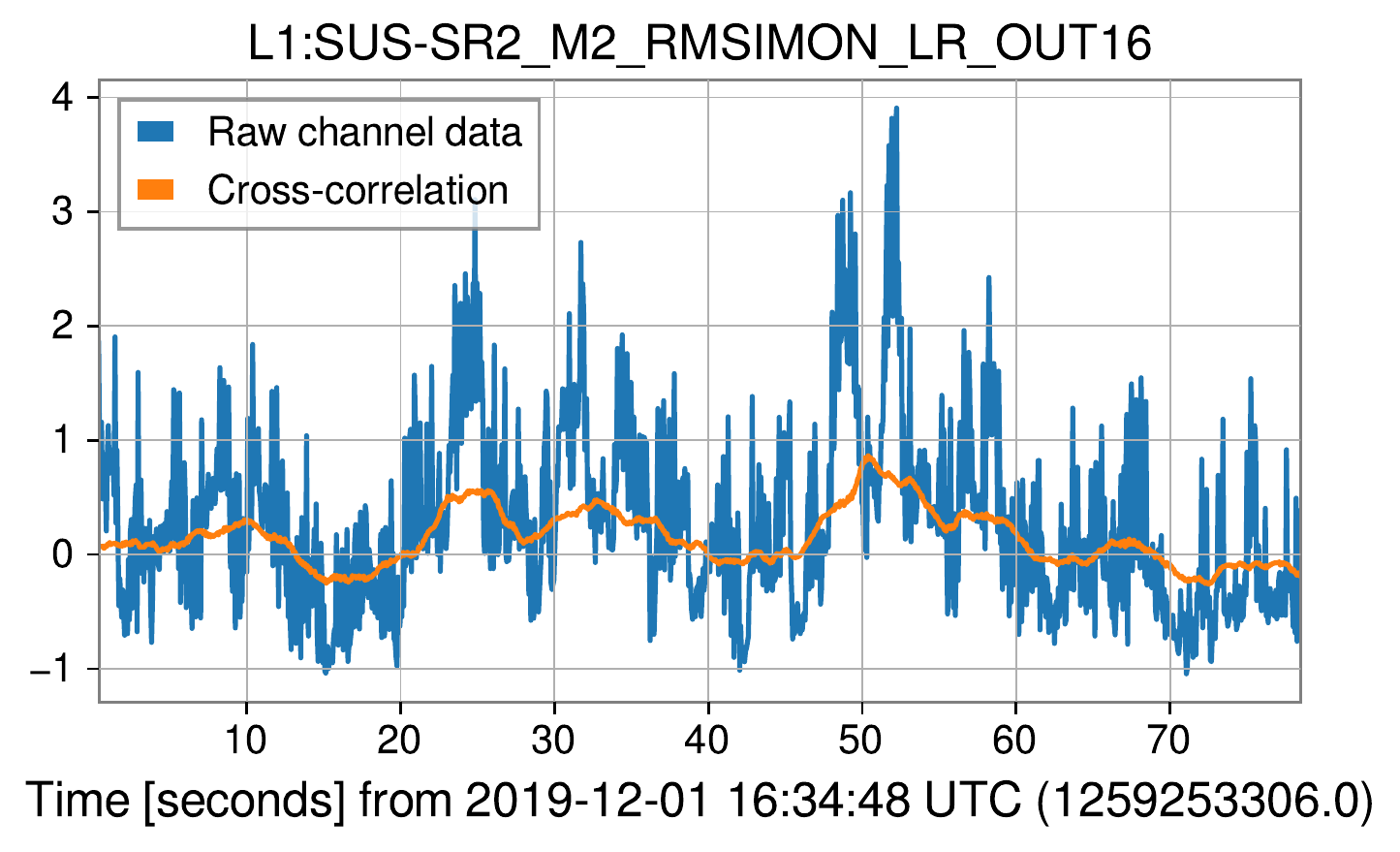}
            \includegraphics[width=.31\linewidth]{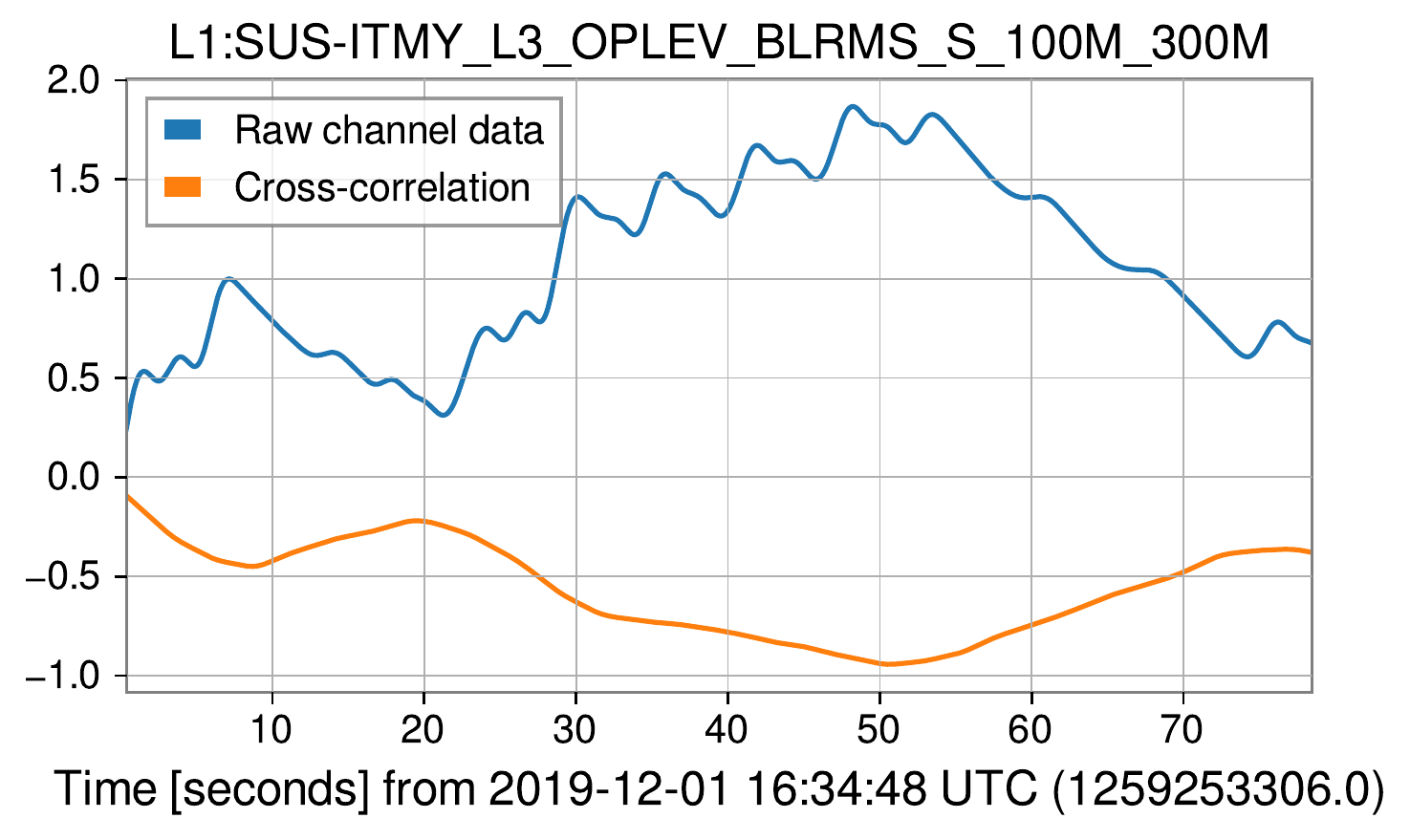}
            \includegraphics[width=.31\linewidth]{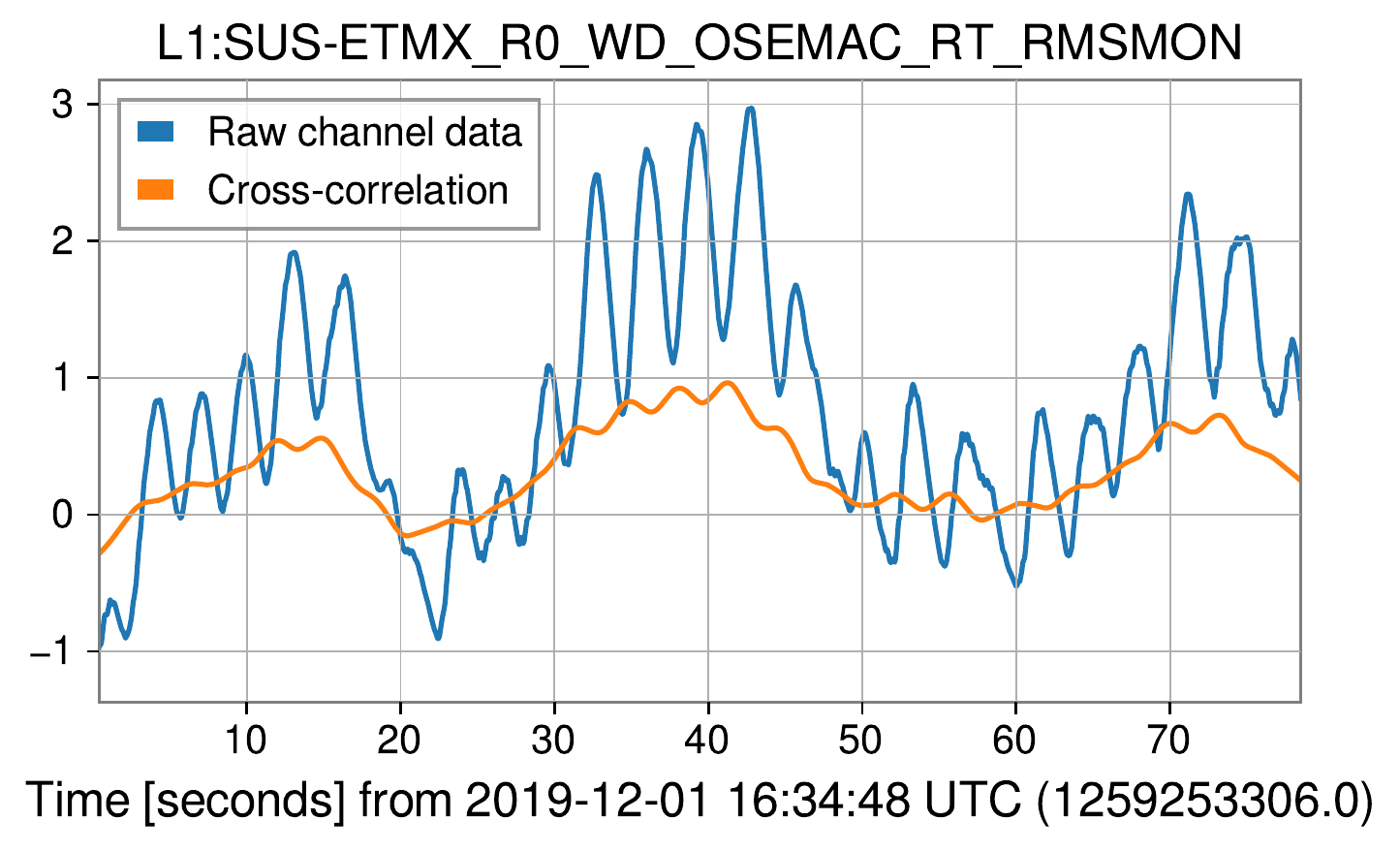}
            }
\centerline{
            \includegraphics[width=.31\linewidth]{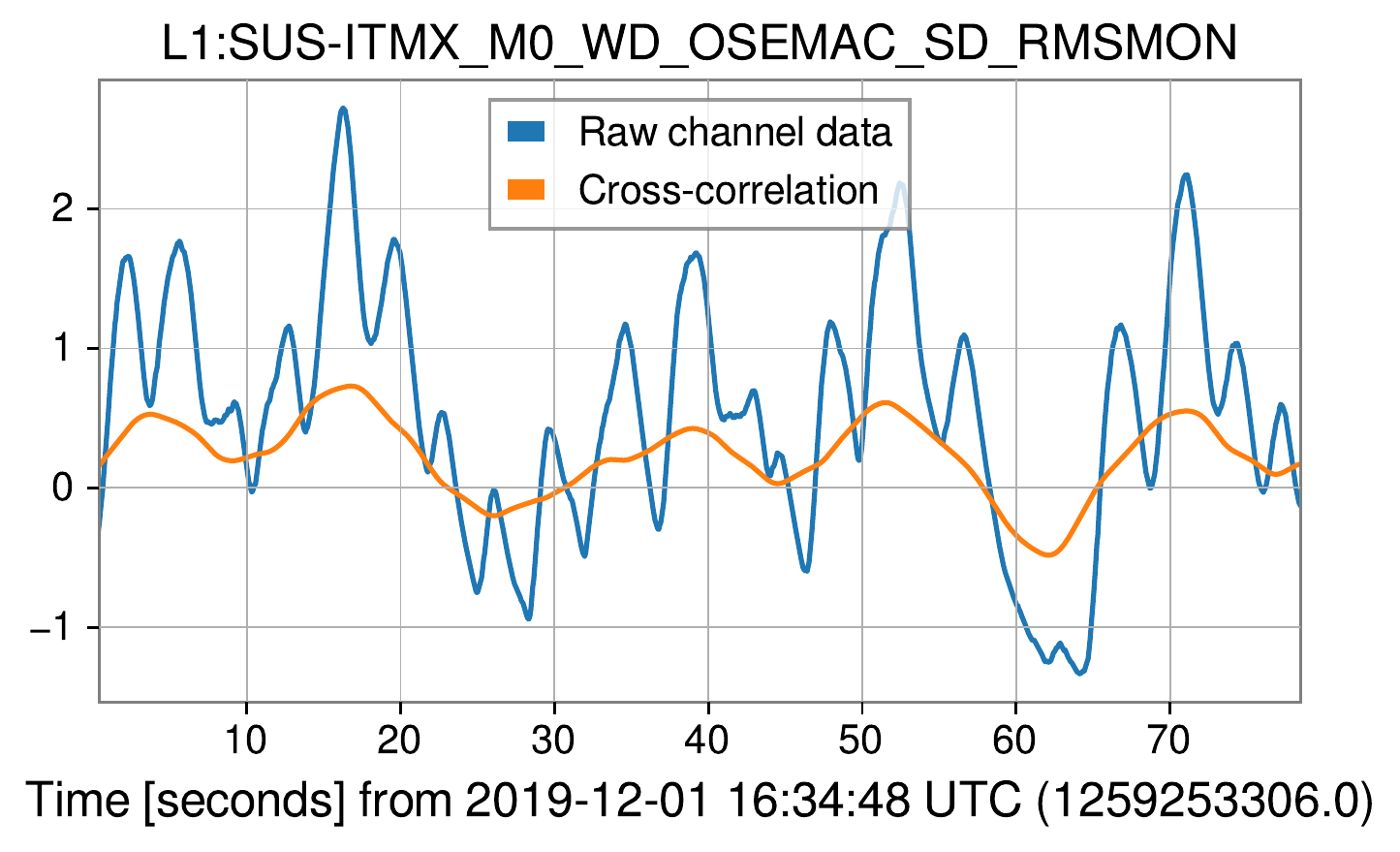}
            \includegraphics[width=.31\linewidth]{figures/raw_and_conv_2021-10-13/1259253345.438_ch16-27345.pdf}
            \includegraphics[width=.31\linewidth]{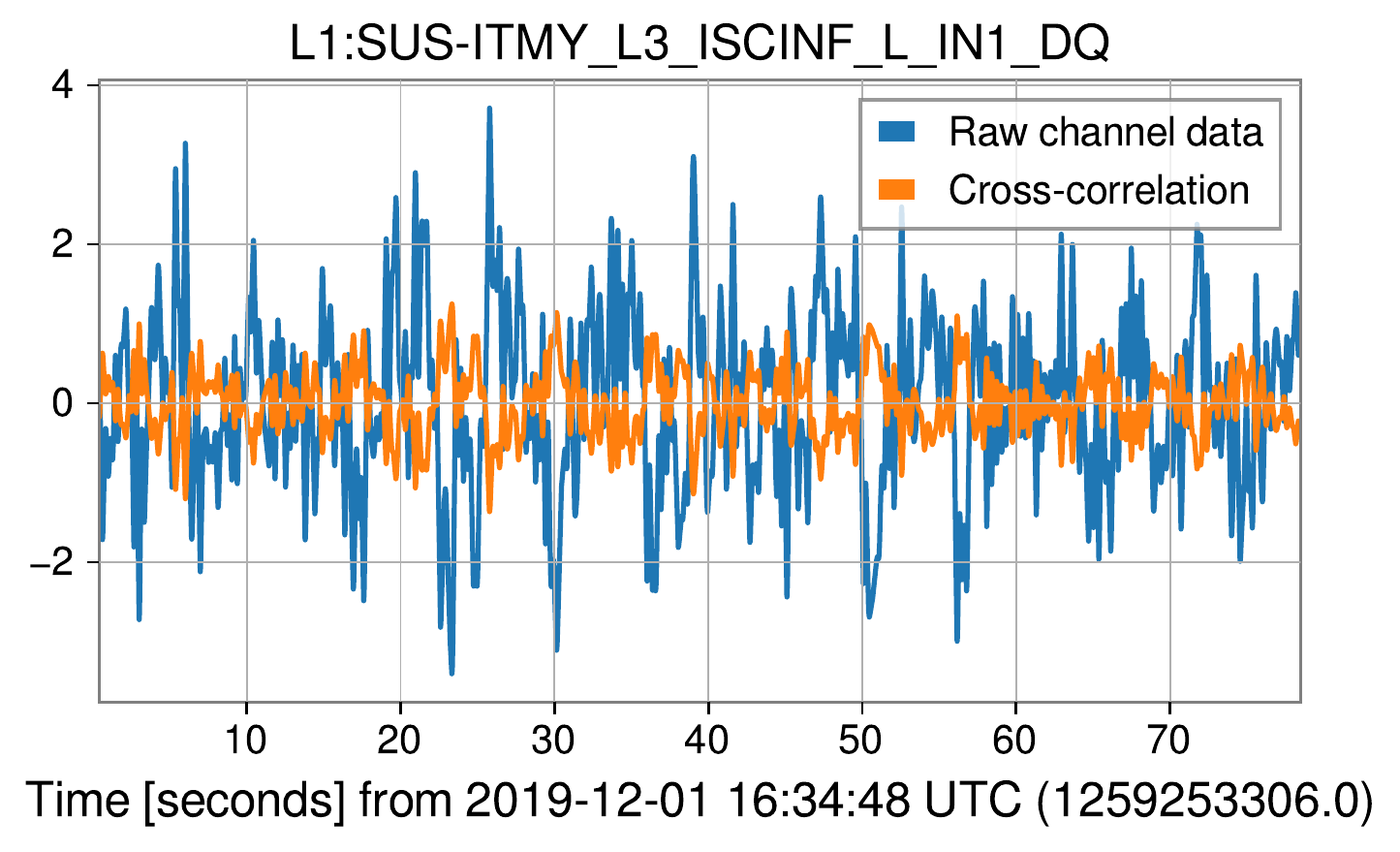}
            }
\centerline{
            \includegraphics[width=.31\linewidth]{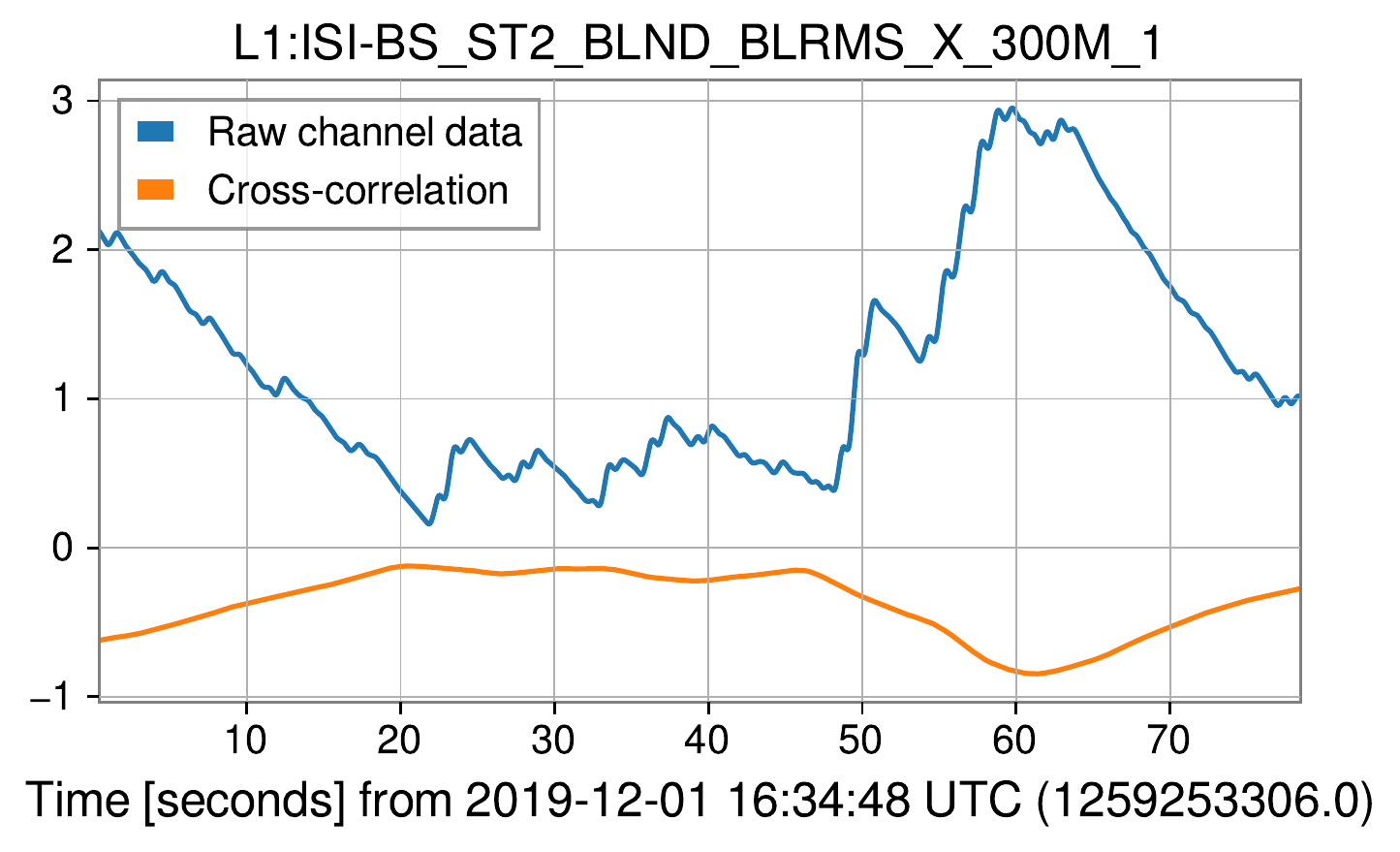}
            \includegraphics[width=.31\linewidth]{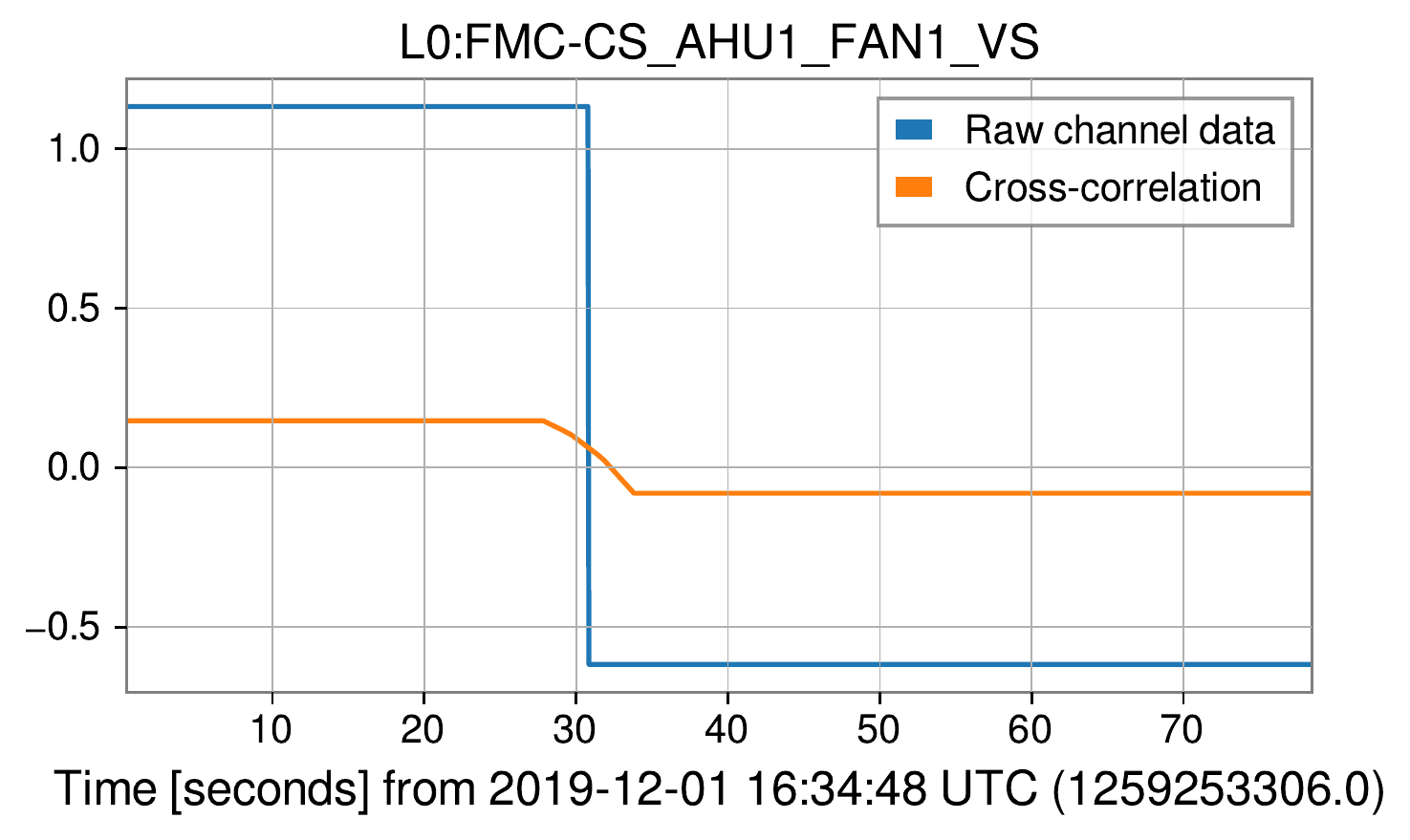}
            \includegraphics[width=.31\linewidth]{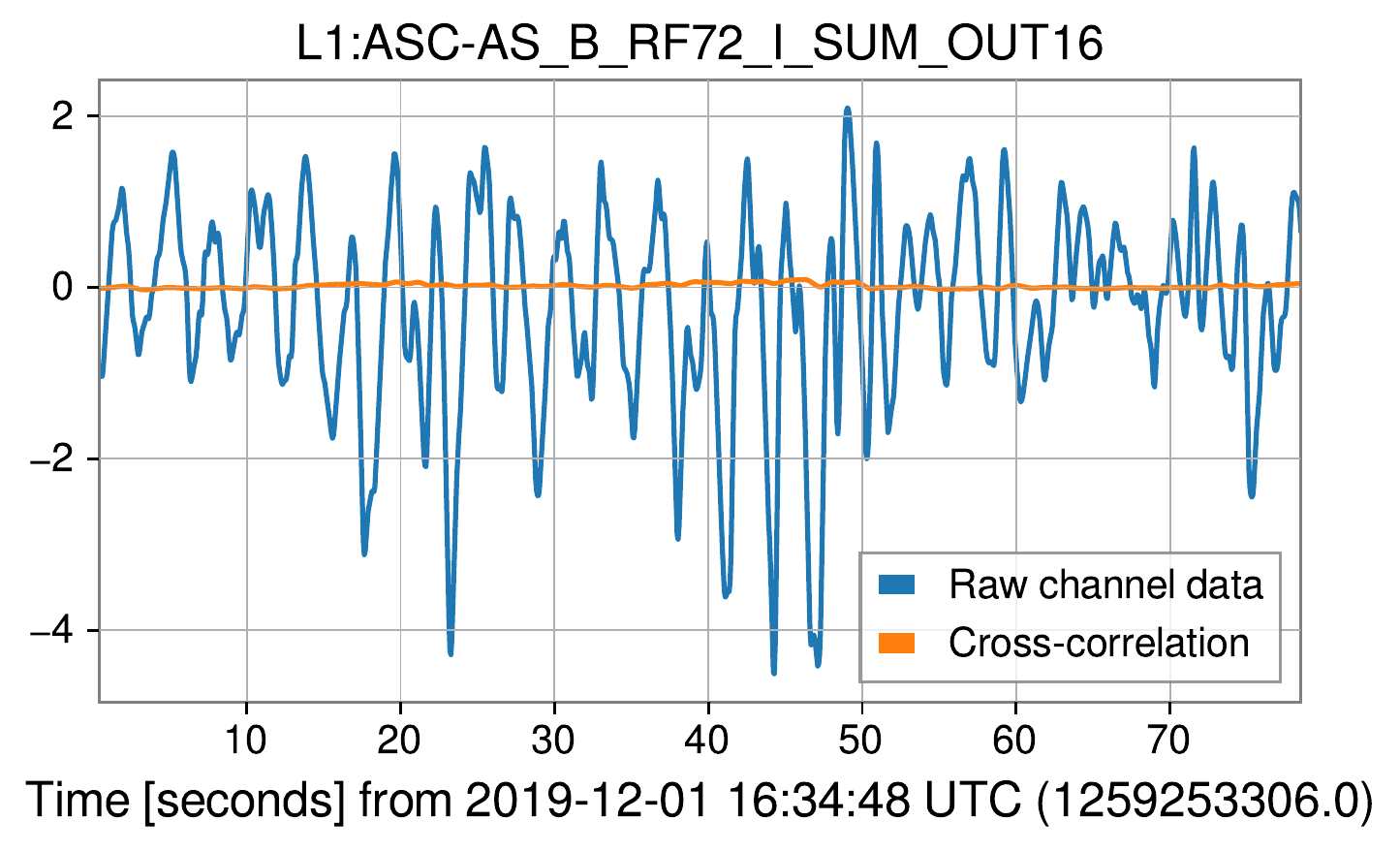}
            }
\caption{Raw channel data (blue) from the same time period illustrated in Fig. \ref{fig:sl_omegagram1_with_prob}, along with (orange) the cross-correlation of the channel and the associated learned filter (i.e., its contribution to the classifier's result).
Note the oscillations in the "L1:SUS-ETMX" channels with peaks spaced approximately 3.5 seconds apart, similar to the spacing between the peaks of the characteristic scattered light arches in Fig. \ref{fig:sl_omegagram1_with_prob}.
}\label{fig:sl_raw_and_conv_2021-10-13}
\end{figure*}

Fig. \ref{fig:sl_raw_and_conv_2021-10-13} shows the normalized value of several auxiliary channels during that span (blue) as well as the correlation between the channel and the model's learned filter (orange) at each time sample during that span.
A positive value for the correlation at a given time indicates the channel is contributing to the model's classification of the time as containing a glitch (the higher the value the more heavily weighted the contribution), and a negative value indicates the channel is indicating no glitch is present.

\begin{figure*}[t]
\centerline{
            \includegraphics[width=.31\linewidth]{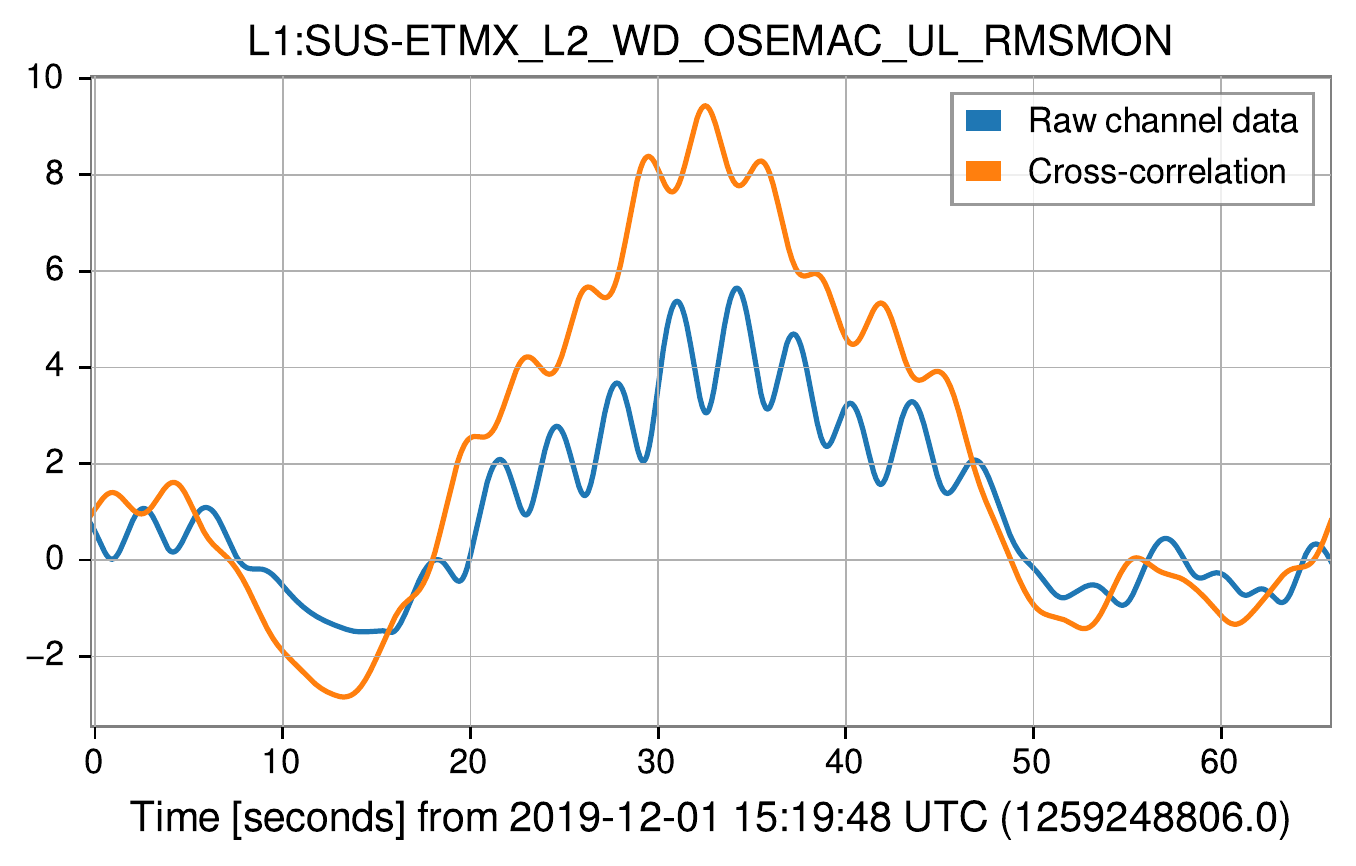}
            \includegraphics[width=.31\linewidth]{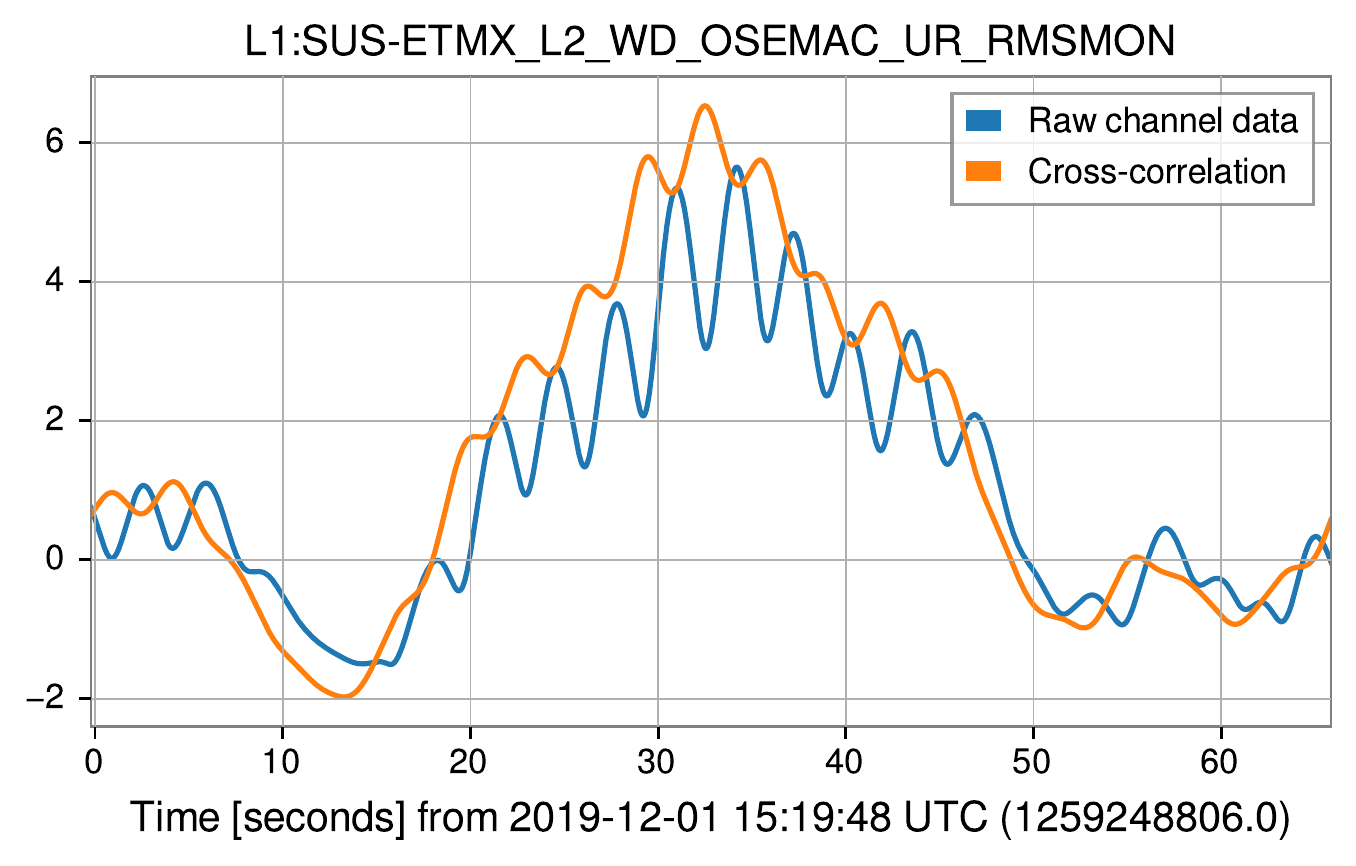}
            \includegraphics[width=.31\linewidth]{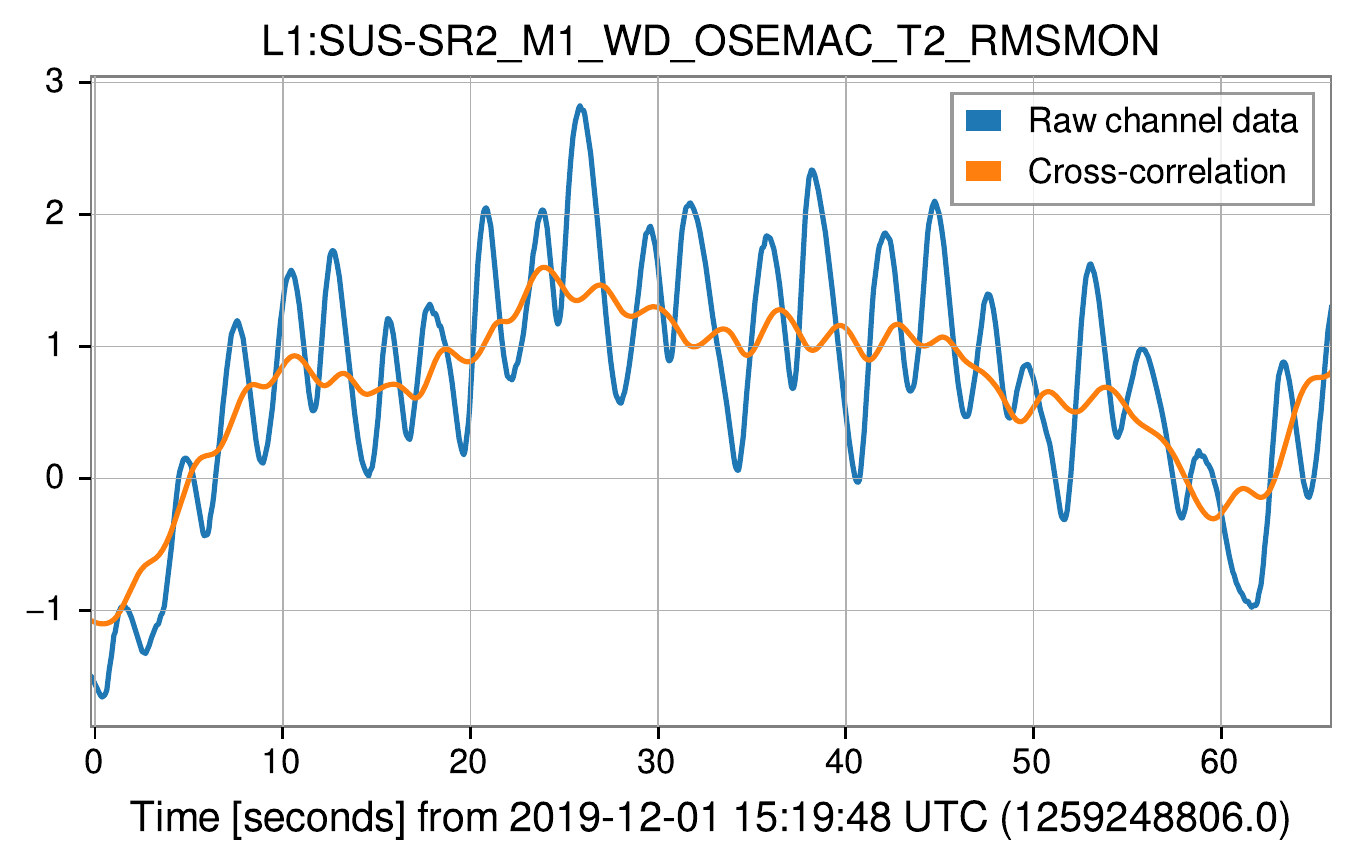}
            }
\caption{Raw channel data (blue) from around the scattered light glitch illustrated in Figs. \ref{fig:sl_omegagram_only} and \ref{fig:sl_omegagram_with_prob_2}, along with (orange) the output of that channel cross-correlated with the associated learned filter (i.e., its contribution to the classifier's result).
Similarly to Fig. \ref{fig:sl_raw_and_conv_2021-10-13}, note the oscillations in the channel data with peaks spaced approximately 3.2 seconds apart, similar to the spacing between the peaks of the characteristic scattered light arches in Fig. \ref{fig:sl_omegagram_only}.
}\label{fig:sl_raw_and_conv_2021-10-13_2}
\end{figure*}

\begin{figure*}[t]
\centerline{
            \includegraphics[width=.31\linewidth]{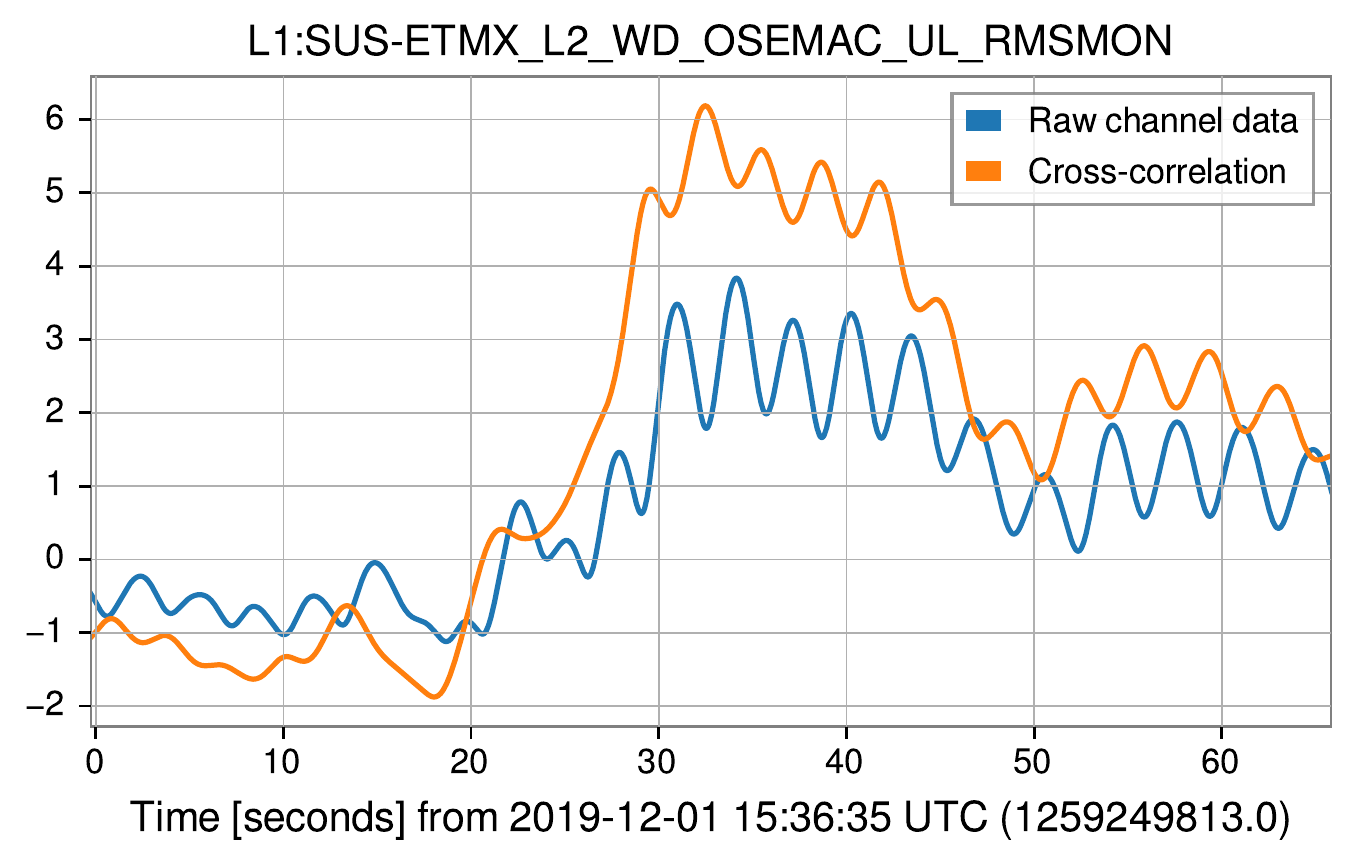}
            \includegraphics[width=.31\linewidth]{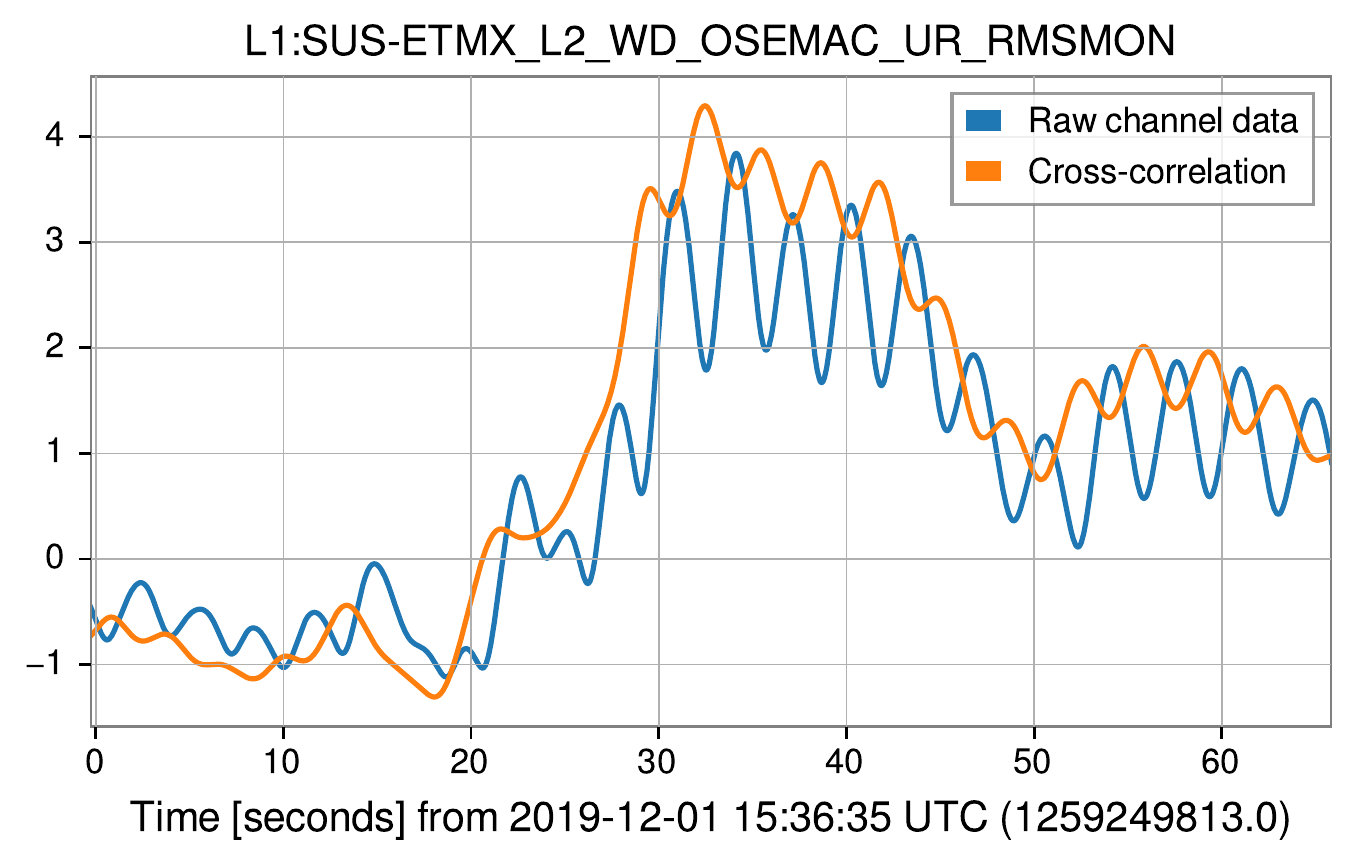}
            \includegraphics[width=.31\linewidth]{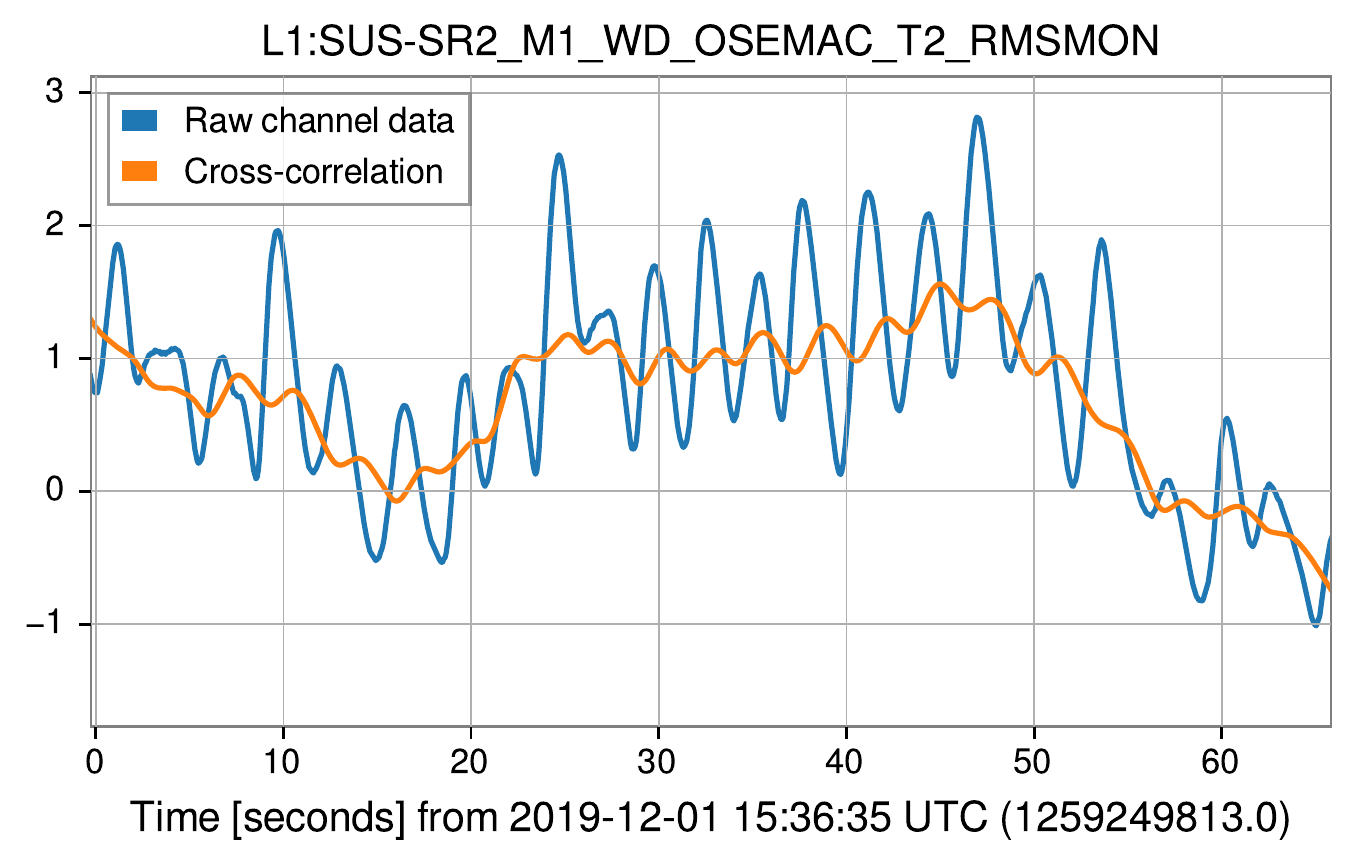}
            }
\caption{Raw channel data (blue) from around the scattered light glitch illustrated in Fig. \ref{fig:sl_omegagram_with_prob_3}, along with (orange) the output of that channel cross-correlated with the associated learned filter (i.e., its contribution to the classifier's result).
}\label{fig:sl_raw_and_conv_2021-10-13_3}
\end{figure*}

\begin{figure*}[t]
\centerline{
            \includegraphics[width=.31\linewidth]{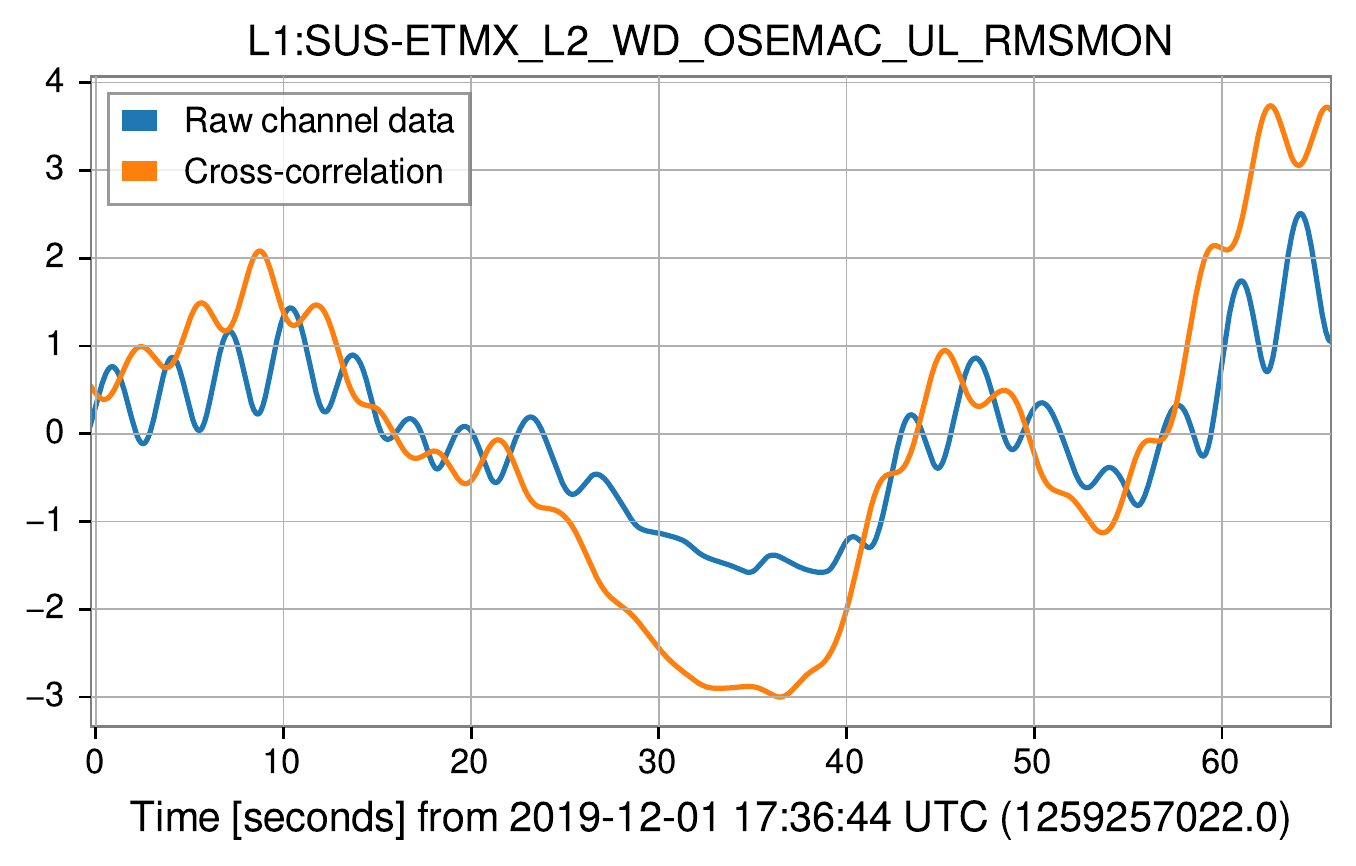}
            \includegraphics[width=.31\linewidth]{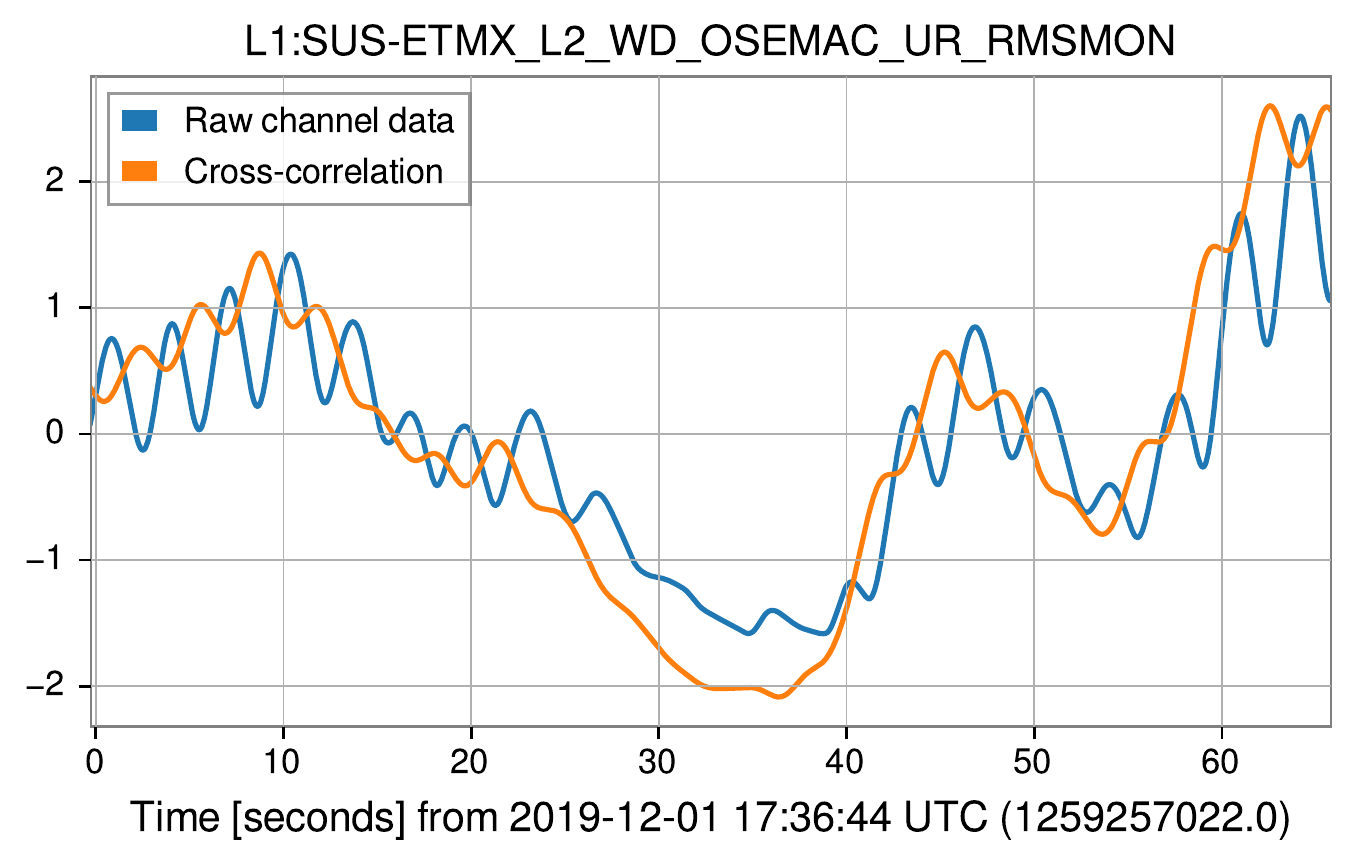}
            \includegraphics[width=.31\linewidth]{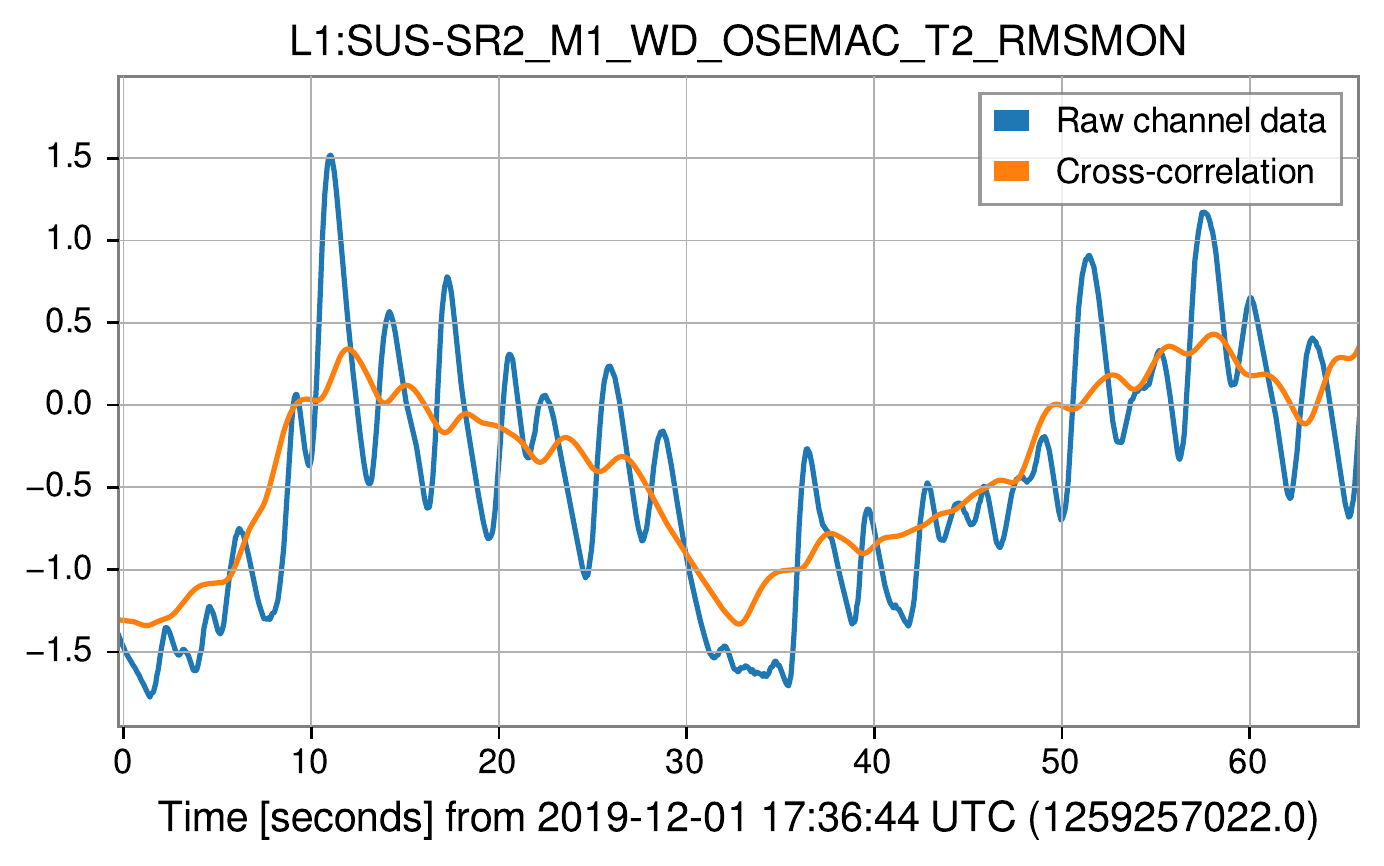}
            }
\centerline{
            \includegraphics[width=.31\linewidth]{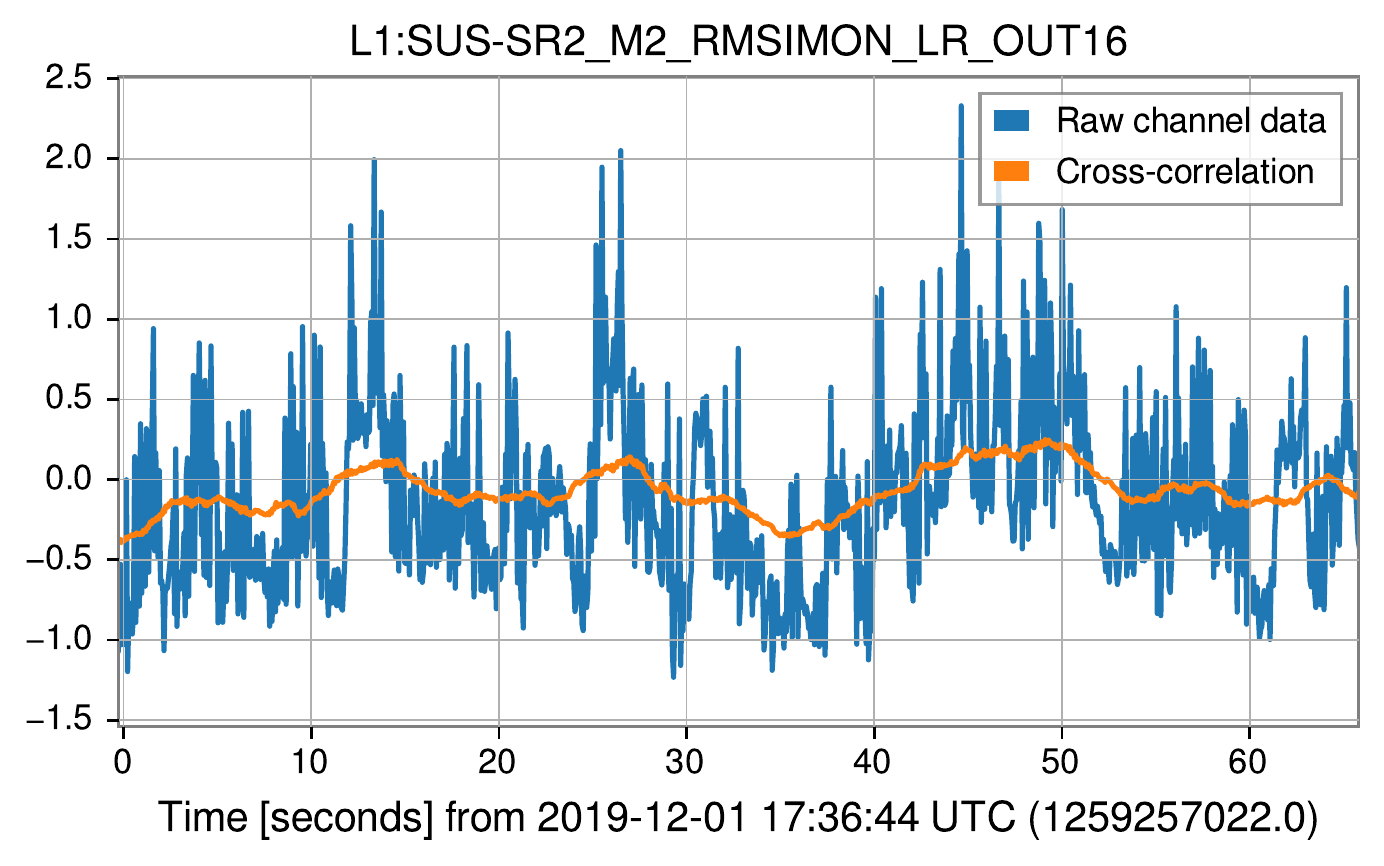}
            \includegraphics[width=.31\linewidth]{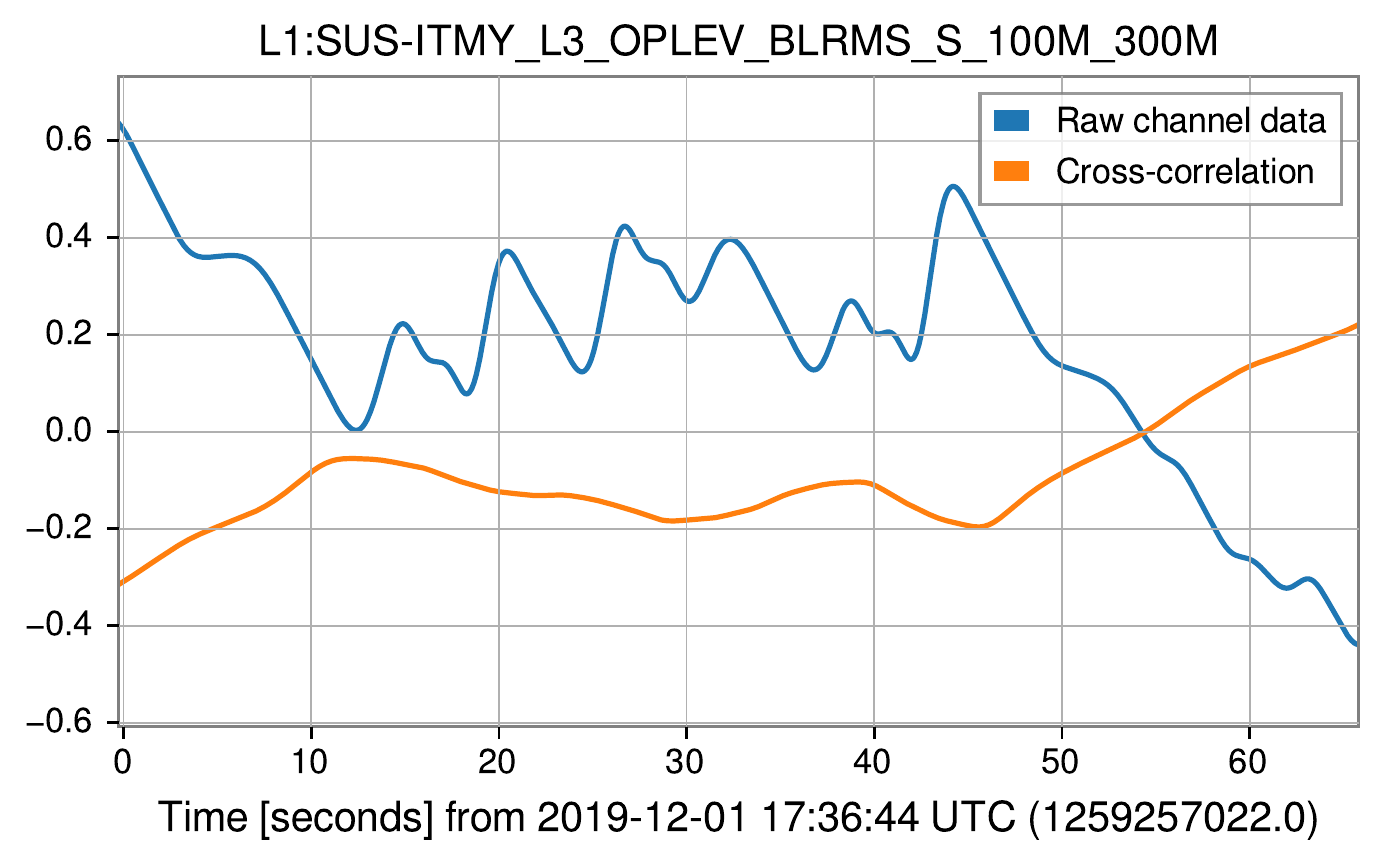}
            \includegraphics[width=.31\linewidth]{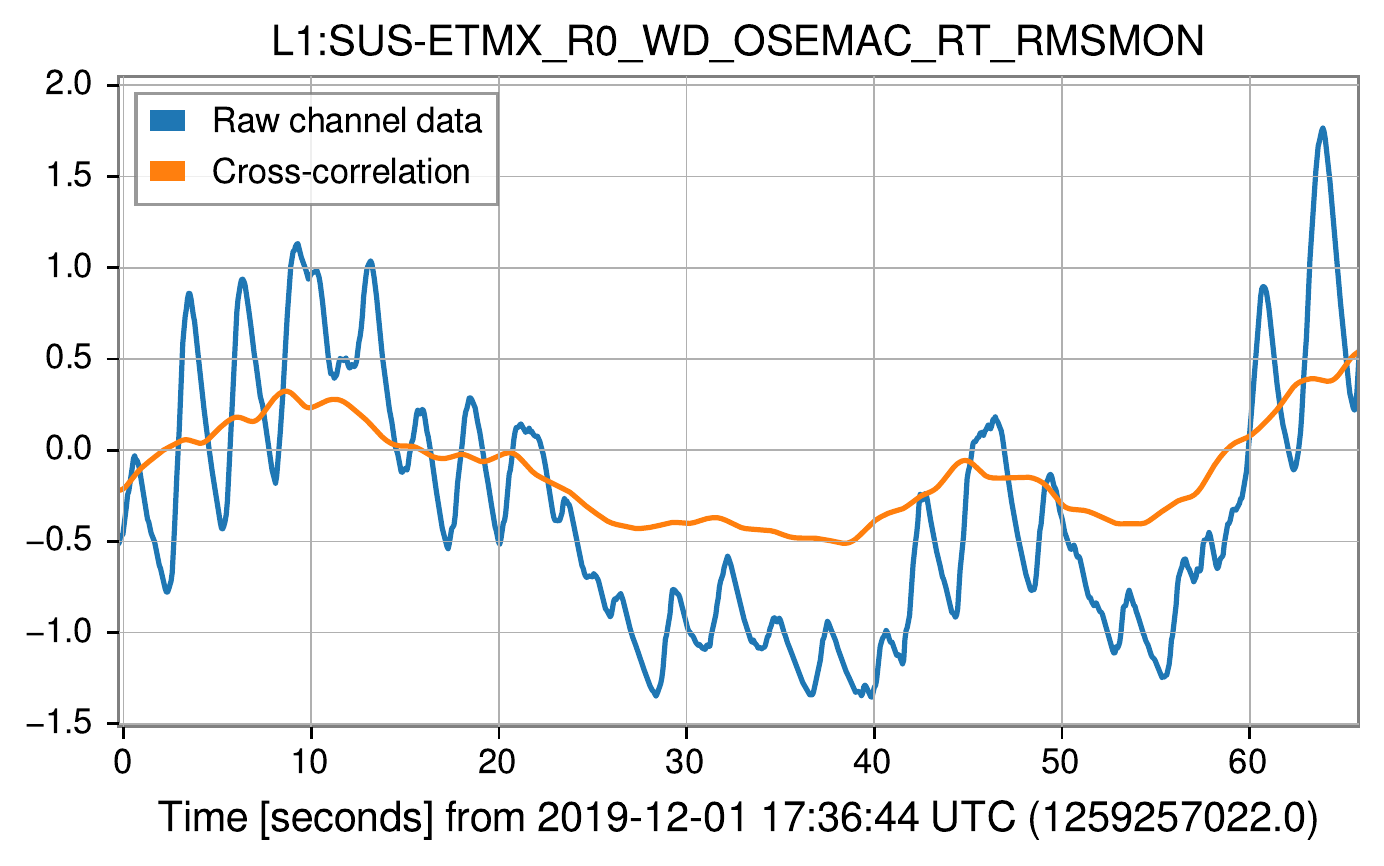}
            }
\centerline{
            \includegraphics[width=.31\linewidth]{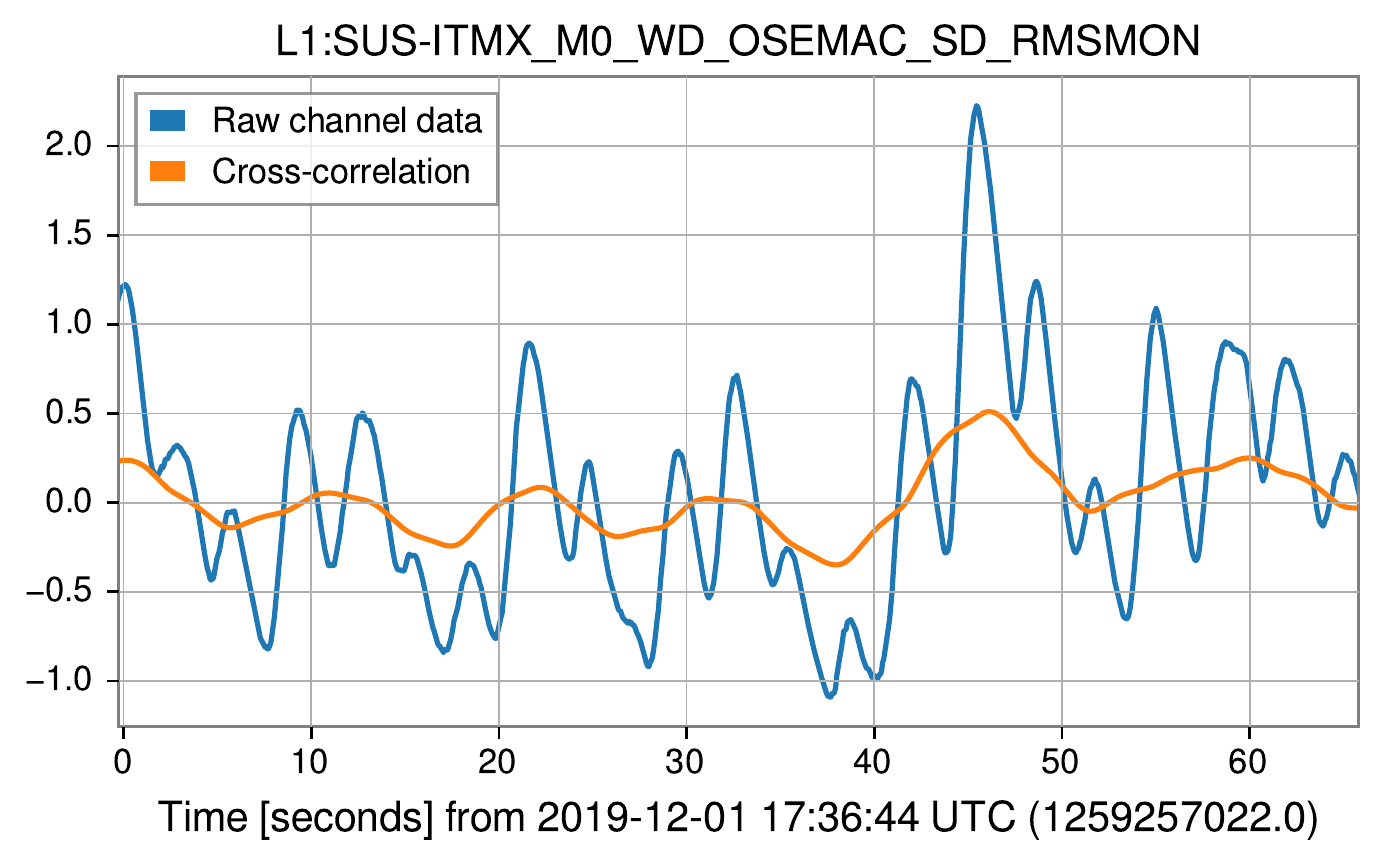}
            \includegraphics[width=.31\linewidth]{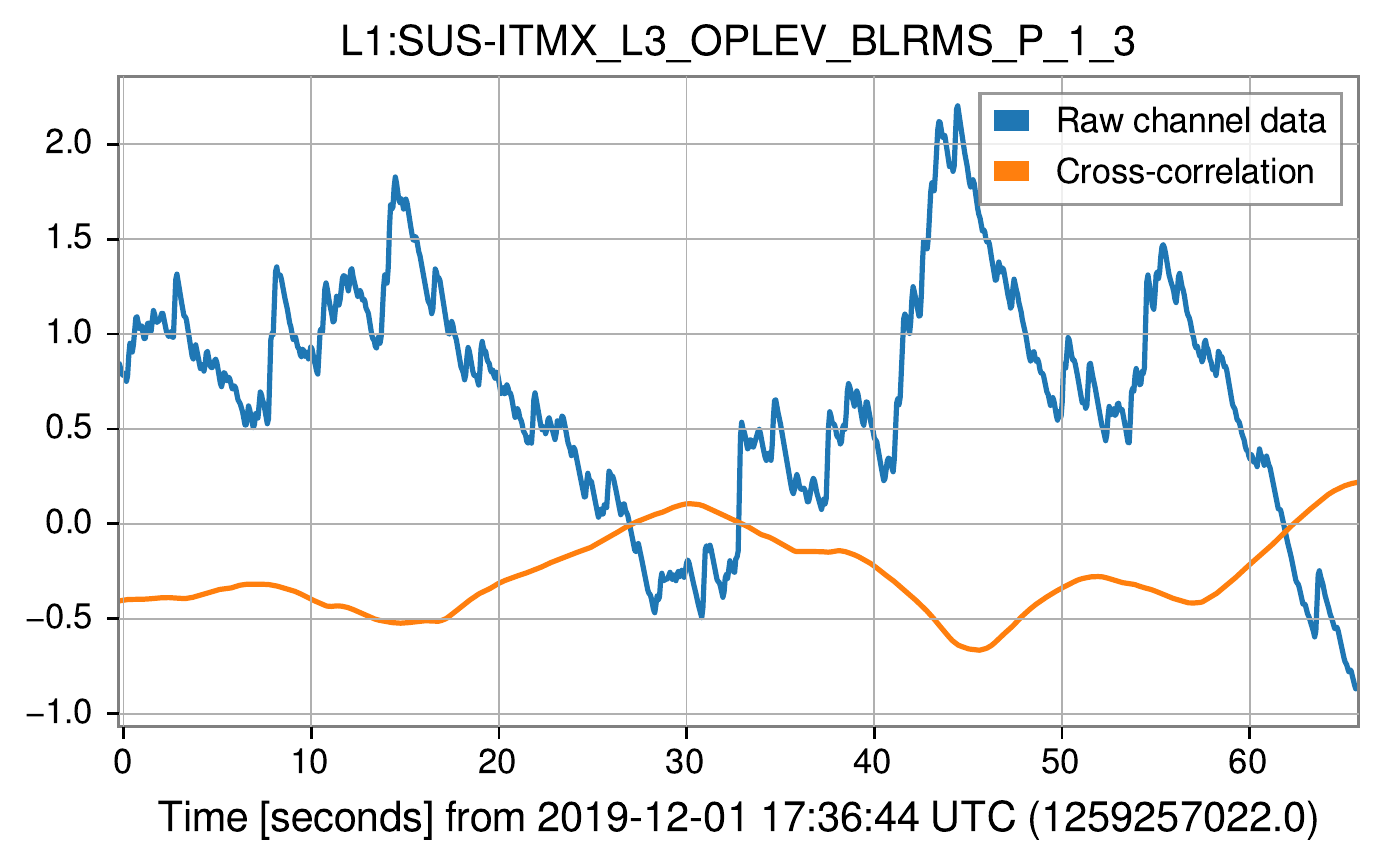}
            \includegraphics[width=.31\linewidth]{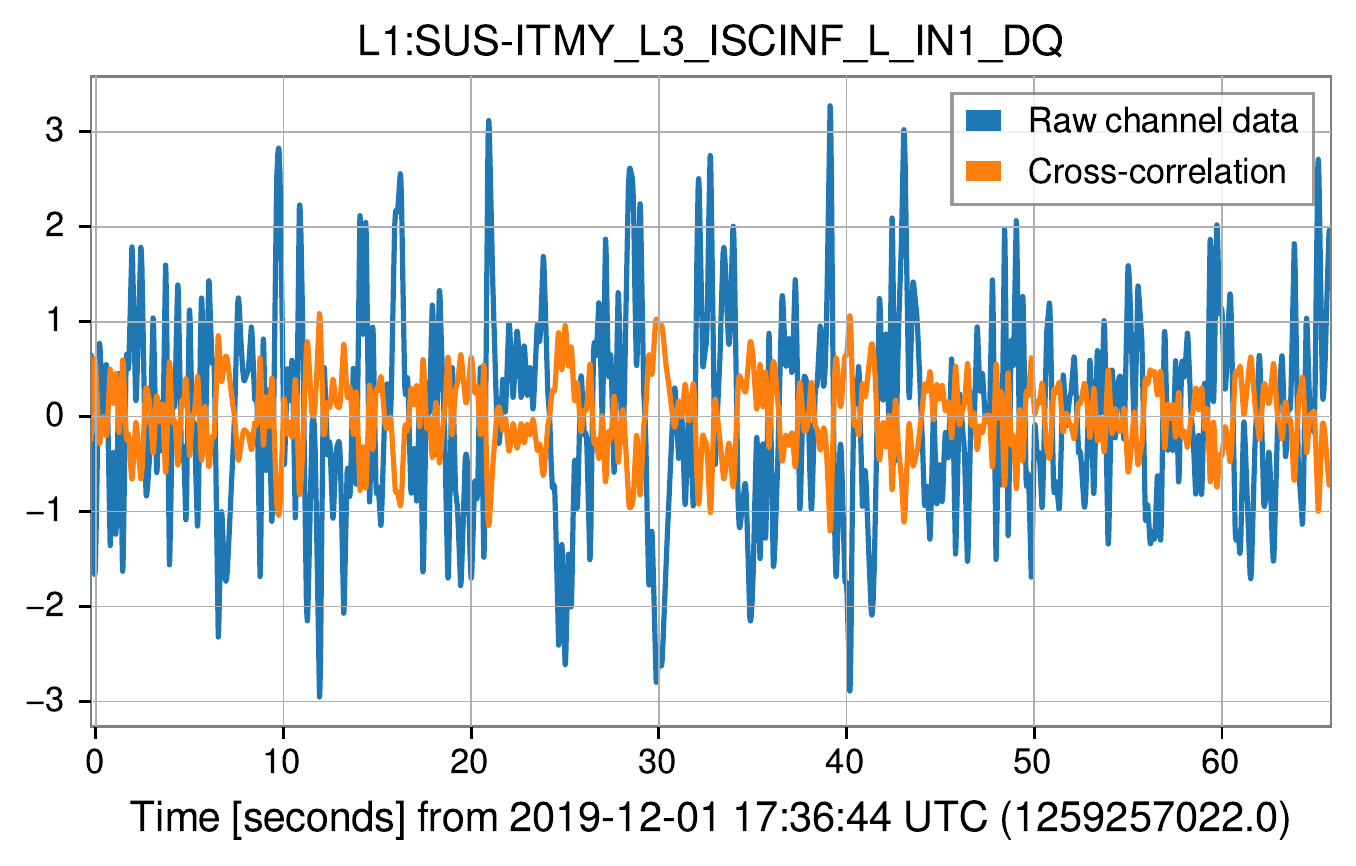}
            }
\centerline{
            \includegraphics[width=.31\linewidth]{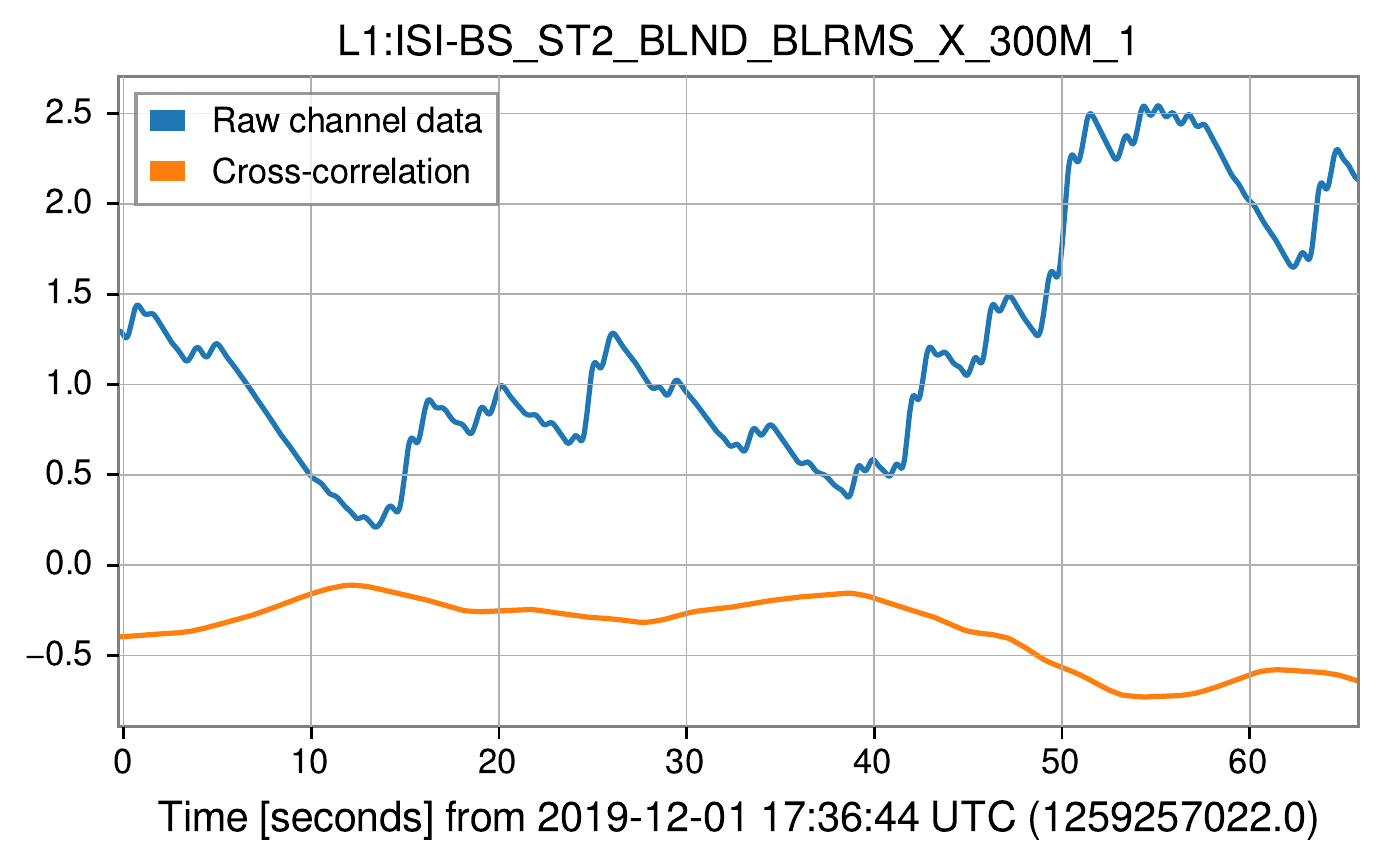}
            \includegraphics[width=.31\linewidth]{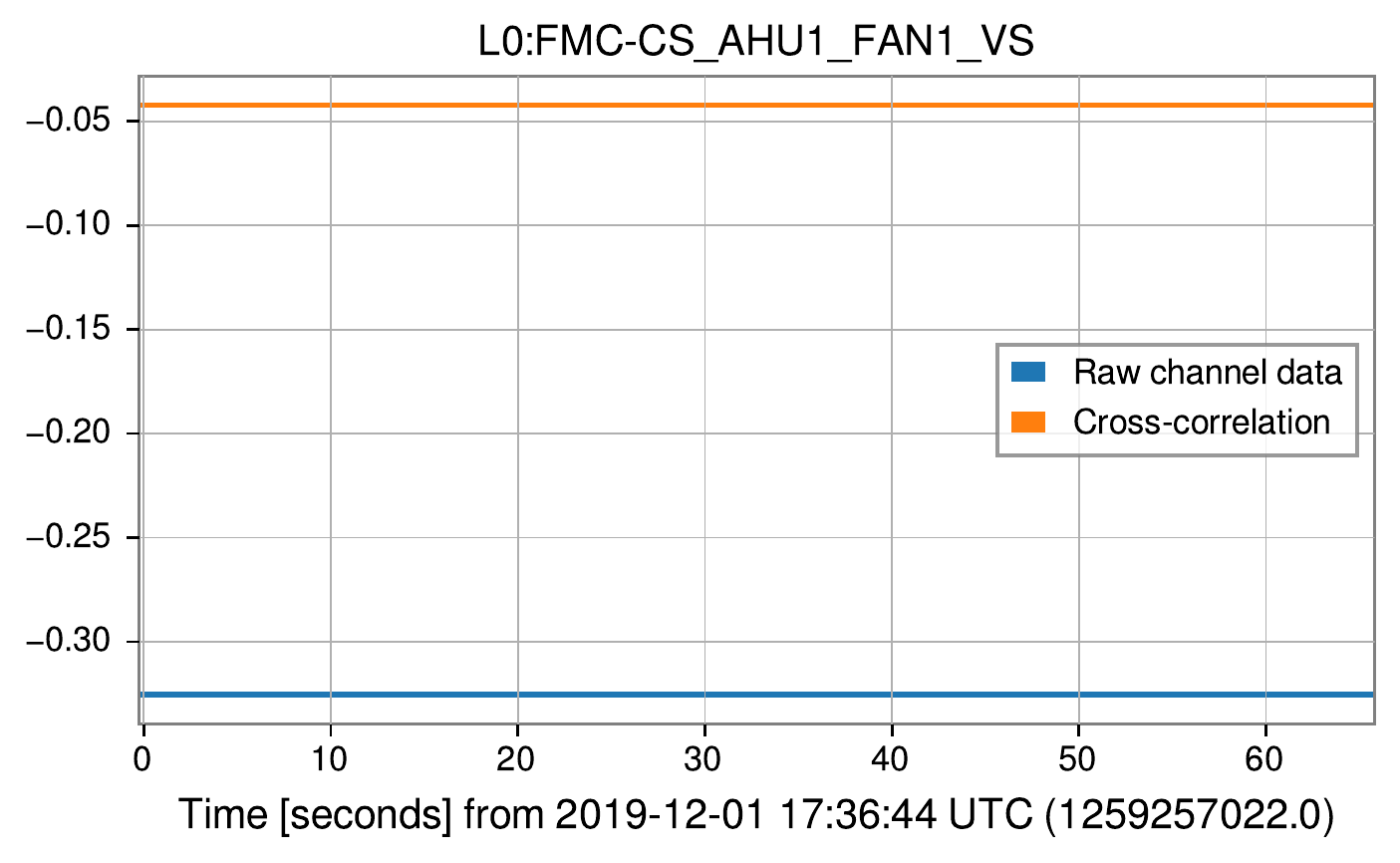}
            \includegraphics[width=.31\linewidth]{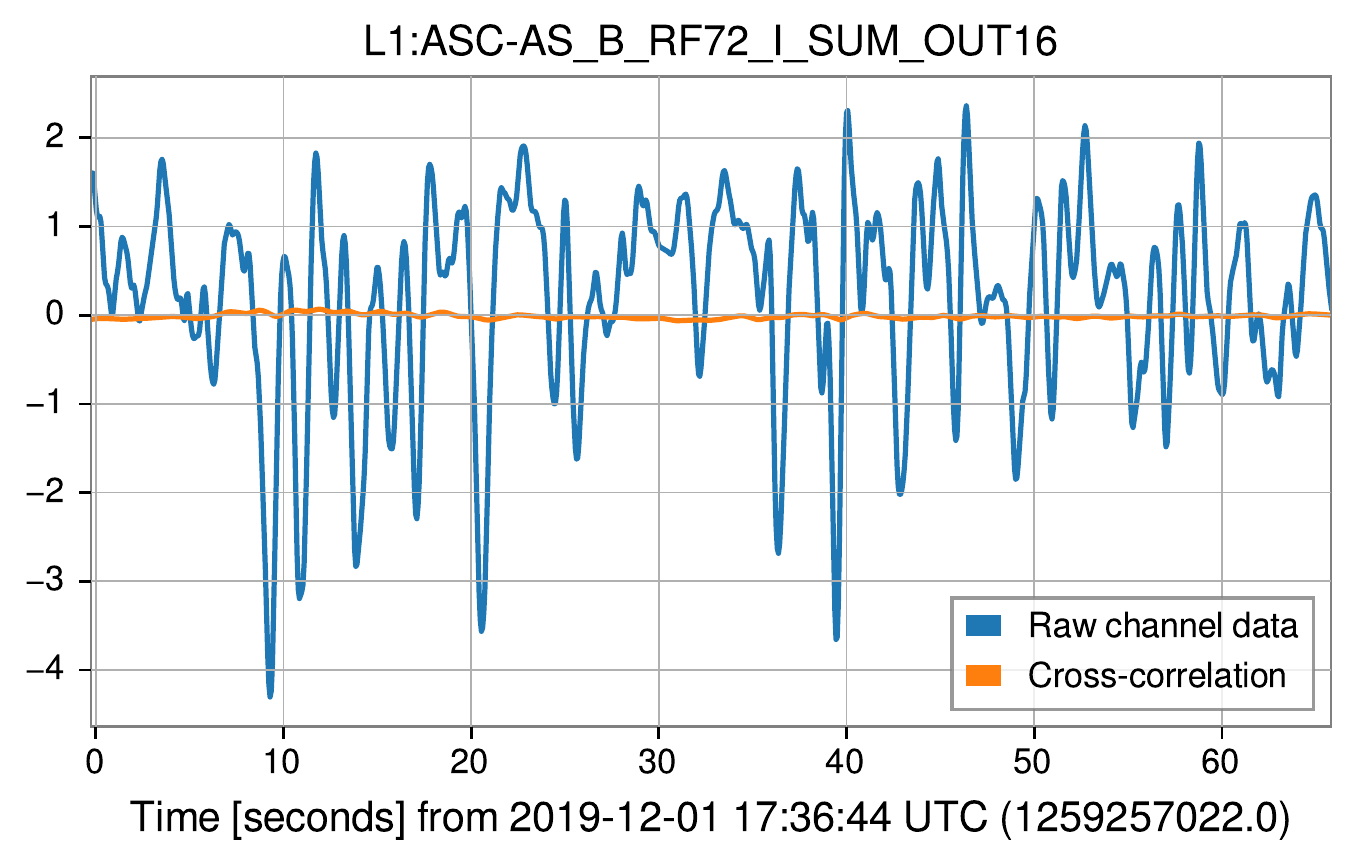}
            }
\caption{Raw channel data (blue) from the glitch-free period about 30 seconds long around GPS time 1,259,257,055 illustrated in Fig. \ref{fig:sl_omegagram_with_prob_gf}, along with (orange) the output of that channel cross-correlated with the associated learned filter (i.e., its contribution to the classifier's result).
}\label{fig:sl_raw_and_conv_2021-10-13_gf}
\end{figure*}

We illustrate similar examples with different glitches in Figs. \ref{fig:sl_raw_and_conv_2021-10-13_2} and \ref{fig:sl_raw_and_conv_2021-10-13_3} and an example with a glitch-free point (the same glitch-free point shown in Fig. \ref{fig:sl_omegagram_with_prob_gf}) in Fig. \ref{fig:sl_raw_and_conv_2021-10-13_gf}.

\subsection{Interpretations}\label{sec:sl_lf_interpretations}

\begin{figure}[t]
    \centering
    \includegraphics[width=\linewidth]{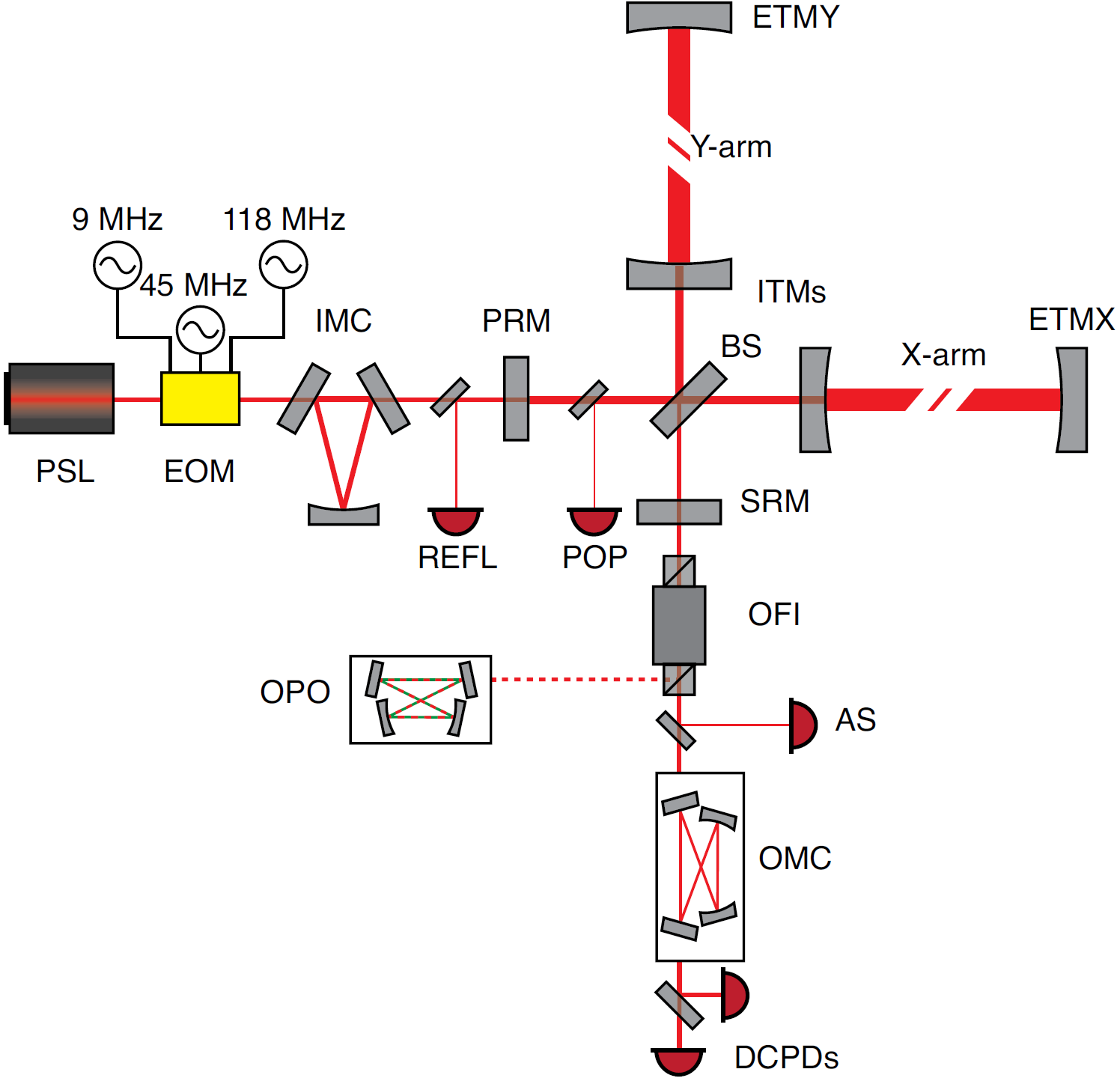}
    \caption{Diagram of the optical layout the \gls{ligo} detectors (from \cite{buikema2020sensitivity}).}
    \label{fig:sl_optical_layout}
\end{figure}

\begin{figure}[t]
    \centering
    \includegraphics[width=.6\linewidth]{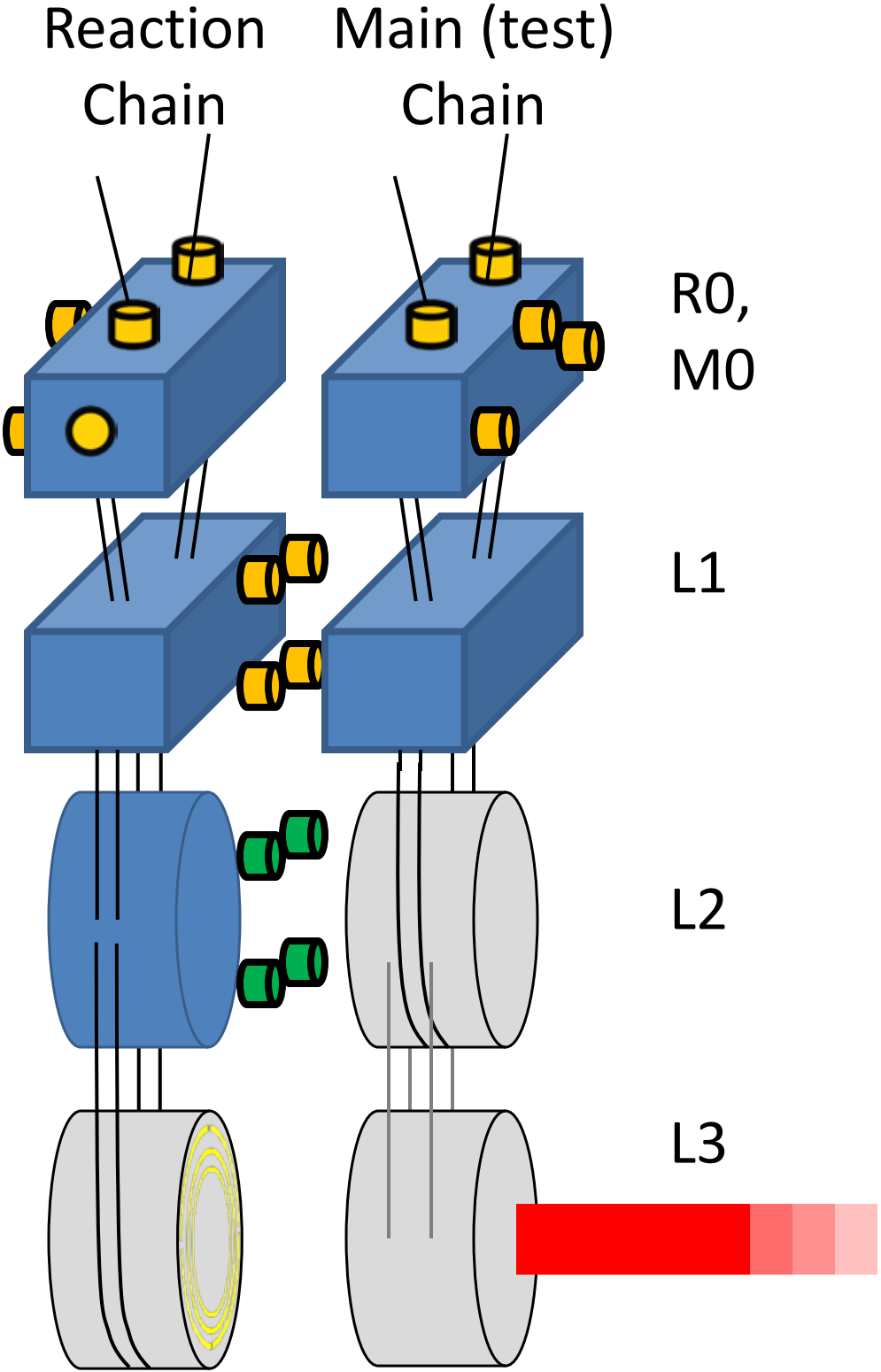}
    \caption{Diagram of the quadruple suspension system (from \cite{G1100866}).}
    \label{fig:sl_quadruple_suspension}
\end{figure}

The most significant channels in Table \ref{tab:sl_scattered_light_EN0NoFC_chans} refer to the third stage (from top) of the quadruple suspension system~\citep{2015CQGra..32a5004S,2012CQGra..29w5004A,2012CQGra..29k5005C,2012CQGra..29c5003C,2019aigw.book..423R,2010CQGra..27h4026B,2004SPIE.5500..194H,2004CQGra..21S.915A} at the X end station (SUS-ETMX\_L2).
Figs. \ref{fig:sl_optical_layout} and \ref{fig:sl_quadruple_suspension} show the test mass chain and the reaction mass chain behind it.
Precision position sensors (so-called OSEMs~\citep{2020CQGra..37w5007S,2017RScI...88l4501W,2012JPhCS.363a2007D,2012CQGra..29k5005C,2012RScI...83d4501S,2009NIMPA.598..737S}) provide the \emph{o}ptical \emph{s}ensing and \emph{e}lectro\emph{m}agnetic actuating capabilities for the upper three stages.
L2 refers to the third (penultimate) stage above the test mass and the bottom mass of the reaction chain. 
UL, UR, LL, LR refer to the sensors at the four quadrants of the penultimate mass of the reaction chain. 
The band-limited RMS motion for these sensors is continuously monitored.
We observe that the channels related to these sensors have the most significant non-zero filters; other channels at SUS-ETMX referring to upper stages of the suspension chain (R0, L1) are also indicated.

Fig. \ref{fig:sl_raw_and_conv_2021-10-13} shows the peaks in the oscillations in the L1:SUS-ETMX channels are spaced approximately 3.5 seconds apart, similar to the spacing between the peaks of the characteristic scattered light arches in the corresponding Omegagram (Fig. \ref{fig:sl_omegagram1_with_prob}).
This suggests that motion indicated by the OSEMs at the end station suspension chain directly gives rise to the observation of scattered light arches in the gravitational-wave strain data channel.
Key to taking strain data is controlling the interferometer arms in unison such that the optical cavities in the arms stay resonant and the differential arm length can be continuously sensed.
The end test mass chain motion is ``locked'' to the motion of the other interferometric optics at the detector's km-scale, and thus its motion relative to other local surfaces in the same chamber can be enhanced, especially during times when high local ground motion is observed.

Fig. \ref{fig:sl_microseismic} shows significantly increased (by approximately a factor of five) microseismic activity in the approximately 0.13-0.15 Hz range for the glitch data used in our studies.

This phenomenon was investigated in \cite{Soni_2021,2013OExpr..2110546C,2012OExpr..20.8329O}, and a scattering mechanism involving the bottom-most reaction mass was verified.
This ring-shaped mass contains five highly reflective circular gold traces (used as electrostatic drive used for interferometer control).
Any incoming stray light will be back-scattered towards the test mass with a phase shift; a fraction of that is transmitted into the interferometer arm through the test mass, where it can interfere with the main beam and manifest as an increase in strain amplitude.
The path between the reaction mass and end test mass can be traversed multiple times (due to multiple reflections), causing the appearance of harmonics of scattering arches on the Omegagrams (see, for example, Fig. \ref{fig:sl_omegagram_only}).
Our study is in agreement with back-scattering from the bottom-most reaction mass being the main source of scattered light glitches in the gravitational-wave data stream.
The four most significant channels associated with scattered light glitches directly measure motion of the penultimate reaction mass right above the indicated scattering surface; its motion is naturally connected to that of the scattering surface that can easily couple back to the interferometer arms.


\section{Conclusion}\label{sec:conclusion}
In this work we have demonstrated the application and capabilities of a machine learning algorithm that employs automatic feature learning to extract relevant information for detecting and analyzing gravitational-wave detector glitches from a vast collection of auxiliary data channels without the need for manual channel selection or feature engineering.
Given tens of thousands of raw time series, the vast majority of which may be irrelevant to the task of glitch analysis, the model automatically selects the few most relevant ones and distills the most pertinent information into several key components that can be further analyzed by detector experts, providing a valuable new tool in their efforts to decrease the incidence of glitches and increase the detectors' sensitivity.
In particular, this method provides accurate strain-independent glitch prediction, automatic channel selection, and per-channel behavioral signatures, as well as a straightforward way to investigate which channels are associated with an individual glitch or set of glitches.

The method performs exceptionally well on a real dataset of scattered light glitches from \gls{llo}, achieving over 97\% accuracy in distinguishing them from glitch-free points.
Beyond glitch identification alone, we show how it selects only 25 out of nearly 40,000 channels as relevant to distinguishing glitches from glitch-free times and how it automatically designs relevant features associated with glitches (or the lack thereof) for each of those channels.
We also illustrate precisely how the model arrives at its prediction for a given time of interest by visualizing multiple real examples.
Unlike some machine learning methods, the method is highly interpretable, allowing detector domain experts to clearly understand how and why it makes a particular individual prediction as well as to probe it for broader understanding of glitch-related behavior over a given time period.
We find physical, detector-specific results with meaningful interpretations consistent with previous research on the possible origins of scattered light glitches and their association with microseismic activity near the detector.

Previous work \cite{learned_features} has also shown that deeper, more complex neural network\textendash based models can improve performance. 
In addition to the \LF{} model employed here, \cite{learned_features} also proposed deeper models incorporating nonlinearities, pooling, and other features and showed that they boost performance on a similar \gls{ligo} auxiliary channel dataset.
However, more complex models are less straightforward to interpret.
For the scattered light study presented here (Sec. \ref{sec:sl_results}), we found improved performance with the \vggoneBN{} model of \cite{learned_features}---test accuracy of 98.2\% vs. 97.1\%---but felt that the performance gain was not significant enough to outweigh the tradeoff to interpretability.
For other applications or datasets, however, one might prefer to sacrifice some interpretability for improved accuracy.
One direction for future work would be to investigate analytical means of interpreting such models applied to this type of data, especially in situations where performance is more critical or the performance gains are more significant.

Additional directions for future work include evaluating the method on other types of glitches and over longer time periods.
Previous work on related models has shown that they generally do not perform as well when asked to analyze data drawn from wider time periods, perhaps because they are not able to account well for long-term distribution shifts in the state of the detector.
One potential avenue to mitigate this would be to identify and test classes of machine learning models that are explicitly designed to handle long-term time dependencies, such as recurrent or long short-term memory neural networks; another would be to retrain or update models like the one employed here online in real time.

We have quantitatively observed here and in previous works that not all glitch morphologies and classes can be equally well predicted based on the auxiliary channels.
One intriguing explanation is that information critical to such glitches is not recorded by any current auxiliary channel.
Absent such necessary information, it would be difficult or even impossible for any auxiliary channel\textendash based method to detect them. 
Therefore, we can treat low-accuracy glitch classes as possible clues to additional auxiliary channels that could be added describing previously unmonitored aspects of the detector.
Therefore the method can also be useful to help us understand what is missing from the auxiliary monitoring network and channels.

Complex systems of humanity have always been plagued with emergent transient issues that can predict or cause eventual failure.
Timely detection, understanding, and repair are paramount to reliable and optimized system operation.
Gravitational-wave detectors can serve as excellent test and development environments for generic solutions to unpredictable issues that inevitably emerge in such extremely complex, sensitive systems.
The solution outlined above is broadly applicable from medical to automotive and space systems, enabling automated anomaly discovery, characterization, and human-in-the-loop anomaly elimination.
The last step of automated anomaly elimination remains a grand challenge of the field, with the long-term goal of accelerating automated understanding of complex systems and eliminating their problems as they emerge.


\begin{acknowledgments}
We acknowledge computing resources from Columbia University's Shared Research Computing Facility project, which is supported by NIH Research Facility Improvement Grant 1G20RR030893-01, and associated funds from the New York State Empire State Development, Division of Science Technology and Innovation (NYSTAR) Contract C090171, both awarded April 15, 2010.

This material is based upon work and data supported by NSF's LIGO Laboratory which is a major facility fully funded by the National Science Foundation. This research has also made use of data obtained from the Gravitational Wave Open Science Center (\url{https://www.gw-openscience.org}), a  service of LIGO Laboratory, the LIGO Scientific Collaboration and the Virgo Collaboration. LIGO is funded by the U.S. National Science Foundation. Virgo is funded by the French Centre National de Recherche Scientifique (CNRS), the Italian Istituto Nazionale della Fisica Nucleare (INFN) and the Dutch Nikhef, with contributions by Polish and Hungarian institutes.

The authors are grateful for the LIGO Scientific Collaboration review of the paper, with special thanks to Gabriele Vajente. This paper is assigned a LIGO DCC number(LIGO-P2200045). The authors acknowledge the LIGO Collaboration for the production of data used in this study and the LIGO Laboratory for enabling Omicron trigger generation on its computing resources (National Science Foundation Grants PHY-0757058 and PHY-0823459). The authors are grateful to the authors and maintainers of the Omicron and Omega pipelines, the LIGO Commissioning and Detector Characterization Teams and LSC domain expert Colleagues whose fundamental work on the LIGO detectors enabled the data used in this paper. The authors would like to thank colleagues of the LIGO Scientific Collaboration and the Virgo Collaboration for their help and useful comments.

The authors thank the University of Florida and Columbia University in the City of New York for their generous support.
The authors are grateful for the generous support of the National Science Foundation under grant CCF-1740391.
I.B. acknowledges the support of the Alfred P. Sloan Foundation and NSF grants PHY-1911796 and PHY-2110060.

\end{acknowledgments}


\appendix

\section{Training and Validation}\label{sec:sl_training}
We follow essentially the same training and validation procedures as \cite{learned_features}, except that instead of classifying between all Omicron glitches and glitch-free points, we classify between glitches classified by GravitySpy as ``scattered light'' and glitch-free points (see Sec. \ref{sec:sl_data}).
We train many models over a grid of hyperparameter settings, evaluate each of them on our validation dataset, and choose the one with the lowest loss.
For more detail we refer the reader to \cite{learned_features}.

\bibliographystyle{apsrev4-1}
\bibliography{references,LIGO_PP_Detector_References,LIGO_ML_References,LIGO_DetChar_References}

\begin{thebibliography}{81}%
\makeatletter
\providecommand \@ifxundefined [1]{%
 \@ifx{#1\undefined}
}%
\providecommand \@ifnum [1]{%
 \ifnum #1\expandafter \@firstoftwo
 \else \expandafter \@secondoftwo
 \fi
}%
\providecommand \@ifx [1]{%
 \ifx #1\expandafter \@firstoftwo
 \else \expandafter \@secondoftwo
 \fi
}%
\providecommand \natexlab [1]{#1}%
\providecommand \enquote  [1]{``#1''}%
\providecommand \bibnamefont  [1]{#1}%
\providecommand \bibfnamefont [1]{#1}%
\providecommand \citenamefont [1]{#1}%
\providecommand \href@noop [0]{\@secondoftwo}%
\providecommand \href [0]{\begingroup \@sanitize@url \@href}%
\providecommand \@href[1]{\@@startlink{#1}\@@href}%
\providecommand \@@href[1]{\endgroup#1\@@endlink}%
\providecommand \@sanitize@url [0]{\catcode `\\12\catcode `\$12\catcode
  `\&12\catcode `\#12\catcode `\^12\catcode `\_12\catcode `\%12\relax}%
\providecommand \@@startlink[1]{}%
\providecommand \@@endlink[0]{}%
\providecommand \url  [0]{\begingroup\@sanitize@url \@url }%
\providecommand \@url [1]{\endgroup\@href {#1}{\urlprefix }}%
\providecommand \urlprefix  [0]{URL }%
\providecommand \Eprint [0]{\href }%
\providecommand \doibase [0]{http://dx.doi.org/}%
\providecommand \selectlanguage [0]{\@gobble}%
\providecommand \bibinfo  [0]{\@secondoftwo}%
\providecommand \bibfield  [0]{\@secondoftwo}%
\providecommand \translation [1]{[#1]}%
\providecommand \BibitemOpen [0]{}%
\providecommand \bibitemStop [0]{}%
\providecommand \bibitemNoStop [0]{.\EOS\space}%
\providecommand \EOS [0]{\spacefactor3000\relax}%
\providecommand \BibitemShut  [1]{\csname bibitem#1\endcsname}%
\let\auto@bib@innerbib\@empty
\bibitem [{\citenamefont {{Akutsu+}}(2021)}]{2021PTEP.2021eA101A}%
  \BibitemOpen
  \bibfield  {author} {\bibinfo {author} {\bibnamefont {{Akutsu+}}},\ }\href
  {\doibase 10.1093/ptep/ptaa125} {\bibfield  {journal} {\bibinfo  {journal}
  {Progress of Theoretical and Experimental Physics}\ }\textbf {\bibinfo
  {volume} {2021}},\ \bibinfo {eid} {05A101} (\bibinfo {year} {2021})},\
  \Eprint {http://arxiv.org/abs/2005.05574} {arXiv:2005.05574
  [physics.ins-det]} \BibitemShut {NoStop}%
\bibitem [{\citenamefont {{Dooley+}}(2016)}]{2016CQGra..33g5009D}%
  \BibitemOpen
  \bibfield  {author} {\bibinfo {author} {\bibnamefont {{Dooley+}}},\ }\href
  {\doibase 10.1088/0264-9381/33/7/075009} {\bibfield  {journal} {\bibinfo
  {journal} {Classical and Quantum Gravity}\ }\textbf {\bibinfo {volume}
  {33}},\ \bibinfo {eid} {075009} (\bibinfo {year} {2016})},\ \Eprint
  {http://arxiv.org/abs/1510.00317} {arXiv:1510.00317 [physics.ins-det]}
  \BibitemShut {NoStop}%
\bibitem [{\citenamefont {Aasi}\ \emph {et~al.}(2015)\citenamefont {Aasi},
  \citenamefont {Abbott}, \citenamefont {Abbott}, \citenamefont {Abbott},
  \citenamefont {Abernathy}, \citenamefont {Ackley}, \citenamefont {Adams},
  \citenamefont {Adams}, \citenamefont {Addesso}, \citenamefont {Adhikari}
  \emph {et~al.}}]{2015CQGra..32g4001L}%
  \BibitemOpen
  \bibfield  {author} {\bibinfo {author} {\bibfnamefont {J.}~\bibnamefont
  {Aasi}}, \bibinfo {author} {\bibfnamefont {B.}~\bibnamefont {Abbott}},
  \bibinfo {author} {\bibfnamefont {R.}~\bibnamefont {Abbott}}, \bibinfo
  {author} {\bibfnamefont {T.}~\bibnamefont {Abbott}}, \bibinfo {author}
  {\bibfnamefont {M.}~\bibnamefont {Abernathy}}, \bibinfo {author}
  {\bibfnamefont {K.}~\bibnamefont {Ackley}}, \bibinfo {author} {\bibfnamefont
  {C.}~\bibnamefont {Adams}}, \bibinfo {author} {\bibfnamefont
  {T.}~\bibnamefont {Adams}}, \bibinfo {author} {\bibfnamefont
  {P.}~\bibnamefont {Addesso}}, \bibinfo {author} {\bibfnamefont
  {R.}~\bibnamefont {Adhikari}},  \emph {et~al.},\ }\href {\doibase
  10.1088/0264-9381/32/7/074001} {\bibfield  {journal} {\bibinfo  {journal}
  {Classical and Quantum Gravity}\ }\textbf {\bibinfo {volume} {32}},\ \bibinfo
  {eid} {074001} (\bibinfo {year} {2015})},\ \Eprint
  {http://arxiv.org/abs/1411.4547} {arXiv:1411.4547 [gr-qc]} \BibitemShut
  {NoStop}%
\bibitem [{\citenamefont {Acernese}\ \emph {et~al.}(2015)\citenamefont
  {Acernese}, \citenamefont {Agathos}, \citenamefont {Agatsuma}, \citenamefont
  {Aisa}, \citenamefont {Allemandou}, \citenamefont {Allocca}, \citenamefont
  {Amarni}, \citenamefont {Astone}, \citenamefont {Balestri}, \citenamefont
  {Ballardin} \emph {et~al.}}]{2015CQGra..32b4001A}%
  \BibitemOpen
  \bibfield  {author} {\bibinfo {author} {\bibfnamefont {F.~a.}\ \bibnamefont
  {Acernese}}, \bibinfo {author} {\bibfnamefont {M.}~\bibnamefont {Agathos}},
  \bibinfo {author} {\bibfnamefont {K.}~\bibnamefont {Agatsuma}}, \bibinfo
  {author} {\bibfnamefont {D.}~\bibnamefont {Aisa}}, \bibinfo {author}
  {\bibfnamefont {N.}~\bibnamefont {Allemandou}}, \bibinfo {author}
  {\bibfnamefont {A.}~\bibnamefont {Allocca}}, \bibinfo {author} {\bibfnamefont
  {J.}~\bibnamefont {Amarni}}, \bibinfo {author} {\bibfnamefont
  {P.}~\bibnamefont {Astone}}, \bibinfo {author} {\bibfnamefont
  {G.}~\bibnamefont {Balestri}}, \bibinfo {author} {\bibfnamefont
  {G.}~\bibnamefont {Ballardin}},  \emph {et~al.},\ }\href {\doibase
  10.1088/0264-9381/32/2/024001} {\bibfield  {journal} {\bibinfo  {journal}
  {Classical and Quantum Gravity}\ }\textbf {\bibinfo {volume} {32}},\ \bibinfo
  {eid} {024001} (\bibinfo {year} {2015})},\ \Eprint
  {http://arxiv.org/abs/1408.3978} {arXiv:1408.3978 [gr-qc]} \BibitemShut
  {NoStop}%
\bibitem [{\citenamefont {{LIGO Scientific Collaboration}}\ and\ \citenamefont
  {{Virgo Collaboration}}(2016)}]{2016PhRvL.116f1102A}%
  \BibitemOpen
  \bibfield  {author} {\bibinfo {author} {\bibnamefont {{LIGO Scientific
  Collaboration}}}\ and\ \bibinfo {author} {\bibnamefont {{Virgo
  Collaboration}}},\ }\href {\doibase 10.1103/PhysRevLett.116.061102}
  {\bibfield  {journal} {\bibinfo  {journal} {\prl}\ }\textbf {\bibinfo
  {volume} {116}},\ \bibinfo {eid} {061102} (\bibinfo {year} {2016})},\ \Eprint
  {http://arxiv.org/abs/1602.03837} {arXiv:1602.03837 [gr-qc]} \BibitemShut
  {NoStop}%
\bibitem [{\citenamefont {Abbott}\ and\ \citenamefont
  {et~al.}(2017{\natexlab{a}})}]{PhysRevLett.119.161101}%
  \BibitemOpen
  \bibfield  {author} {\bibinfo {author} {\bibfnamefont {B.~P.}\ \bibnamefont
  {Abbott}}\ and\ \bibinfo {author} {\bibnamefont {et~al.}},\ }\href {\doibase
  10.1103/PhysRevLett.119.161101} {\bibfield  {journal} {\bibinfo  {journal}
  {Phys. Rev. Lett.}\ }\textbf {\bibinfo {volume} {119}},\ \bibinfo {pages}
  {161101} (\bibinfo {year} {2017}{\natexlab{a}})}\BibitemShut {NoStop}%
\bibitem [{\citenamefont {Abbott}\ and\ \citenamefont
  {et~al.}(2017{\natexlab{b}})}]{2017ApJ...848L..12A}%
  \BibitemOpen
  \bibfield  {author} {\bibinfo {author} {\bibfnamefont {B.~P.}\ \bibnamefont
  {Abbott}}\ and\ \bibinfo {author} {\bibnamefont {et~al.}},\ }\href {\doibase
  10.3847/2041-8213/aa91c9} {\bibfield  {journal} {\bibinfo  {journal}
  {Astrophysical Journal Letters}\ }\textbf {\bibinfo {volume} {848}},\
  \bibinfo {eid} {L12} (\bibinfo {year} {2017}{\natexlab{b}})},\ \Eprint
  {http://arxiv.org/abs/1710.05833} {arXiv:1710.05833 [astro-ph.HE]}
  \BibitemShut {NoStop}%
\bibitem [{\citenamefont {Abbott}\ \emph {et~al.}(2019)\citenamefont {Abbott},
  \citenamefont {Abbott}, \citenamefont {Abbott}, \citenamefont {Abraham},
  \citenamefont {Acernese}, \citenamefont {Ackley}, \citenamefont {Adams},
  \citenamefont {Adhikari}, \citenamefont {Adya}, \citenamefont {Affeldt},\
  and\ \citenamefont {et~al.}}]{Abbott_2019}%
  \BibitemOpen
  \bibfield  {author} {\bibinfo {author} {\bibfnamefont {B.}~\bibnamefont
  {Abbott}}, \bibinfo {author} {\bibfnamefont {R.}~\bibnamefont {Abbott}},
  \bibinfo {author} {\bibfnamefont {T.}~\bibnamefont {Abbott}}, \bibinfo
  {author} {\bibfnamefont {S.}~\bibnamefont {Abraham}}, \bibinfo {author}
  {\bibfnamefont {F.}~\bibnamefont {Acernese}}, \bibinfo {author}
  {\bibfnamefont {K.}~\bibnamefont {Ackley}}, \bibinfo {author} {\bibfnamefont
  {C.}~\bibnamefont {Adams}}, \bibinfo {author} {\bibfnamefont
  {R.}~\bibnamefont {Adhikari}}, \bibinfo {author} {\bibfnamefont
  {V.}~\bibnamefont {Adya}}, \bibinfo {author} {\bibfnamefont {C.}~\bibnamefont
  {Affeldt}}, \ and\ \bibinfo {author} {\bibnamefont {et~al.}},\ }\href
  {\doibase 10.1103/physrevx.9.031040} {\bibfield  {journal} {\bibinfo
  {journal} {Physical Review X}\ }\textbf {\bibinfo {volume} {9}} (\bibinfo
  {year} {2019}),\ 10.1103/physrevx.9.031040}\BibitemShut {NoStop}%
\bibitem [{\citenamefont {Abbott}\ \emph {et~al.}(2020)\citenamefont {Abbott},
  \citenamefont {Abbott}, \citenamefont {Abraham}, \citenamefont {Acernese},
  \citenamefont {Ackley}, \citenamefont {Adams}, \citenamefont {Adams},
  \citenamefont {Adhikari}, \citenamefont {Adya}, \citenamefont {Affeldt} \emph
  {et~al.}}]{2020arXiv201014527A}%
  \BibitemOpen
  \bibfield  {author} {\bibinfo {author} {\bibfnamefont {R.}~\bibnamefont
  {Abbott}}, \bibinfo {author} {\bibfnamefont {T.~D.}\ \bibnamefont {Abbott}},
  \bibinfo {author} {\bibfnamefont {S.}~\bibnamefont {Abraham}}, \bibinfo
  {author} {\bibfnamefont {F.}~\bibnamefont {Acernese}}, \bibinfo {author}
  {\bibfnamefont {K.}~\bibnamefont {Ackley}}, \bibinfo {author} {\bibfnamefont
  {A.}~\bibnamefont {Adams}}, \bibinfo {author} {\bibfnamefont
  {C.}~\bibnamefont {Adams}}, \bibinfo {author} {\bibfnamefont {R.~X.}\
  \bibnamefont {Adhikari}}, \bibinfo {author} {\bibfnamefont {V.~B.}\
  \bibnamefont {Adya}}, \bibinfo {author} {\bibfnamefont {C.}~\bibnamefont
  {Affeldt}},  \emph {et~al.} (\bibinfo {collaboration} {{The LIGO Scientific
  Collaboration} and {the Virgo Collaboration} and {the KAGRA
  Collaboration}}),\ }\href@noop {} {\bibfield  {journal} {\bibinfo  {journal}
  {arXiv e-prints}\ ,\ \bibinfo {eid} {arXiv:2010.14527}} (\bibinfo {year}
  {2020})},\ \Eprint {http://arxiv.org/abs/2010.14527} {arXiv:2010.14527
  [gr-qc]} \BibitemShut {NoStop}%
\bibitem [{\citenamefont {{The LIGO Scientific Collaboration}}\ and\
  \citenamefont {{the Virgo Collaboration}}(2021)}]{2021arXiv210801045T}%
  \BibitemOpen
  \bibfield  {author} {\bibinfo {author} {\bibnamefont {{The LIGO Scientific
  Collaboration}}}\ and\ \bibinfo {author} {\bibnamefont {{the Virgo
  Collaboration}}},\ }\href@noop {} {\bibfield  {journal} {\bibinfo  {journal}
  {arXiv e-prints}\ ,\ \bibinfo {eid} {arXiv:2108.01045}} (\bibinfo {year}
  {2021})},\ \Eprint {http://arxiv.org/abs/2108.01045} {arXiv:2108.01045
  [gr-qc]} \BibitemShut {NoStop}%
\bibitem [{\citenamefont {{The LIGO Scientific Collaboration}}\ \emph
  {et~al.}(2021)\citenamefont {{The LIGO Scientific Collaboration}},
  \citenamefont {{the Virgo Collaboration}},\ and\ \citenamefont {{the KAGRA
  Collaboration}}}]{2021arXiv211103606T}%
  \BibitemOpen
  \bibfield  {author} {\bibinfo {author} {\bibnamefont {{The LIGO Scientific
  Collaboration}}}, \bibinfo {author} {\bibnamefont {{the Virgo
  Collaboration}}}, \ and\ \bibinfo {author} {\bibnamefont {{the KAGRA
  Collaboration}}},\ }\href@noop {} {\bibfield  {journal} {\bibinfo  {journal}
  {arXiv e-prints}\ ,\ \bibinfo {eid} {arXiv:2111.03606}} (\bibinfo {year}
  {2021})},\ \Eprint {http://arxiv.org/abs/2111.03606} {arXiv:2111.03606
  [gr-qc]} \BibitemShut {NoStop}%
\bibitem [{\citenamefont {{Nitz}}\ \emph {et~al.}(2019)\citenamefont {{Nitz}},
  \citenamefont {{Capano}}, \citenamefont {{Nielsen}}, \citenamefont {{Reyes}},
  \citenamefont {{White}}, \citenamefont {{Brown}},\ and\ \citenamefont
  {{Krishnan}}}]{2019ApJ...872..195N}%
  \BibitemOpen
  \bibfield  {author} {\bibinfo {author} {\bibfnamefont {A.~H.}\ \bibnamefont
  {{Nitz}}}, \bibinfo {author} {\bibfnamefont {C.}~\bibnamefont {{Capano}}},
  \bibinfo {author} {\bibfnamefont {A.~B.}\ \bibnamefont {{Nielsen}}}, \bibinfo
  {author} {\bibfnamefont {S.}~\bibnamefont {{Reyes}}}, \bibinfo {author}
  {\bibfnamefont {R.}~\bibnamefont {{White}}}, \bibinfo {author} {\bibfnamefont
  {D.~A.}\ \bibnamefont {{Brown}}}, \ and\ \bibinfo {author} {\bibfnamefont
  {B.}~\bibnamefont {{Krishnan}}},\ }\href {\doibase 10.3847/1538-4357/ab0108}
  {\bibfield  {journal} {\bibinfo  {journal} {\apj}\ }\textbf {\bibinfo
  {volume} {872}},\ \bibinfo {eid} {195} (\bibinfo {year} {2019})},\ \Eprint
  {http://arxiv.org/abs/1811.01921} {arXiv:1811.01921 [gr-qc]} \BibitemShut
  {NoStop}%
\bibitem [{\citenamefont {{Nitz}}\ \emph {et~al.}(2020)\citenamefont {{Nitz}},
  \citenamefont {{Dent}}, \citenamefont {{Davies}}, \citenamefont {{Kumar}},
  \citenamefont {{Capano}}, \citenamefont {{Harry}}, \citenamefont {{Mozzon}},
  \citenamefont {{Nuttall}}, \citenamefont {{Lundgren}},\ and\ \citenamefont
  {{T{\'a}pai}}}]{2020ApJ...891..123N}%
  \BibitemOpen
  \bibfield  {author} {\bibinfo {author} {\bibfnamefont {A.~H.}\ \bibnamefont
  {{Nitz}}}, \bibinfo {author} {\bibfnamefont {T.}~\bibnamefont {{Dent}}},
  \bibinfo {author} {\bibfnamefont {G.~S.}\ \bibnamefont {{Davies}}}, \bibinfo
  {author} {\bibfnamefont {S.}~\bibnamefont {{Kumar}}}, \bibinfo {author}
  {\bibfnamefont {C.~D.}\ \bibnamefont {{Capano}}}, \bibinfo {author}
  {\bibfnamefont {I.}~\bibnamefont {{Harry}}}, \bibinfo {author} {\bibfnamefont
  {S.}~\bibnamefont {{Mozzon}}}, \bibinfo {author} {\bibfnamefont
  {L.}~\bibnamefont {{Nuttall}}}, \bibinfo {author} {\bibfnamefont
  {A.}~\bibnamefont {{Lundgren}}}, \ and\ \bibinfo {author} {\bibfnamefont
  {M.}~\bibnamefont {{T{\'a}pai}}},\ }\href {\doibase 10.3847/1538-4357/ab733f}
  {\bibfield  {journal} {\bibinfo  {journal} {\apj}\ }\textbf {\bibinfo
  {volume} {891}},\ \bibinfo {eid} {123} (\bibinfo {year} {2020})},\ \Eprint
  {http://arxiv.org/abs/1910.05331} {arXiv:1910.05331 [astro-ph.HE]}
  \BibitemShut {NoStop}%
\bibitem [{\citenamefont {{Nitz}}\ \emph {et~al.}(2021)\citenamefont {{Nitz}},
  \citenamefont {{Capano}}, \citenamefont {{Kumar}}, \citenamefont {{Wang}},
  \citenamefont {{Kastha}}, \citenamefont {{Sch{\"a}fer}}, \citenamefont
  {{Dhurkunde}},\ and\ \citenamefont {{Cabero}}}]{2021arXiv210509151N}%
  \BibitemOpen
  \bibfield  {author} {\bibinfo {author} {\bibfnamefont {A.~H.}\ \bibnamefont
  {{Nitz}}}, \bibinfo {author} {\bibfnamefont {C.~D.}\ \bibnamefont
  {{Capano}}}, \bibinfo {author} {\bibfnamefont {S.}~\bibnamefont {{Kumar}}},
  \bibinfo {author} {\bibfnamefont {Y.-F.}\ \bibnamefont {{Wang}}}, \bibinfo
  {author} {\bibfnamefont {S.}~\bibnamefont {{Kastha}}}, \bibinfo {author}
  {\bibfnamefont {M.}~\bibnamefont {{Sch{\"a}fer}}}, \bibinfo {author}
  {\bibfnamefont {R.}~\bibnamefont {{Dhurkunde}}}, \ and\ \bibinfo {author}
  {\bibfnamefont {M.}~\bibnamefont {{Cabero}}},\ }\href@noop {} {\bibfield
  {journal} {\bibinfo  {journal} {arXiv e-prints}\ ,\ \bibinfo {eid}
  {arXiv:2105.09151}} (\bibinfo {year} {2021})},\ \Eprint
  {http://arxiv.org/abs/2105.09151} {arXiv:2105.09151 [astro-ph.HE]}
  \BibitemShut {NoStop}%
\bibitem [{\citenamefont {Venumadhav}\ \emph {et~al.}(2019)\citenamefont
  {Venumadhav}, \citenamefont {Zackay}, \citenamefont {Roulet}, \citenamefont
  {Dai},\ and\ \citenamefont {Zaldarriaga}}]{venumadhav2019new}%
  \BibitemOpen
  \bibfield  {author} {\bibinfo {author} {\bibfnamefont {T.}~\bibnamefont
  {Venumadhav}}, \bibinfo {author} {\bibfnamefont {B.}~\bibnamefont {Zackay}},
  \bibinfo {author} {\bibfnamefont {J.}~\bibnamefont {Roulet}}, \bibinfo
  {author} {\bibfnamefont {L.}~\bibnamefont {Dai}}, \ and\ \bibinfo {author}
  {\bibfnamefont {M.}~\bibnamefont {Zaldarriaga}},\ }\href@noop {} {\enquote
  {\bibinfo {title} {New binary black hole mergers in the second observing run
  of advanced ligo and advanced virgo},}\ } (\bibinfo {year} {2019}),\ \Eprint
  {http://arxiv.org/abs/1904.07214} {arXiv:1904.07214 [astro-ph.HE]}
  \BibitemShut {NoStop}%
\bibitem [{\citenamefont {Zackay}\ \emph
  {et~al.}(2019{\natexlab{a}})\citenamefont {Zackay}, \citenamefont {Dai},
  \citenamefont {Venumadhav}, \citenamefont {Roulet},\ and\ \citenamefont
  {Zaldarriaga}}]{zackay2019detecting}%
  \BibitemOpen
  \bibfield  {author} {\bibinfo {author} {\bibfnamefont {B.}~\bibnamefont
  {Zackay}}, \bibinfo {author} {\bibfnamefont {L.}~\bibnamefont {Dai}},
  \bibinfo {author} {\bibfnamefont {T.}~\bibnamefont {Venumadhav}}, \bibinfo
  {author} {\bibfnamefont {J.}~\bibnamefont {Roulet}}, \ and\ \bibinfo {author}
  {\bibfnamefont {M.}~\bibnamefont {Zaldarriaga}},\ }\href@noop {} {\enquote
  {\bibinfo {title} {Detecting gravitational waves with disparate detector
  responses: Two new binary black hole mergers},}\ } (\bibinfo {year}
  {2019}{\natexlab{a}}),\ \Eprint {http://arxiv.org/abs/1910.09528}
  {arXiv:1910.09528 [astro-ph.HE]} \BibitemShut {NoStop}%
\bibitem [{\citenamefont {Zackay}\ \emph
  {et~al.}(2019{\natexlab{b}})\citenamefont {Zackay}, \citenamefont
  {Venumadhav}, \citenamefont {Dai}, \citenamefont {Roulet},\ and\
  \citenamefont {Zaldarriaga}}]{Zackay_2019}%
  \BibitemOpen
  \bibfield  {author} {\bibinfo {author} {\bibfnamefont {B.}~\bibnamefont
  {Zackay}}, \bibinfo {author} {\bibfnamefont {T.}~\bibnamefont {Venumadhav}},
  \bibinfo {author} {\bibfnamefont {L.}~\bibnamefont {Dai}}, \bibinfo {author}
  {\bibfnamefont {J.}~\bibnamefont {Roulet}}, \ and\ \bibinfo {author}
  {\bibfnamefont {M.}~\bibnamefont {Zaldarriaga}},\ }\href {\doibase
  10.1103/physrevd.100.023007} {\bibfield  {journal} {\bibinfo  {journal}
  {Physical Review D}\ }\textbf {\bibinfo {volume} {100}} (\bibinfo {year}
  {2019}{\natexlab{b}}),\ 10.1103/physrevd.100.023007}\BibitemShut {NoStop}%
\bibitem [{\citenamefont {Gayathri}\ \emph {et~al.}(2020)\citenamefont
  {Gayathri}, \citenamefont {Healy}, \citenamefont {Lange}, \citenamefont
  {O'Brien}, \citenamefont {Szczepanczyk}, \citenamefont {Bartos},
  \citenamefont {Campanelli}, \citenamefont {Klimenko}, \citenamefont
  {Lousto},\ and\ \citenamefont {O'Shaughnessy}}]{gayathri2020gw190521}%
  \BibitemOpen
  \bibfield  {author} {\bibinfo {author} {\bibfnamefont {V.}~\bibnamefont
  {Gayathri}}, \bibinfo {author} {\bibfnamefont {J.}~\bibnamefont {Healy}},
  \bibinfo {author} {\bibfnamefont {J.}~\bibnamefont {Lange}}, \bibinfo
  {author} {\bibfnamefont {B.}~\bibnamefont {O'Brien}}, \bibinfo {author}
  {\bibfnamefont {M.}~\bibnamefont {Szczepanczyk}}, \bibinfo {author}
  {\bibfnamefont {I.}~\bibnamefont {Bartos}}, \bibinfo {author} {\bibfnamefont
  {M.}~\bibnamefont {Campanelli}}, \bibinfo {author} {\bibfnamefont
  {S.}~\bibnamefont {Klimenko}}, \bibinfo {author} {\bibfnamefont
  {C.}~\bibnamefont {Lousto}}, \ and\ \bibinfo {author} {\bibfnamefont
  {R.}~\bibnamefont {O'Shaughnessy}},\ }\href@noop {} {\  (\bibinfo {year}
  {2020})},\ \Eprint {http://arxiv.org/abs/2009.05461} {arXiv:2009.05461
  [astro-ph.HE]} \BibitemShut {NoStop}%
\bibitem [{\citenamefont {{Davis}}\ \emph {et~al.}(2019)\citenamefont
  {{Davis}}, \citenamefont {{Massinger}}, \citenamefont {{Lundgren}},
  \citenamefont {{Driggers}}, \citenamefont {{Urban}},\ and\ \citenamefont
  {{Nuttall}}}]{2019CQGra..36e5011D}%
  \BibitemOpen
  \bibfield  {author} {\bibinfo {author} {\bibfnamefont {D.}~\bibnamefont
  {{Davis}}}, \bibinfo {author} {\bibfnamefont {T.}~\bibnamefont
  {{Massinger}}}, \bibinfo {author} {\bibfnamefont {A.}~\bibnamefont
  {{Lundgren}}}, \bibinfo {author} {\bibfnamefont {J.~C.}\ \bibnamefont
  {{Driggers}}}, \bibinfo {author} {\bibfnamefont {A.~L.}\ \bibnamefont
  {{Urban}}}, \ and\ \bibinfo {author} {\bibfnamefont {L.}~\bibnamefont
  {{Nuttall}}},\ }\href {\doibase 10.1088/1361-6382/ab01c5} {\bibfield
  {journal} {\bibinfo  {journal} {Classical and Quantum Gravity}\ }\textbf
  {\bibinfo {volume} {36}},\ \bibinfo {eid} {055011} (\bibinfo {year}
  {2019})},\ \Eprint {http://arxiv.org/abs/1809.05348} {arXiv:1809.05348
  [astro-ph.IM]} \BibitemShut {NoStop}%
\bibitem [{\citenamefont {{Vajente}}\ \emph {et~al.}(2020)\citenamefont
  {{Vajente}}, \citenamefont {{Huang}}, \citenamefont {{Isi}}, \citenamefont
  {{Driggers}}, \citenamefont {{Kissel}}, \citenamefont {{Szczepa{\'n}czyk}},\
  and\ \citenamefont {{Vitale}}}]{2020PhRvD.101d2003V}%
  \BibitemOpen
  \bibfield  {author} {\bibinfo {author} {\bibfnamefont {G.}~\bibnamefont
  {{Vajente}}}, \bibinfo {author} {\bibfnamefont {Y.}~\bibnamefont {{Huang}}},
  \bibinfo {author} {\bibfnamefont {M.}~\bibnamefont {{Isi}}}, \bibinfo
  {author} {\bibfnamefont {J.~C.}\ \bibnamefont {{Driggers}}}, \bibinfo
  {author} {\bibfnamefont {J.~S.}\ \bibnamefont {{Kissel}}}, \bibinfo {author}
  {\bibfnamefont {M.~J.}\ \bibnamefont {{Szczepa{\'n}czyk}}}, \ and\ \bibinfo
  {author} {\bibfnamefont {S.}~\bibnamefont {{Vitale}}},\ }\href {\doibase
  10.1103/PhysRevD.101.042003} {\bibfield  {journal} {\bibinfo  {journal}
  {\prd}\ }\textbf {\bibinfo {volume} {101}},\ \bibinfo {eid} {042003}
  (\bibinfo {year} {2020})},\ \Eprint {http://arxiv.org/abs/1911.09083}
  {arXiv:1911.09083 [gr-qc]} \BibitemShut {NoStop}%
\bibitem [{\citenamefont {{Zevin}}\ \emph {et~al.}(2017)\citenamefont
  {{Zevin}}, \citenamefont {{Coughlin}}, \citenamefont {{Bahaadini}},
  \citenamefont {{Besler}}, \citenamefont {{Rohani}}, \citenamefont {{Allen}},
  \citenamefont {{Cabero}}, \citenamefont {{Crowston}}, \citenamefont
  {{Katsaggelos}}, \citenamefont {{Larson}}, \citenamefont {{Lee}},
  \citenamefont {{Lintott}}, \citenamefont {{Littenberg}}, \citenamefont
  {{Lundgren}}, \citenamefont {{{\O}sterlund}}, \citenamefont {{Smith}},
  \citenamefont {{Trouille}},\ and\ \citenamefont {{Kalogera}}}]{GSpy}%
  \BibitemOpen
  \bibfield  {author} {\bibinfo {author} {\bibfnamefont {M.}~\bibnamefont
  {{Zevin}}}, \bibinfo {author} {\bibfnamefont {S.}~\bibnamefont {{Coughlin}}},
  \bibinfo {author} {\bibfnamefont {S.}~\bibnamefont {{Bahaadini}}}, \bibinfo
  {author} {\bibfnamefont {E.}~\bibnamefont {{Besler}}}, \bibinfo {author}
  {\bibfnamefont {N.}~\bibnamefont {{Rohani}}}, \bibinfo {author}
  {\bibfnamefont {S.}~\bibnamefont {{Allen}}}, \bibinfo {author} {\bibfnamefont
  {M.}~\bibnamefont {{Cabero}}}, \bibinfo {author} {\bibfnamefont
  {K.}~\bibnamefont {{Crowston}}}, \bibinfo {author} {\bibfnamefont {A.~K.}\
  \bibnamefont {{Katsaggelos}}}, \bibinfo {author} {\bibfnamefont {S.~L.}\
  \bibnamefont {{Larson}}}, \bibinfo {author} {\bibfnamefont {T.~K.}\
  \bibnamefont {{Lee}}}, \bibinfo {author} {\bibfnamefont {C.}~\bibnamefont
  {{Lintott}}}, \bibinfo {author} {\bibfnamefont {T.~B.}\ \bibnamefont
  {{Littenberg}}}, \bibinfo {author} {\bibfnamefont {A.}~\bibnamefont
  {{Lundgren}}}, \bibinfo {author} {\bibfnamefont {C.}~\bibnamefont
  {{{\O}sterlund}}}, \bibinfo {author} {\bibfnamefont {J.~R.}\ \bibnamefont
  {{Smith}}}, \bibinfo {author} {\bibfnamefont {L.}~\bibnamefont {{Trouille}}},
  \ and\ \bibinfo {author} {\bibfnamefont {V.}~\bibnamefont {{Kalogera}}},\
  }\href {\doibase 10.1088/1361-6382/aa5cea} {\bibfield  {journal} {\bibinfo
  {journal} {Classical and Quantum Gravity}\ }\textbf {\bibinfo {volume}
  {34}},\ \bibinfo {eid} {064003} (\bibinfo {year} {2017})},\ \Eprint
  {http://arxiv.org/abs/1611.04596} {arXiv:1611.04596 [gr-qc]} \BibitemShut
  {NoStop}%
\bibitem [{\citenamefont {{Merritt}}\ \emph
  {et~al.}(2021{\natexlab{a}})\citenamefont {{Merritt}}, \citenamefont
  {{Farr}}, \citenamefont {{Hur}}, \citenamefont {{Edelman}},\ and\
  \citenamefont {{Doctor}}}]{2021arXiv210812044M}%
  \BibitemOpen
  \bibfield  {author} {\bibinfo {author} {\bibfnamefont {J.}~\bibnamefont
  {{Merritt}}}, \bibinfo {author} {\bibfnamefont {B.}~\bibnamefont {{Farr}}},
  \bibinfo {author} {\bibfnamefont {R.}~\bibnamefont {{Hur}}}, \bibinfo
  {author} {\bibfnamefont {B.}~\bibnamefont {{Edelman}}}, \ and\ \bibinfo
  {author} {\bibfnamefont {Z.}~\bibnamefont {{Doctor}}},\ }\href@noop {}
  {\bibfield  {journal} {\bibinfo  {journal} {arXiv e-prints}\ ,\ \bibinfo
  {eid} {arXiv:2108.12044}} (\bibinfo {year} {2021}{\natexlab{a}})},\ \Eprint
  {http://arxiv.org/abs/2108.12044} {arXiv:2108.12044 [gr-qc]} \BibitemShut
  {NoStop}%
\bibitem [{\citenamefont {{Davis}}(2021)}]{2021CQGra..38m5014D}%
  \BibitemOpen
  \bibfield  {author} {\bibinfo {author} {\bibfnamefont {D.~e.~a.}\
  \bibnamefont {{Davis}}},\ }\href {\doibase 10.1088/1361-6382/abfd85}
  {\bibfield  {journal} {\bibinfo  {journal} {Classical and Quantum Gravity}\
  }\textbf {\bibinfo {volume} {38}},\ \bibinfo {eid} {135014} (\bibinfo {year}
  {2021})},\ \Eprint {http://arxiv.org/abs/2101.11673} {arXiv:2101.11673
  [astro-ph.IM]} \BibitemShut {NoStop}%
\bibitem [{\citenamefont {{Stachie}}\ \emph {et~al.}(2020)\citenamefont
  {{Stachie}}, \citenamefont {{Canton}}, \citenamefont {{Burns}}, \citenamefont
  {{Christensen}}, \citenamefont {{Hamburg}}, \citenamefont {{Briggs}},
  \citenamefont {{Broida}}, \citenamefont {{Goldstein}}, \citenamefont
  {{Hayes}}, \citenamefont {{Littenberg}}, \citenamefont {{Shawhan}},
  \citenamefont {{Veitch}}, \citenamefont {{Veres}},\ and\ \citenamefont
  {{Wilson-Hodge}}}]{2020CQGra..37q5001S}%
  \BibitemOpen
  \bibfield  {author} {\bibinfo {author} {\bibfnamefont {C.}~\bibnamefont
  {{Stachie}}}, \bibinfo {author} {\bibfnamefont {T.~D.}\ \bibnamefont
  {{Canton}}}, \bibinfo {author} {\bibfnamefont {E.}~\bibnamefont {{Burns}}},
  \bibinfo {author} {\bibfnamefont {N.}~\bibnamefont {{Christensen}}}, \bibinfo
  {author} {\bibfnamefont {R.}~\bibnamefont {{Hamburg}}}, \bibinfo {author}
  {\bibfnamefont {M.}~\bibnamefont {{Briggs}}}, \bibinfo {author}
  {\bibfnamefont {J.}~\bibnamefont {{Broida}}}, \bibinfo {author}
  {\bibfnamefont {A.}~\bibnamefont {{Goldstein}}}, \bibinfo {author}
  {\bibfnamefont {F.}~\bibnamefont {{Hayes}}}, \bibinfo {author} {\bibfnamefont
  {T.}~\bibnamefont {{Littenberg}}}, \bibinfo {author} {\bibfnamefont
  {P.}~\bibnamefont {{Shawhan}}}, \bibinfo {author} {\bibfnamefont
  {J.}~\bibnamefont {{Veitch}}}, \bibinfo {author} {\bibfnamefont
  {P.}~\bibnamefont {{Veres}}}, \ and\ \bibinfo {author} {\bibfnamefont
  {C.~A.}\ \bibnamefont {{Wilson-Hodge}}},\ }\href {\doibase
  10.1088/1361-6382/aba28a} {\bibfield  {journal} {\bibinfo  {journal}
  {Classical and Quantum Gravity}\ }\textbf {\bibinfo {volume} {37}},\ \bibinfo
  {eid} {175001} (\bibinfo {year} {2020})},\ \Eprint
  {http://arxiv.org/abs/2001.01462} {arXiv:2001.01462 [gr-qc]} \BibitemShut
  {NoStop}%
\bibitem [{\citenamefont {{Davis}}\ \emph {et~al.}(2020)\citenamefont
  {{Davis}}, \citenamefont {{White}},\ and\ \citenamefont
  {{Saulson}}}]{2020CQGra..37n5001D}%
  \BibitemOpen
  \bibfield  {author} {\bibinfo {author} {\bibfnamefont {D.}~\bibnamefont
  {{Davis}}}, \bibinfo {author} {\bibfnamefont {L.~V.}\ \bibnamefont
  {{White}}}, \ and\ \bibinfo {author} {\bibfnamefont {P.~R.}\ \bibnamefont
  {{Saulson}}},\ }\href {\doibase 10.1088/1361-6382/ab91e6} {\bibfield
  {journal} {\bibinfo  {journal} {Classical and Quantum Gravity}\ }\textbf
  {\bibinfo {volume} {37}},\ \bibinfo {eid} {145001} (\bibinfo {year}
  {2020})},\ \Eprint {http://arxiv.org/abs/2002.09429} {arXiv:2002.09429
  [gr-qc]} \BibitemShut {NoStop}%
\bibitem [{\citenamefont {{Essick}}\ \emph {et~al.}(2020)\citenamefont
  {{Essick}}, \citenamefont {{Godwin}}, \citenamefont {{Hanna}}, \citenamefont
  {{Blackburn}},\ and\ \citenamefont {{Katsavounidis}}}]{2020arXiv200512761E}%
  \BibitemOpen
  \bibfield  {author} {\bibinfo {author} {\bibfnamefont {R.}~\bibnamefont
  {{Essick}}}, \bibinfo {author} {\bibfnamefont {P.}~\bibnamefont {{Godwin}}},
  \bibinfo {author} {\bibfnamefont {C.}~\bibnamefont {{Hanna}}}, \bibinfo
  {author} {\bibfnamefont {L.}~\bibnamefont {{Blackburn}}}, \ and\ \bibinfo
  {author} {\bibfnamefont {E.}~\bibnamefont {{Katsavounidis}}},\ }\href@noop {}
  {\bibfield  {journal} {\bibinfo  {journal} {arXiv e-prints}\ ,\ \bibinfo
  {eid} {arXiv:2005.12761}} (\bibinfo {year} {2020})},\ \Eprint
  {http://arxiv.org/abs/2005.12761} {arXiv:2005.12761 [astro-ph.IM]}
  \BibitemShut {NoStop}%
\bibitem [{\citenamefont {{Cuoco}}(2020)}]{2020arXiv200503745C}%
  \BibitemOpen
  \bibfield  {author} {\bibinfo {author} {\bibfnamefont {E.~e.~a.}\
  \bibnamefont {{Cuoco}}},\ }\href@noop {} {\bibfield  {journal} {\bibinfo
  {journal} {arXiv e-prints}\ ,\ \bibinfo {eid} {arXiv:2005.03745}} (\bibinfo
  {year} {2020})},\ \Eprint {http://arxiv.org/abs/2005.03745} {arXiv:2005.03745
  [astro-ph.HE]} \BibitemShut {NoStop}%
\bibitem [{\citenamefont {{Razzano}}\ and\ \citenamefont
  {{Cuoco}}(2018)}]{2018CQGra..35i5016R}%
  \BibitemOpen
  \bibfield  {author} {\bibinfo {author} {\bibfnamefont {M.}~\bibnamefont
  {{Razzano}}}\ and\ \bibinfo {author} {\bibfnamefont {E.}~\bibnamefont
  {{Cuoco}}},\ }\href {\doibase 10.1088/1361-6382/aab793} {\bibfield  {journal}
  {\bibinfo  {journal} {Classical and Quantum Gravity}\ }\textbf {\bibinfo
  {volume} {35}},\ \bibinfo {eid} {095016} (\bibinfo {year} {2018})},\ \Eprint
  {http://arxiv.org/abs/1803.09933} {arXiv:1803.09933 [gr-qc]} \BibitemShut
  {NoStop}%
\bibitem [{\citenamefont {Mukund}\ \emph {et~al.}(2017)\citenamefont {Mukund},
  \citenamefont {Abraham}, \citenamefont {Kandhasamy}, \citenamefont {Mitra},\
  and\ \citenamefont {Philip}}]{2017PhRvD..95j4059M}%
  \BibitemOpen
  \bibfield  {author} {\bibinfo {author} {\bibfnamefont {N.}~\bibnamefont
  {Mukund}}, \bibinfo {author} {\bibfnamefont {S.}~\bibnamefont {Abraham}},
  \bibinfo {author} {\bibfnamefont {S.}~\bibnamefont {Kandhasamy}}, \bibinfo
  {author} {\bibfnamefont {S.}~\bibnamefont {Mitra}}, \ and\ \bibinfo {author}
  {\bibfnamefont {N.~S.}\ \bibnamefont {Philip}},\ }\href {\doibase
  10.1103/PhysRevD.95.104059} {\bibfield  {journal} {\bibinfo  {journal}
  {Physical Review D}\ }\textbf {\bibinfo {volume} {95}},\ \bibinfo {eid}
  {104059} (\bibinfo {year} {2017})},\ \Eprint
  {http://arxiv.org/abs/1609.07259} {arXiv:1609.07259 [astro-ph.IM]}
  \BibitemShut {NoStop}%
\bibitem [{\citenamefont {{Valdes Sanchez}}(2017)}]{2017PhDT........25V}%
  \BibitemOpen
  \bibfield  {author} {\bibinfo {author} {\bibfnamefont {G.~A.}\ \bibnamefont
  {{Valdes Sanchez}}},\ }\emph {\bibinfo {title} {{Data Analysis Techniques for
  Ligo Detector Characterization}}},\ \href@noop {} {Ph.D. thesis},\ \bibinfo
  {school} {The University of Texas at San Antonio} (\bibinfo {year}
  {2017})\BibitemShut {NoStop}%
\bibitem [{\citenamefont {{Massinger}}(2016)}]{2016PhDT.......149M}%
  \BibitemOpen
  \bibfield  {author} {\bibinfo {author} {\bibfnamefont {T.~J.}\ \bibnamefont
  {{Massinger}}},\ }\emph {\bibinfo {title} {{Detector characterization for
  advanced LIGO}}},\ \href@noop {} {Ph.D. thesis},\ \bibinfo  {school}
  {Syracuse University} (\bibinfo {year} {2016})\BibitemShut {NoStop}%
\bibitem [{\citenamefont {{Nuttall}}\ \emph {et~al.}(2015)\citenamefont
  {{Nuttall}}, \citenamefont {{Massinger}}, \citenamefont {{Areeda}},
  \citenamefont {{Betzwieser}}, \citenamefont {{Dwyer}}, \citenamefont
  {{Effler}}, \citenamefont {{Fisher}}, \citenamefont {{Fritschel}},
  \citenamefont {{Kissel}}, \citenamefont {{Lundgren}}, \citenamefont
  {{Macleod}}, \citenamefont {{Martynov}}, \citenamefont {{McIver}},
  \citenamefont {{Mullavey}}, \citenamefont {{Sigg}}, \citenamefont {{Smith}},
  \citenamefont {{Vajente}}, \citenamefont {{Williamson}},\ and\ \citenamefont
  {{Wipf}}}]{2015CQGra..32x5005N}%
  \BibitemOpen
  \bibfield  {author} {\bibinfo {author} {\bibfnamefont {L.~K.}\ \bibnamefont
  {{Nuttall}}}, \bibinfo {author} {\bibfnamefont {T.~J.}\ \bibnamefont
  {{Massinger}}}, \bibinfo {author} {\bibfnamefont {J.}~\bibnamefont
  {{Areeda}}}, \bibinfo {author} {\bibfnamefont {J.}~\bibnamefont
  {{Betzwieser}}}, \bibinfo {author} {\bibfnamefont {S.}~\bibnamefont
  {{Dwyer}}}, \bibinfo {author} {\bibfnamefont {A.}~\bibnamefont {{Effler}}},
  \bibinfo {author} {\bibfnamefont {R.~P.}\ \bibnamefont {{Fisher}}}, \bibinfo
  {author} {\bibfnamefont {P.}~\bibnamefont {{Fritschel}}}, \bibinfo {author}
  {\bibfnamefont {J.~S.}\ \bibnamefont {{Kissel}}}, \bibinfo {author}
  {\bibfnamefont {A.~P.}\ \bibnamefont {{Lundgren}}}, \bibinfo {author}
  {\bibfnamefont {D.~M.}\ \bibnamefont {{Macleod}}}, \bibinfo {author}
  {\bibfnamefont {D.}~\bibnamefont {{Martynov}}}, \bibinfo {author}
  {\bibfnamefont {J.}~\bibnamefont {{McIver}}}, \bibinfo {author}
  {\bibfnamefont {A.}~\bibnamefont {{Mullavey}}}, \bibinfo {author}
  {\bibfnamefont {D.}~\bibnamefont {{Sigg}}}, \bibinfo {author} {\bibfnamefont
  {J.~R.}\ \bibnamefont {{Smith}}}, \bibinfo {author} {\bibfnamefont
  {G.}~\bibnamefont {{Vajente}}}, \bibinfo {author} {\bibfnamefont {A.~R.}\
  \bibnamefont {{Williamson}}}, \ and\ \bibinfo {author} {\bibfnamefont
  {C.~C.}\ \bibnamefont {{Wipf}}},\ }\href {\doibase
  10.1088/0264-9381/32/24/245005} {\bibfield  {journal} {\bibinfo  {journal}
  {Classical and Quantum Gravity}\ }\textbf {\bibinfo {volume} {32}},\ \bibinfo
  {eid} {245005} (\bibinfo {year} {2015})},\ \Eprint
  {http://arxiv.org/abs/1508.07316} {arXiv:1508.07316 [gr-qc]} \BibitemShut
  {NoStop}%
\bibitem [{\citenamefont {{Biswas}}\ \emph {et~al.}(2013)\citenamefont
  {{Biswas}}, \citenamefont {{Blackburn}}, \citenamefont {{Cao}}, \citenamefont
  {{Essick}}, \citenamefont {{Hodge}}, \citenamefont {{Katsavounidis}},
  \citenamefont {{Kim}}, \citenamefont {{Kim}}, \citenamefont {{Le Bigot}},
  \citenamefont {{Lee}}, \citenamefont {{Oh}}, \citenamefont {{Oh}},
  \citenamefont {{Son}}, \citenamefont {{Tao}}, \citenamefont {{Vaulin}},\ and\
  \citenamefont {{Wang}}}]{2013PhRvD..88f2003B}%
  \BibitemOpen
  \bibfield  {author} {\bibinfo {author} {\bibfnamefont {R.}~\bibnamefont
  {{Biswas}}}, \bibinfo {author} {\bibfnamefont {L.}~\bibnamefont
  {{Blackburn}}}, \bibinfo {author} {\bibfnamefont {J.}~\bibnamefont {{Cao}}},
  \bibinfo {author} {\bibfnamefont {R.}~\bibnamefont {{Essick}}}, \bibinfo
  {author} {\bibfnamefont {K.~A.}\ \bibnamefont {{Hodge}}}, \bibinfo {author}
  {\bibfnamefont {E.}~\bibnamefont {{Katsavounidis}}}, \bibinfo {author}
  {\bibfnamefont {K.}~\bibnamefont {{Kim}}}, \bibinfo {author} {\bibfnamefont
  {Y.-M.}\ \bibnamefont {{Kim}}}, \bibinfo {author} {\bibfnamefont {E.-O.}\
  \bibnamefont {{Le Bigot}}}, \bibinfo {author} {\bibfnamefont {C.-H.}\
  \bibnamefont {{Lee}}}, \bibinfo {author} {\bibfnamefont {J.~J.}\ \bibnamefont
  {{Oh}}}, \bibinfo {author} {\bibfnamefont {S.~H.}\ \bibnamefont {{Oh}}},
  \bibinfo {author} {\bibfnamefont {E.~J.}\ \bibnamefont {{Son}}}, \bibinfo
  {author} {\bibfnamefont {Y.}~\bibnamefont {{Tao}}}, \bibinfo {author}
  {\bibfnamefont {R.}~\bibnamefont {{Vaulin}}}, \ and\ \bibinfo {author}
  {\bibfnamefont {X.}~\bibnamefont {{Wang}}},\ }\href {\doibase
  10.1103/PhysRevD.88.062003} {\bibfield  {journal} {\bibinfo  {journal}
  {Physical Review D}\ }\textbf {\bibinfo {volume} {88}},\ \bibinfo {eid}
  {062003} (\bibinfo {year} {2013})},\ \Eprint {http://arxiv.org/abs/1303.6984}
  {arXiv:1303.6984 [astro-ph.IM]} \BibitemShut {NoStop}%
\bibitem [{\citenamefont {{MacLeod}}(2013)}]{2013PhDT.......555M}%
  \BibitemOpen
  \bibfield  {author} {\bibinfo {author} {\bibfnamefont {D.}~\bibnamefont
  {{MacLeod}}},\ }\emph {\bibinfo {title} {{Improving the sensitivity of
  searches for gravitational waves from compact binary coalescences}}},\
  \href@noop {} {Ph.D. thesis},\ \bibinfo  {school} {Cardiff University (United
  Kingdom)} (\bibinfo {year} {2013})\BibitemShut {NoStop}%
\bibitem [{\citenamefont {{Aasi}}(2012)}]{2012CQGra..29o5002A}%
  \BibitemOpen
  \bibfield  {author} {\bibinfo {author} {\bibfnamefont {J.~e.~a.}\
  \bibnamefont {{Aasi}}},\ }\href {\doibase 10.1088/0264-9381/29/15/155002}
  {\bibfield  {journal} {\bibinfo  {journal} {Classical and Quantum Gravity}\
  }\textbf {\bibinfo {volume} {29}},\ \bibinfo {eid} {155002} (\bibinfo {year}
  {2012})},\ \Eprint {http://arxiv.org/abs/1203.5613} {arXiv:1203.5613 [gr-qc]}
  \BibitemShut {NoStop}%
\bibitem [{\citenamefont {{Christensen}}\ \emph {et~al.}(2010)\citenamefont
  {{Christensen}}, \citenamefont {{LIGO Scientific Collaboration}},\ and\
  \citenamefont {{Virgo Collaboration}}}]{2010CQGra..27s4010C}%
  \BibitemOpen
  \bibfield  {author} {\bibinfo {author} {\bibfnamefont {N.}~\bibnamefont
  {{Christensen}}}, \bibinfo {author} {\bibnamefont {{LIGO Scientific
  Collaboration}}}, \ and\ \bibinfo {author} {\bibnamefont {{Virgo
  Collaboration}}},\ }\href {\doibase 10.1088/0264-9381/27/19/194010}
  {\bibfield  {journal} {\bibinfo  {journal} {Classical and Quantum Gravity}\
  }\textbf {\bibinfo {volume} {27}},\ \bibinfo {eid} {194010} (\bibinfo {year}
  {2010})}\BibitemShut {NoStop}%
\bibitem [{\citenamefont {Isogai}\ \emph {et~al.}(2010)\citenamefont {Isogai},
  \citenamefont {Collaboration}, \citenamefont {Collaboration} \emph
  {et~al.}}]{2010JPhCS.243a2005I}%
  \BibitemOpen
  \bibfield  {author} {\bibinfo {author} {\bibfnamefont {T.}~\bibnamefont
  {Isogai}}, \bibinfo {author} {\bibfnamefont {L.~S.}\ \bibnamefont
  {Collaboration}}, \bibinfo {author} {\bibfnamefont {V.}~\bibnamefont
  {Collaboration}},  \emph {et~al.},\ }in\ \href {\doibase
  10.1088/1742-6596/243/1/012005} {\emph {\bibinfo {booktitle} {Journal of
  Physics Conference Series}}},\ Vol.\ \bibinfo {volume} {243}\ (\bibinfo
  {organization} {IOP Publishing},\ \bibinfo {year} {2010})\ p.\ \bibinfo
  {pages} {012005}\BibitemShut {NoStop}%
\bibitem [{\citenamefont {{Mukherjee}}\ \emph {et~al.}(2010)\citenamefont
  {{Mukherjee}}, \citenamefont {{Obaid}},\ and\ \citenamefont
  {{Matkarimov}}}]{2010JPhCS.243a2006M}%
  \BibitemOpen
  \bibfield  {author} {\bibinfo {author} {\bibfnamefont {S.}~\bibnamefont
  {{Mukherjee}}}, \bibinfo {author} {\bibfnamefont {R.}~\bibnamefont
  {{Obaid}}}, \ and\ \bibinfo {author} {\bibfnamefont {B.}~\bibnamefont
  {{Matkarimov}}},\ }in\ \href {\doibase 10.1088/1742-6596/243/1/012006} {\emph
  {\bibinfo {booktitle} {Journal of Physics Conference Series}}},\ \bibinfo
  {series} {Journal of Physics Conference Series}, Vol.\ \bibinfo {volume}
  {243}\ (\bibinfo {year} {2010})\ p.\ \bibinfo {pages} {012006}\BibitemShut
  {NoStop}%
\bibitem [{\citenamefont {{Blackburn}}(2008)}]{2008CQGra..25r4004B}%
  \BibitemOpen
  \bibfield  {author} {\bibinfo {author} {\bibfnamefont {L.~e.~a.}\
  \bibnamefont {{Blackburn}}},\ }\href {\doibase
  10.1088/0264-9381/25/18/184004} {\bibfield  {journal} {\bibinfo  {journal}
  {Classical and Quantum Gravity}\ }\textbf {\bibinfo {volume} {25}},\ \bibinfo
  {eid} {184004} (\bibinfo {year} {2008})},\ \Eprint
  {http://arxiv.org/abs/0804.0800} {arXiv:0804.0800 [gr-qc]} \BibitemShut
  {NoStop}%
\bibitem [{\citenamefont {{Sigg}}\ \emph {et~al.}(2002)\citenamefont {{Sigg}},
  \citenamefont {{Bork}},\ and\ \citenamefont
  {{Zweizig}}}]{2002nmgm.meet.1841S}%
  \BibitemOpen
  \bibfield  {author} {\bibinfo {author} {\bibfnamefont {D.}~\bibnamefont
  {{Sigg}}}, \bibinfo {author} {\bibfnamefont {R.}~\bibnamefont {{Bork}}}, \
  and\ \bibinfo {author} {\bibfnamefont {J.}~\bibnamefont {{Zweizig}}},\ }in\
  \href {\doibase 10.1142/9789812777386\_0401} {\emph {\bibinfo {booktitle}
  {The Ninth Marcel Grossmann Meeting}}},\ \bibinfo {editor} {edited by\
  \bibinfo {editor} {\bibfnamefont {V.~G.}\ \bibnamefont {{Gurzadyan}}},
  \bibinfo {editor} {\bibfnamefont {R.~T.}\ \bibnamefont {{Jantzen}}}, \ and\
  \bibinfo {editor} {\bibfnamefont {R.}~\bibnamefont {{Ruffini}}}}\ (\bibinfo
  {year} {2002})\ pp.\ \bibinfo {pages} {1841--1842}\BibitemShut {NoStop}%
\bibitem [{\citenamefont {Gurav}\ \emph {et~al.}(2020)\citenamefont {Gurav},
  \citenamefont {Barish}, \citenamefont {Vajente},\ and\ \citenamefont
  {Papalexakis}}]{gurav2020unsupervised}%
  \BibitemOpen
  \bibfield  {author} {\bibinfo {author} {\bibfnamefont {R.}~\bibnamefont
  {Gurav}}, \bibinfo {author} {\bibfnamefont {B.}~\bibnamefont {Barish}},
  \bibinfo {author} {\bibfnamefont {G.}~\bibnamefont {Vajente}}, \ and\
  \bibinfo {author} {\bibfnamefont {E.~E.}\ \bibnamefont {Papalexakis}},\ }in\
  \href@noop {} {\emph {\bibinfo {booktitle} {AAI 2020 Fall Symposium on
  Physics-Guided AI to Accelerate Scientific Discovery}}}\ (\bibinfo {year}
  {2020})\BibitemShut {NoStop}%
\bibitem [{\citenamefont {Colgan}\ \emph {et~al.}()\citenamefont {Colgan},
  \citenamefont {Corley}, \citenamefont {Lau}, \citenamefont {Bartos},
  \citenamefont {Wright}, \citenamefont {M\'arka},\ and\ \citenamefont
  {M\'arka}}]{EMU}%
  \BibitemOpen
  \bibfield  {author} {\bibinfo {author} {\bibfnamefont {R.~E.}\ \bibnamefont
  {Colgan}}, \bibinfo {author} {\bibfnamefont {K.~R.}\ \bibnamefont {Corley}},
  \bibinfo {author} {\bibfnamefont {Y.}~\bibnamefont {Lau}}, \bibinfo {author}
  {\bibfnamefont {I.}~\bibnamefont {Bartos}}, \bibinfo {author} {\bibfnamefont
  {J.~N.}\ \bibnamefont {Wright}}, \bibinfo {author} {\bibfnamefont
  {Z.}~\bibnamefont {M\'arka}}, \ and\ \bibinfo {author} {\bibfnamefont
  {S.}~\bibnamefont {M\'arka}},\ }\href {\doibase 10.1103/PhysRevD.101.102003}
  {\ \textbf {\bibinfo {volume} {101}},\ \bibinfo {pages} {102003}}\BibitemShut
  {NoStop}%
\bibitem [{\citenamefont {Colgan}\ \emph {et~al.}(2022)\citenamefont {Colgan},
  \citenamefont {Yan}, \citenamefont {Márka}, \citenamefont {Bartos},
  \citenamefont {Márka},\ and\ \citenamefont {Wright}}]{learned_features}%
  \BibitemOpen
  \bibfield  {author} {\bibinfo {author} {\bibfnamefont {R.~E.}\ \bibnamefont
  {Colgan}}, \bibinfo {author} {\bibfnamefont {J.}~\bibnamefont {Yan}},
  \bibinfo {author} {\bibfnamefont {Z.}~\bibnamefont {Márka}}, \bibinfo
  {author} {\bibfnamefont {I.}~\bibnamefont {Bartos}}, \bibinfo {author}
  {\bibfnamefont {S.}~\bibnamefont {Márka}}, \ and\ \bibinfo {author}
  {\bibfnamefont {J.~N.}\ \bibnamefont {Wright}},\ }\href@noop {} {\enquote
  {\bibinfo {title} {Architectural optimization and feature learning for
  high-dimensional time series datasets},}\ } (\bibinfo {year} {2022}),\
  \Eprint {http://arxiv.org/abs/2202.13486} {arXiv:2202.13486 [cs.LG]}
  \BibitemShut {NoStop}%
\bibitem [{\citenamefont {Abbott}\ \emph {et~al.}(2016)\citenamefont {Abbott},
  \citenamefont {Abbott}, \citenamefont {Abbott}, \citenamefont {Abernathy},
  \citenamefont {Acernese}, \citenamefont {Ackley}, \citenamefont {Adamo},
  \citenamefont {Adams}, \citenamefont {Adams}, \citenamefont {Addesso} \emph
  {et~al.}}]{2016CQGra..33m4001A}%
  \BibitemOpen
  \bibfield  {author} {\bibinfo {author} {\bibfnamefont {B.~P.}\ \bibnamefont
  {Abbott}}, \bibinfo {author} {\bibfnamefont {R.}~\bibnamefont {Abbott}},
  \bibinfo {author} {\bibfnamefont {T.}~\bibnamefont {Abbott}}, \bibinfo
  {author} {\bibfnamefont {M.}~\bibnamefont {Abernathy}}, \bibinfo {author}
  {\bibfnamefont {F.}~\bibnamefont {Acernese}}, \bibinfo {author}
  {\bibfnamefont {K.}~\bibnamefont {Ackley}}, \bibinfo {author} {\bibfnamefont
  {M.}~\bibnamefont {Adamo}}, \bibinfo {author} {\bibfnamefont
  {C.}~\bibnamefont {Adams}}, \bibinfo {author} {\bibfnamefont
  {T.}~\bibnamefont {Adams}}, \bibinfo {author} {\bibfnamefont
  {P.}~\bibnamefont {Addesso}},  \emph {et~al.},\ }\href {\doibase
  10.1088/0264-9381/33/13/134001} {\bibfield  {journal} {\bibinfo  {journal}
  {Classical and Quantum Gravity}\ }\textbf {\bibinfo {volume} {33}},\ \bibinfo
  {eid} {134001} (\bibinfo {year} {2016})},\ \Eprint
  {http://arxiv.org/abs/1602.03844} {arXiv:1602.03844 [gr-qc]} \BibitemShut
  {NoStop}%
\bibitem [{\citenamefont {Davis}\ and\ \citenamefont
  {Walker}(2022)}]{galaxies10010012}%
  \BibitemOpen
  \bibfield  {author} {\bibinfo {author} {\bibfnamefont {D.}~\bibnamefont
  {Davis}}\ and\ \bibinfo {author} {\bibfnamefont {M.}~\bibnamefont {Walker}},\
  }\href {\doibase 10.3390/galaxies10010012} {\bibfield  {journal} {\bibinfo
  {journal} {Galaxies}\ }\textbf {\bibinfo {volume} {10}} (\bibinfo {year}
  {2022}),\ 10.3390/galaxies10010012}\BibitemShut {NoStop}%
\bibitem [{\citenamefont {{Merritt}}\ \emph
  {et~al.}(2021{\natexlab{b}})\citenamefont {{Merritt}}, \citenamefont
  {{Farr}}, \citenamefont {{Hur}}, \citenamefont {{Edelman}},\ and\
  \citenamefont {{Doctor}}}]{2021PhRvD.104j2004M}%
  \BibitemOpen
  \bibfield  {author} {\bibinfo {author} {\bibfnamefont {J.~D.}\ \bibnamefont
  {{Merritt}}}, \bibinfo {author} {\bibfnamefont {B.}~\bibnamefont {{Farr}}},
  \bibinfo {author} {\bibfnamefont {R.}~\bibnamefont {{Hur}}}, \bibinfo
  {author} {\bibfnamefont {B.}~\bibnamefont {{Edelman}}}, \ and\ \bibinfo
  {author} {\bibfnamefont {Z.}~\bibnamefont {{Doctor}}},\ }\href {\doibase
  10.1103/PhysRevD.104.102004} {\bibfield  {journal} {\bibinfo  {journal}
  {\prd}\ }\textbf {\bibinfo {volume} {104}},\ \bibinfo {eid} {102004}
  (\bibinfo {year} {2021}{\natexlab{b}})},\ \Eprint
  {http://arxiv.org/abs/2108.12044} {arXiv:2108.12044 [gr-qc]} \BibitemShut
  {NoStop}%
\bibitem [{\citenamefont {{Nguyen}}(2021)}]{2021CQGra..38n5001N}%
  \BibitemOpen
  \bibfield  {author} {\bibinfo {author} {\bibfnamefont {P.~e.~a.}\
  \bibnamefont {{Nguyen}}},\ }\href {\doibase 10.1088/1361-6382/ac011a}
  {\bibfield  {journal} {\bibinfo  {journal} {Classical and Quantum Gravity}\
  }\textbf {\bibinfo {volume} {38}},\ \bibinfo {eid} {145001} (\bibinfo {year}
  {2021})},\ \Eprint {http://arxiv.org/abs/2101.09935} {arXiv:2101.09935
  [astro-ph.IM]} \BibitemShut {NoStop}%
\bibitem [{\citenamefont {{White}}\ and\ \citenamefont {{LIGO
  Team}}(2019)}]{2019APS..APRK01063W}%
  \BibitemOpen
  \bibfield  {author} {\bibinfo {author} {\bibfnamefont {L.}~\bibnamefont
  {{White}}}\ and\ \bibinfo {author} {\bibnamefont {{LIGO Team}}},\ }in\
  \href@noop {} {\emph {\bibinfo {booktitle} {APS April Meeting Abstracts}}},\
  \bibinfo {series} {APS Meeting Abstracts}, Vol.\ \bibinfo {volume} {2019}\
  (\bibinfo {year} {2019})\ p.\ \bibinfo {pages} {K01.063}\BibitemShut
  {NoStop}%
\bibitem [{\citenamefont {{Davis}}(2019)}]{2019PhDT.......117D}%
  \BibitemOpen
  \bibfield  {author} {\bibinfo {author} {\bibfnamefont {D.}~\bibnamefont
  {{Davis}}},\ }\emph {\bibinfo {title} {{Improving the sensitivity of Advanced
  LIGO through detector characterization}}},\ \href@noop {} {Ph.D. thesis},\
  \bibinfo  {school} {Syracuse University, United States} (\bibinfo {year}
  {2019})\BibitemShut {NoStop}%
\bibitem [{\citenamefont {{Robinet}}(2015)}]{Omicron}%
  \BibitemOpen
  \bibfield  {author} {\bibinfo {author} {\bibfnamefont {F.}~\bibnamefont
  {{Robinet}}},\ }\href@noop {} {\enquote {\bibinfo {title} {{Omicron: An
  Algorithm to Detect and Characterize Transient Noise in Gravitational-Wave
  Detectors}},}\ }\bibinfo {howpublished}
  {\url{https://tds.ego-gw.it/ql/?c=10651}} (\bibinfo {year}
  {2015})\BibitemShut {NoStop}%
\bibitem [{\citenamefont {Zou}\ and\ \citenamefont {Hastie}(2005)}]{elastic}%
  \BibitemOpen
  \bibfield  {author} {\bibinfo {author} {\bibfnamefont {H.}~\bibnamefont
  {Zou}}\ and\ \bibinfo {author} {\bibfnamefont {T.}~\bibnamefont {Hastie}},\
  }\href {http://www.jstor.org/stable/3647580} {\bibfield  {journal} {\bibinfo
  {journal} {Journal of the Royal Statistical Society. Series B (Statistical
  Methodology)}\ }\textbf {\bibinfo {volume} {67}},\ \bibinfo {pages} {301}
  (\bibinfo {year} {2005})}\BibitemShut {NoStop}%
\bibitem [{\citenamefont {Rollins}(2011)}]{rollins_thesis}%
  \BibitemOpen
  \bibfield  {author} {\bibinfo {author} {\bibfnamefont {J.}~\bibnamefont
  {Rollins}},\ }\emph {\bibinfo {title} {Multimessenger Astronomy with
  Low-Latency Searches for Transient Gravitational Waves}},\ \href@noop {}
  {Ph.D. thesis},\ \bibinfo  {school} {Columbia University} (\bibinfo {year}
  {2011})\BibitemShut {NoStop}%
\bibitem [{\citenamefont {Chatterji}\ \emph {et~al.}(2004)\citenamefont
  {Chatterji}, \citenamefont {Blackburn}, \citenamefont {Martin},\ and\
  \citenamefont {Katsavounidis}}]{2004CQGra..21S1809C}%
  \BibitemOpen
  \bibfield  {author} {\bibinfo {author} {\bibfnamefont {S.}~\bibnamefont
  {Chatterji}}, \bibinfo {author} {\bibfnamefont {L.}~\bibnamefont
  {Blackburn}}, \bibinfo {author} {\bibfnamefont {G.}~\bibnamefont {Martin}}, \
  and\ \bibinfo {author} {\bibfnamefont {E.}~\bibnamefont {Katsavounidis}},\
  }\href {\doibase 10.1088/0264-9381/21/20/024} {\bibfield  {journal} {\bibinfo
   {journal} {Classical and Quantum Gravity}\ }\textbf {\bibinfo {volume}
  {21}},\ \bibinfo {pages} {S1809} (\bibinfo {year} {2004})},\ \Eprint
  {http://arxiv.org/abs/gr-qc/0412119} {arXiv:gr-qc/0412119 [gr-qc]}
  \BibitemShut {NoStop}%
\bibitem [{\citenamefont {Soni}\ and\ \citenamefont
  {Collaboration)}()}]{Soni_2021}%
  \BibitemOpen
  \bibfield  {author} {\bibinfo {author} {\bibfnamefont {S.}~\bibnamefont
  {Soni}}\ and\ \bibinfo {author} {\bibfnamefont {T.~L.~S.}\ \bibnamefont
  {Collaboration)}},\ }\href {\doibase 10.1088/1361-6382/abc906} {\ \textbf
  {\bibinfo {volume} {38}},\ \bibinfo {pages} {025016}}\BibitemShut {NoStop}%
\bibitem [{\citenamefont {{Bianchi}}\ \emph {et~al.}(2021)\citenamefont
  {{Bianchi}}, \citenamefont {{Longo}}, \citenamefont {{Valdes}}, \citenamefont
  {{Gonz{\'a}lez}},\ and\ \citenamefont {{Plastino}}}]{2021arXiv210707565B}%
  \BibitemOpen
  \bibfield  {author} {\bibinfo {author} {\bibfnamefont {S.}~\bibnamefont
  {{Bianchi}}}, \bibinfo {author} {\bibfnamefont {A.}~\bibnamefont {{Longo}}},
  \bibinfo {author} {\bibfnamefont {G.}~\bibnamefont {{Valdes}}}, \bibinfo
  {author} {\bibfnamefont {G.}~\bibnamefont {{Gonz{\'a}lez}}}, \ and\ \bibinfo
  {author} {\bibfnamefont {W.}~\bibnamefont {{Plastino}}},\ }\href@noop {}
  {\bibfield  {journal} {\bibinfo  {journal} {arXiv e-prints}\ ,\ \bibinfo
  {eid} {arXiv:2107.07565}} (\bibinfo {year} {2021})},\ \Eprint
  {http://arxiv.org/abs/2107.07565} {arXiv:2107.07565 [astro-ph.IM]}
  \BibitemShut {NoStop}%
\bibitem [{\citenamefont {{Austin}}(2020)}]{2020PhDT........41A}%
  \BibitemOpen
  \bibfield  {author} {\bibinfo {author} {\bibfnamefont {C.~D.}\ \bibnamefont
  {{Austin}}},\ }\emph {\bibinfo {title} {{Measurements and mitigation of
  scattered light noise in LIGO}}},\ \href@noop {} {Ph.D. thesis},\ \bibinfo
  {school} {Louisiana State University, United States} (\bibinfo {year}
  {2020})\BibitemShut {NoStop}%
\bibitem [{\citenamefont {{Maga{\~n}a-Sandoval}}\ \emph
  {et~al.}(2012)\citenamefont {{Maga{\~n}a-Sandoval}}, \citenamefont
  {{Adhikari}}, \citenamefont {{Frolov}}, \citenamefont {{Harms}},
  \citenamefont {{Lee}}, \citenamefont {{Sankar}}, \citenamefont {{Saulson}},\
  and\ \citenamefont {{Smith}}}]{2012JOSAA..29.1722M}%
  \BibitemOpen
  \bibfield  {author} {\bibinfo {author} {\bibfnamefont {F.}~\bibnamefont
  {{Maga{\~n}a-Sandoval}}}, \bibinfo {author} {\bibfnamefont {R.~X.}\
  \bibnamefont {{Adhikari}}}, \bibinfo {author} {\bibfnamefont
  {V.}~\bibnamefont {{Frolov}}}, \bibinfo {author} {\bibfnamefont
  {J.}~\bibnamefont {{Harms}}}, \bibinfo {author} {\bibfnamefont
  {J.}~\bibnamefont {{Lee}}}, \bibinfo {author} {\bibfnamefont
  {S.}~\bibnamefont {{Sankar}}}, \bibinfo {author} {\bibfnamefont {P.~R.}\
  \bibnamefont {{Saulson}}}, \ and\ \bibinfo {author} {\bibfnamefont {J.~R.}\
  \bibnamefont {{Smith}}},\ }\href {\doibase 10.1364/JOSAA.29.001722}
  {\bibfield  {journal} {\bibinfo  {journal} {Journal of the Optical Society of
  America A}\ }\textbf {\bibinfo {volume} {29}},\ \bibinfo {pages} {1722}
  (\bibinfo {year} {2012})},\ \Eprint {http://arxiv.org/abs/1204.2528}
  {arXiv:1204.2528 [physics.optics]} \BibitemShut {NoStop}%
\bibitem [{\citenamefont {{The Virgo
  Collaboration}}(2011)}]{2011arXiv1108.1598T}%
  \BibitemOpen
  \bibfield  {author} {\bibinfo {author} {\bibnamefont {{The Virgo
  Collaboration}}},\ }\href@noop {} {\bibfield  {journal} {\bibinfo  {journal}
  {arXiv e-prints}\ ,\ \bibinfo {eid} {arXiv:1108.1598}} (\bibinfo {year}
  {2011})},\ \Eprint {http://arxiv.org/abs/1108.1598} {arXiv:1108.1598 [gr-qc]}
  \BibitemShut {NoStop}%
\bibitem [{\citenamefont {{Vinet}}\ \emph {et~al.}(1997)\citenamefont
  {{Vinet}}, \citenamefont {{Brisson}}, \citenamefont {{Braccini}},
  \citenamefont {{Ferrante}}, \citenamefont {{Pinard}}, \citenamefont
  {{Bondu}},\ and\ \citenamefont {{Tourni{\'e}}}}]{1997PhRvD..56.6085V}%
  \BibitemOpen
  \bibfield  {author} {\bibinfo {author} {\bibfnamefont {J.-Y.}\ \bibnamefont
  {{Vinet}}}, \bibinfo {author} {\bibfnamefont {V.}~\bibnamefont {{Brisson}}},
  \bibinfo {author} {\bibfnamefont {S.}~\bibnamefont {{Braccini}}}, \bibinfo
  {author} {\bibfnamefont {I.}~\bibnamefont {{Ferrante}}}, \bibinfo {author}
  {\bibfnamefont {L.}~\bibnamefont {{Pinard}}}, \bibinfo {author}
  {\bibfnamefont {F.}~\bibnamefont {{Bondu}}}, \ and\ \bibinfo {author}
  {\bibfnamefont {E.}~\bibnamefont {{Tourni{\'e}}}},\ }\href {\doibase
  10.1103/PhysRevD.56.6085} {\bibfield  {journal} {\bibinfo  {journal} {\prd}\
  }\textbf {\bibinfo {volume} {56}},\ \bibinfo {pages} {6085} (\bibinfo {year}
  {1997})}\BibitemShut {NoStop}%
\bibitem [{\citenamefont {{Saha}}(1997)}]{1997PhDT.......130S}%
  \BibitemOpen
  \bibfield  {author} {\bibinfo {author} {\bibfnamefont {P.}~\bibnamefont
  {{Saha}}},\ }\emph {\bibinfo {title} {{Noise analysis of a suspended high
  power Michelson interferometer}}},\ \href@noop {} {Ph.D. thesis},\ \bibinfo
  {school} {MASSACHUSETTS INSTITUTE OF TECHNOLOGY} (\bibinfo {year}
  {1997})\BibitemShut {NoStop}%
\bibitem [{\citenamefont {Asali}\ and\ \citenamefont {Marka}()}]{T2000052}%
  \BibitemOpen
  \bibfield  {author} {\bibinfo {author} {\bibfnamefont {Y.}~\bibnamefont
  {Asali}}\ and\ \bibinfo {author} {\bibfnamefont {Z.}~\bibnamefont {Marka}},\
  }\href@noop {} {\enquote {\bibinfo {title} {Effect of commissioning break on
  scattering glitches, ligo document t2000052},}\ }\BibitemShut {NoStop}%
\bibitem [{\citenamefont {{Matichard}}\ \emph {et~al.}(2015)\citenamefont
  {{Matichard}}, \citenamefont {{Lantz}}, \citenamefont {{Mittleman}},
  \citenamefont {{Mason}}, \citenamefont {{Kissel}}, \citenamefont {{Abbott}},
  \citenamefont {{Biscans}}, \citenamefont {{McIver}}, \citenamefont
  {{Abbott}}, \citenamefont {{Abbott}}, \citenamefont {{Allwine}},
  \citenamefont {{Barnum}}, \citenamefont {{Birch}}, \citenamefont
  {{Celerier}}, \citenamefont {{Clark}}, \citenamefont {{Coyne}}, \citenamefont
  {{DeBra}}, \citenamefont {{DeRosa}}, \citenamefont {{Evans}}, \citenamefont
  {{Foley}}, \citenamefont {{Fritschel}}, \citenamefont {{Giaime}},
  \citenamefont {{Gray}}, \citenamefont {{Grabeel}}, \citenamefont {{Hanson}},
  \citenamefont {{Hardham}}, \citenamefont {{Hillard}}, \citenamefont {{Hua}},
  \citenamefont {{Kucharczyk}}, \citenamefont {{Landry}}, \citenamefont {{Le
  Roux}}, \citenamefont {{Lhuillier}}, \citenamefont {{Macleod}}, \citenamefont
  {{Macinnis}}, \citenamefont {{Mitchell}}, \citenamefont {{O'Reilly}},
  \citenamefont {{Ottaway}}, \citenamefont {{Paris}}, \citenamefont {{Pele}},
  \citenamefont {{Puma}}, \citenamefont {{Radkins}}, \citenamefont {{Ramet}},
  \citenamefont {{Robinson}}, \citenamefont {{Ruet}}, \citenamefont {{Sarin}},
  \citenamefont {{Shoemaker}}, \citenamefont {{Stein}}, \citenamefont
  {{Thomas}}, \citenamefont {{Vargas}}, \citenamefont {{Venkateswara}},
  \citenamefont {{Warner}},\ and\ \citenamefont {{Wen}}}]{2015CQGra..32r5003M}%
  \BibitemOpen
  \bibfield  {author} {\bibinfo {author} {\bibfnamefont {F.}~\bibnamefont
  {{Matichard}}}, \bibinfo {author} {\bibfnamefont {B.}~\bibnamefont
  {{Lantz}}}, \bibinfo {author} {\bibfnamefont {R.}~\bibnamefont
  {{Mittleman}}}, \bibinfo {author} {\bibfnamefont {K.}~\bibnamefont
  {{Mason}}}, \bibinfo {author} {\bibfnamefont {J.}~\bibnamefont {{Kissel}}},
  \bibinfo {author} {\bibfnamefont {B.}~\bibnamefont {{Abbott}}}, \bibinfo
  {author} {\bibfnamefont {S.}~\bibnamefont {{Biscans}}}, \bibinfo {author}
  {\bibfnamefont {J.}~\bibnamefont {{McIver}}}, \bibinfo {author}
  {\bibfnamefont {R.}~\bibnamefont {{Abbott}}}, \bibinfo {author}
  {\bibfnamefont {S.}~\bibnamefont {{Abbott}}}, \bibinfo {author}
  {\bibfnamefont {E.}~\bibnamefont {{Allwine}}}, \bibinfo {author}
  {\bibfnamefont {S.}~\bibnamefont {{Barnum}}}, \bibinfo {author}
  {\bibfnamefont {J.}~\bibnamefont {{Birch}}}, \bibinfo {author} {\bibfnamefont
  {C.}~\bibnamefont {{Celerier}}}, \bibinfo {author} {\bibfnamefont
  {D.}~\bibnamefont {{Clark}}}, \bibinfo {author} {\bibfnamefont
  {D.}~\bibnamefont {{Coyne}}}, \bibinfo {author} {\bibfnamefont
  {D.}~\bibnamefont {{DeBra}}}, \bibinfo {author} {\bibfnamefont
  {R.}~\bibnamefont {{DeRosa}}}, \bibinfo {author} {\bibfnamefont
  {M.}~\bibnamefont {{Evans}}}, \bibinfo {author} {\bibfnamefont
  {S.}~\bibnamefont {{Foley}}}, \bibinfo {author} {\bibfnamefont
  {P.}~\bibnamefont {{Fritschel}}}, \bibinfo {author} {\bibfnamefont {J.~A.}\
  \bibnamefont {{Giaime}}}, \bibinfo {author} {\bibfnamefont {C.}~\bibnamefont
  {{Gray}}}, \bibinfo {author} {\bibfnamefont {G.}~\bibnamefont {{Grabeel}}},
  \bibinfo {author} {\bibfnamefont {J.}~\bibnamefont {{Hanson}}}, \bibinfo
  {author} {\bibfnamefont {C.}~\bibnamefont {{Hardham}}}, \bibinfo {author}
  {\bibfnamefont {M.}~\bibnamefont {{Hillard}}}, \bibinfo {author}
  {\bibfnamefont {W.}~\bibnamefont {{Hua}}}, \bibinfo {author} {\bibfnamefont
  {C.}~\bibnamefont {{Kucharczyk}}}, \bibinfo {author} {\bibfnamefont
  {M.}~\bibnamefont {{Landry}}}, \bibinfo {author} {\bibfnamefont
  {A.}~\bibnamefont {{Le Roux}}}, \bibinfo {author} {\bibfnamefont
  {V.}~\bibnamefont {{Lhuillier}}}, \bibinfo {author} {\bibfnamefont
  {D.}~\bibnamefont {{Macleod}}}, \bibinfo {author} {\bibfnamefont
  {M.}~\bibnamefont {{Macinnis}}}, \bibinfo {author} {\bibfnamefont
  {R.}~\bibnamefont {{Mitchell}}}, \bibinfo {author} {\bibfnamefont
  {B.}~\bibnamefont {{O'Reilly}}}, \bibinfo {author} {\bibfnamefont
  {D.}~\bibnamefont {{Ottaway}}}, \bibinfo {author} {\bibfnamefont
  {H.}~\bibnamefont {{Paris}}}, \bibinfo {author} {\bibfnamefont
  {A.}~\bibnamefont {{Pele}}}, \bibinfo {author} {\bibfnamefont
  {M.}~\bibnamefont {{Puma}}}, \bibinfo {author} {\bibfnamefont
  {H.}~\bibnamefont {{Radkins}}}, \bibinfo {author} {\bibfnamefont
  {C.}~\bibnamefont {{Ramet}}}, \bibinfo {author} {\bibfnamefont
  {M.}~\bibnamefont {{Robinson}}}, \bibinfo {author} {\bibfnamefont
  {L.}~\bibnamefont {{Ruet}}}, \bibinfo {author} {\bibfnamefont
  {P.}~\bibnamefont {{Sarin}}}, \bibinfo {author} {\bibfnamefont
  {D.}~\bibnamefont {{Shoemaker}}}, \bibinfo {author} {\bibfnamefont
  {A.}~\bibnamefont {{Stein}}}, \bibinfo {author} {\bibfnamefont
  {J.}~\bibnamefont {{Thomas}}}, \bibinfo {author} {\bibfnamefont
  {M.}~\bibnamefont {{Vargas}}}, \bibinfo {author} {\bibfnamefont
  {K.}~\bibnamefont {{Venkateswara}}}, \bibinfo {author} {\bibfnamefont
  {J.}~\bibnamefont {{Warner}}}, \ and\ \bibinfo {author} {\bibfnamefont
  {S.}~\bibnamefont {{Wen}}},\ }\href {\doibase 10.1088/0264-9381/32/18/185003}
  {\bibfield  {journal} {\bibinfo  {journal} {Classical and Quantum Gravity}\
  }\textbf {\bibinfo {volume} {32}},\ \bibinfo {eid} {185003} (\bibinfo {year}
  {2015})},\ \Eprint {http://arxiv.org/abs/1502.06300} {arXiv:1502.06300
  [physics.ins-det]} \BibitemShut {NoStop}%
\bibitem [{\citenamefont {{Sigg}}(2016)}]{2016SPIE.9960E..09S}%
  \BibitemOpen
  \bibfield  {author} {\bibinfo {author} {\bibfnamefont {D.}~\bibnamefont
  {{Sigg}}},\ }in\ \href {\doibase 10.1117/12.2243115} {\emph {\bibinfo
  {booktitle} {Interferometry XVIII}}},\ \bibinfo {series} {Society of
  Photo-Optical Instrumentation Engineers (SPIE) Conference Series}, Vol.\
  \bibinfo {volume} {9960},\ \bibinfo {editor} {edited by\ \bibinfo {editor}
  {\bibfnamefont {K.}~\bibnamefont {{Creath}}}, \bibinfo {editor}
  {\bibfnamefont {J.}~\bibnamefont {{Burke}}}, \ and\ \bibinfo {editor}
  {\bibfnamefont {A.}~\bibnamefont {{Albertazzi Gon{\c{c}}alves}}}}\ (\bibinfo
  {year} {2016})\ p.\ \bibinfo {pages} {996009}\BibitemShut {NoStop}%
\bibitem [{\citenamefont {{Rollins}}(2016)}]{2016RScI...87i4502R}%
  \BibitemOpen
  \bibfield  {author} {\bibinfo {author} {\bibfnamefont {J.~G.}\ \bibnamefont
  {{Rollins}}},\ }\href {\doibase 10.1063/1.4961665} {\bibfield  {journal}
  {\bibinfo  {journal} {Review of Scientific Instruments}\ }\textbf {\bibinfo
  {volume} {87}},\ \bibinfo {eid} {094502} (\bibinfo {year}
  {2016})}\BibitemShut {NoStop}%
\bibitem [{\citenamefont {Buikema}\ \emph {et~al.}()\citenamefont {Buikema},
  \citenamefont {Cahillane}, \citenamefont {Mansell}, \citenamefont {Blair},
  \citenamefont {Abbott}, \citenamefont {Adams}, \citenamefont {Adhikari},
  \citenamefont {Ananyeva}, \citenamefont {Appert}, \citenamefont {Arai} \emph
  {et~al.}}]{buikema2020sensitivity}%
  \BibitemOpen
  \bibfield  {author} {\bibinfo {author} {\bibfnamefont {A.}~\bibnamefont
  {Buikema}}, \bibinfo {author} {\bibfnamefont {C.}~\bibnamefont {Cahillane}},
  \bibinfo {author} {\bibfnamefont {G.}~\bibnamefont {Mansell}}, \bibinfo
  {author} {\bibfnamefont {C.}~\bibnamefont {Blair}}, \bibinfo {author}
  {\bibfnamefont {R.}~\bibnamefont {Abbott}}, \bibinfo {author} {\bibfnamefont
  {C.}~\bibnamefont {Adams}}, \bibinfo {author} {\bibfnamefont
  {R.}~\bibnamefont {Adhikari}}, \bibinfo {author} {\bibfnamefont
  {A.}~\bibnamefont {Ananyeva}}, \bibinfo {author} {\bibfnamefont
  {S.}~\bibnamefont {Appert}}, \bibinfo {author} {\bibfnamefont
  {K.}~\bibnamefont {Arai}},  \emph {et~al.},\ }\href {\doibase
  10.1103/PhysRevD.102.062003} {\ \textbf {\bibinfo {volume} {102}},\ \bibinfo
  {eid} {062003}},\ \Eprint {http://arxiv.org/abs/2008.01301} {2008.01301}
  \BibitemShut {NoStop}%
\bibitem [{\citenamefont {Shapiro}()}]{G1100866}%
  \BibitemOpen
  \bibfield  {author} {\bibinfo {author} {\bibfnamefont {B.}~\bibnamefont
  {Shapiro}},\ }\href@noop {} {\enquote {\bibinfo {title} {Overview of advanced
  ligo suspensions, ligo document g1100866},}\ }\BibitemShut {NoStop}%
\bibitem [{\citenamefont {{Shapiro}}\ \emph {et~al.}(2015)\citenamefont
  {{Shapiro}}, \citenamefont {{Adhikari}}, \citenamefont {{Driggers}},
  \citenamefont {{Kissel}}, \citenamefont {{Lantz}}, \citenamefont
  {{Rollins}},\ and\ \citenamefont {{Youcef-Toumi}}}]{2015CQGra..32a5004S}%
  \BibitemOpen
  \bibfield  {author} {\bibinfo {author} {\bibfnamefont {B.~N.}\ \bibnamefont
  {{Shapiro}}}, \bibinfo {author} {\bibfnamefont {R.}~\bibnamefont
  {{Adhikari}}}, \bibinfo {author} {\bibfnamefont {J.}~\bibnamefont
  {{Driggers}}}, \bibinfo {author} {\bibfnamefont {J.}~\bibnamefont
  {{Kissel}}}, \bibinfo {author} {\bibfnamefont {B.}~\bibnamefont {{Lantz}}},
  \bibinfo {author} {\bibfnamefont {J.}~\bibnamefont {{Rollins}}}, \ and\
  \bibinfo {author} {\bibfnamefont {K.}~\bibnamefont {{Youcef-Toumi}}},\ }\href
  {\doibase 10.1088/0264-9381/32/1/015004} {\bibfield  {journal} {\bibinfo
  {journal} {Classical and Quantum Gravity}\ }\textbf {\bibinfo {volume}
  {32}},\ \bibinfo {eid} {015004} (\bibinfo {year} {2015})}\BibitemShut
  {NoStop}%
\bibitem [{\citenamefont {{Aston}}\ \emph {et~al.}(2012)\citenamefont
  {{Aston}}, \citenamefont {{Barton}}, \citenamefont {{Bell}}, \citenamefont
  {{Beveridge}}, \citenamefont {{Bland}}, \citenamefont {{Brummitt}},
  \citenamefont {{Cagnoli}}, \citenamefont {{Cantley}}, \citenamefont
  {{Carbone}}, \citenamefont {{Cumming}}, \citenamefont {{Cunningham}},
  \citenamefont {{Cutler}}, \citenamefont {{Greenhalgh}}, \citenamefont
  {{Hammond}}, \citenamefont {{Haughian}}, \citenamefont {{Hayler}},
  \citenamefont {{Heptonstall}}, \citenamefont {{Heefner}}, \citenamefont
  {{Hoyland}}, \citenamefont {{Hough}}, \citenamefont {{Jones}}, \citenamefont
  {{Kissel}}, \citenamefont {{Kumar}}, \citenamefont {{Lockerbie}},
  \citenamefont {{Lodhia}}, \citenamefont {{Martin}}, \citenamefont {{Murray}},
  \citenamefont {{O'Dell}}, \citenamefont {{Plissi}}, \citenamefont {{Reid}},
  \citenamefont {{Romie}}, \citenamefont {{Robertson}}, \citenamefont
  {{Rowan}}, \citenamefont {{Shapiro}}, \citenamefont {{Speake}}, \citenamefont
  {{Strain}}, \citenamefont {{Tokmakov}}, \citenamefont {{Torrie}},
  \citenamefont {{van Veggel}}, \citenamefont {{Vecchio}},\ and\ \citenamefont
  {{Wilmut}}}]{2012CQGra..29w5004A}%
  \BibitemOpen
  \bibfield  {author} {\bibinfo {author} {\bibfnamefont {S.~M.}\ \bibnamefont
  {{Aston}}}, \bibinfo {author} {\bibfnamefont {M.~A.}\ \bibnamefont
  {{Barton}}}, \bibinfo {author} {\bibfnamefont {A.~S.}\ \bibnamefont
  {{Bell}}}, \bibinfo {author} {\bibfnamefont {N.}~\bibnamefont {{Beveridge}}},
  \bibinfo {author} {\bibfnamefont {B.}~\bibnamefont {{Bland}}}, \bibinfo
  {author} {\bibfnamefont {A.~J.}\ \bibnamefont {{Brummitt}}}, \bibinfo
  {author} {\bibfnamefont {G.}~\bibnamefont {{Cagnoli}}}, \bibinfo {author}
  {\bibfnamefont {C.~A.}\ \bibnamefont {{Cantley}}}, \bibinfo {author}
  {\bibfnamefont {L.}~\bibnamefont {{Carbone}}}, \bibinfo {author}
  {\bibfnamefont {A.~V.}\ \bibnamefont {{Cumming}}}, \bibinfo {author}
  {\bibfnamefont {L.}~\bibnamefont {{Cunningham}}}, \bibinfo {author}
  {\bibfnamefont {R.~M.}\ \bibnamefont {{Cutler}}}, \bibinfo {author}
  {\bibfnamefont {R.~J.~S.}\ \bibnamefont {{Greenhalgh}}}, \bibinfo {author}
  {\bibfnamefont {G.~D.}\ \bibnamefont {{Hammond}}}, \bibinfo {author}
  {\bibfnamefont {K.}~\bibnamefont {{Haughian}}}, \bibinfo {author}
  {\bibfnamefont {T.~M.}\ \bibnamefont {{Hayler}}}, \bibinfo {author}
  {\bibfnamefont {A.}~\bibnamefont {{Heptonstall}}}, \bibinfo {author}
  {\bibfnamefont {J.}~\bibnamefont {{Heefner}}}, \bibinfo {author}
  {\bibfnamefont {D.}~\bibnamefont {{Hoyland}}}, \bibinfo {author}
  {\bibfnamefont {J.}~\bibnamefont {{Hough}}}, \bibinfo {author} {\bibfnamefont
  {R.}~\bibnamefont {{Jones}}}, \bibinfo {author} {\bibfnamefont {J.~S.}\
  \bibnamefont {{Kissel}}}, \bibinfo {author} {\bibfnamefont {R.}~\bibnamefont
  {{Kumar}}}, \bibinfo {author} {\bibfnamefont {N.~A.}\ \bibnamefont
  {{Lockerbie}}}, \bibinfo {author} {\bibfnamefont {D.}~\bibnamefont
  {{Lodhia}}}, \bibinfo {author} {\bibfnamefont {I.~W.}\ \bibnamefont
  {{Martin}}}, \bibinfo {author} {\bibfnamefont {P.~G.}\ \bibnamefont
  {{Murray}}}, \bibinfo {author} {\bibfnamefont {J.}~\bibnamefont {{O'Dell}}},
  \bibinfo {author} {\bibfnamefont {M.~V.}\ \bibnamefont {{Plissi}}}, \bibinfo
  {author} {\bibfnamefont {S.}~\bibnamefont {{Reid}}}, \bibinfo {author}
  {\bibfnamefont {J.}~\bibnamefont {{Romie}}}, \bibinfo {author} {\bibfnamefont
  {N.~A.}\ \bibnamefont {{Robertson}}}, \bibinfo {author} {\bibfnamefont
  {S.}~\bibnamefont {{Rowan}}}, \bibinfo {author} {\bibfnamefont
  {B.}~\bibnamefont {{Shapiro}}}, \bibinfo {author} {\bibfnamefont {C.~C.}\
  \bibnamefont {{Speake}}}, \bibinfo {author} {\bibfnamefont {K.~A.}\
  \bibnamefont {{Strain}}}, \bibinfo {author} {\bibfnamefont {K.~V.}\
  \bibnamefont {{Tokmakov}}}, \bibinfo {author} {\bibfnamefont
  {C.}~\bibnamefont {{Torrie}}}, \bibinfo {author} {\bibfnamefont {A.~A.}\
  \bibnamefont {{van Veggel}}}, \bibinfo {author} {\bibfnamefont
  {A.}~\bibnamefont {{Vecchio}}}, \ and\ \bibinfo {author} {\bibfnamefont
  {I.}~\bibnamefont {{Wilmut}}},\ }\href {\doibase
  10.1088/0264-9381/29/23/235004} {\bibfield  {journal} {\bibinfo  {journal}
  {Classical and Quantum Gravity}\ }\textbf {\bibinfo {volume} {29}},\ \bibinfo
  {eid} {235004} (\bibinfo {year} {2012})}\BibitemShut {NoStop}%
\bibitem [{\citenamefont {{Carbone}}\ \emph {et~al.}(2012)\citenamefont
  {{Carbone}}, \citenamefont {{Aston}}, \citenamefont {{Cutler}}, \citenamefont
  {{Freise}}, \citenamefont {{Greenhalgh}}, \citenamefont {{Heefner}},
  \citenamefont {{Hoyland}}, \citenamefont {{Lockerbie}}, \citenamefont
  {{Lodhia}}, \citenamefont {{Robertson}}, \citenamefont {{Speake}},
  \citenamefont {{Strain}},\ and\ \citenamefont
  {{Vecchio}}}]{2012CQGra..29k5005C}%
  \BibitemOpen
  \bibfield  {author} {\bibinfo {author} {\bibfnamefont {L.}~\bibnamefont
  {{Carbone}}}, \bibinfo {author} {\bibfnamefont {S.~M.}\ \bibnamefont
  {{Aston}}}, \bibinfo {author} {\bibfnamefont {R.~M.}\ \bibnamefont
  {{Cutler}}}, \bibinfo {author} {\bibfnamefont {A.}~\bibnamefont {{Freise}}},
  \bibinfo {author} {\bibfnamefont {J.}~\bibnamefont {{Greenhalgh}}}, \bibinfo
  {author} {\bibfnamefont {J.}~\bibnamefont {{Heefner}}}, \bibinfo {author}
  {\bibfnamefont {D.}~\bibnamefont {{Hoyland}}}, \bibinfo {author}
  {\bibfnamefont {N.~A.}\ \bibnamefont {{Lockerbie}}}, \bibinfo {author}
  {\bibfnamefont {D.}~\bibnamefont {{Lodhia}}}, \bibinfo {author}
  {\bibfnamefont {N.~A.}\ \bibnamefont {{Robertson}}}, \bibinfo {author}
  {\bibfnamefont {C.~C.}\ \bibnamefont {{Speake}}}, \bibinfo {author}
  {\bibfnamefont {K.~A.}\ \bibnamefont {{Strain}}}, \ and\ \bibinfo {author}
  {\bibfnamefont {A.}~\bibnamefont {{Vecchio}}},\ }\href {\doibase
  10.1088/0264-9381/29/11/115005} {\bibfield  {journal} {\bibinfo  {journal}
  {Classical and Quantum Gravity}\ }\textbf {\bibinfo {volume} {29}},\ \bibinfo
  {eid} {115005} (\bibinfo {year} {2012})},\ \Eprint
  {http://arxiv.org/abs/1205.5643} {arXiv:1205.5643 [gr-qc]} \BibitemShut
  {NoStop}%
\bibitem [{\citenamefont {{Cumming}}\ \emph {et~al.}(2012)\citenamefont
  {{Cumming}}, \citenamefont {{Bell}}, \citenamefont {{Barsotti}},
  \citenamefont {{Barton}}, \citenamefont {{Cagnoli}}, \citenamefont {{Cook}},
  \citenamefont {{Cunningham}}, \citenamefont {{Evans}}, \citenamefont
  {{Hammond}}, \citenamefont {{Harry}}, \citenamefont {{Heptonstall}},
  \citenamefont {{Hough}}, \citenamefont {{Jones}}, \citenamefont {{Kumar}},
  \citenamefont {{Mittleman}}, \citenamefont {{Robertson}}, \citenamefont
  {{Rowan}}, \citenamefont {{Shapiro}}, \citenamefont {{Strain}}, \citenamefont
  {{Tokmakov}}, \citenamefont {{Torrie}},\ and\ \citenamefont {{van
  Veggel}}}]{2012CQGra..29c5003C}%
  \BibitemOpen
  \bibfield  {author} {\bibinfo {author} {\bibfnamefont {A.~V.}\ \bibnamefont
  {{Cumming}}}, \bibinfo {author} {\bibfnamefont {A.~S.}\ \bibnamefont
  {{Bell}}}, \bibinfo {author} {\bibfnamefont {L.}~\bibnamefont {{Barsotti}}},
  \bibinfo {author} {\bibfnamefont {M.~A.}\ \bibnamefont {{Barton}}}, \bibinfo
  {author} {\bibfnamefont {G.}~\bibnamefont {{Cagnoli}}}, \bibinfo {author}
  {\bibfnamefont {D.}~\bibnamefont {{Cook}}}, \bibinfo {author} {\bibfnamefont
  {L.}~\bibnamefont {{Cunningham}}}, \bibinfo {author} {\bibfnamefont
  {M.}~\bibnamefont {{Evans}}}, \bibinfo {author} {\bibfnamefont {G.~D.}\
  \bibnamefont {{Hammond}}}, \bibinfo {author} {\bibfnamefont {G.~M.}\
  \bibnamefont {{Harry}}}, \bibinfo {author} {\bibfnamefont {A.}~\bibnamefont
  {{Heptonstall}}}, \bibinfo {author} {\bibfnamefont {J.}~\bibnamefont
  {{Hough}}}, \bibinfo {author} {\bibfnamefont {R.}~\bibnamefont {{Jones}}},
  \bibinfo {author} {\bibfnamefont {R.}~\bibnamefont {{Kumar}}}, \bibinfo
  {author} {\bibfnamefont {R.}~\bibnamefont {{Mittleman}}}, \bibinfo {author}
  {\bibfnamefont {N.~A.}\ \bibnamefont {{Robertson}}}, \bibinfo {author}
  {\bibfnamefont {S.}~\bibnamefont {{Rowan}}}, \bibinfo {author} {\bibfnamefont
  {B.}~\bibnamefont {{Shapiro}}}, \bibinfo {author} {\bibfnamefont {K.~A.}\
  \bibnamefont {{Strain}}}, \bibinfo {author} {\bibfnamefont {K.}~\bibnamefont
  {{Tokmakov}}}, \bibinfo {author} {\bibfnamefont {C.}~\bibnamefont
  {{Torrie}}}, \ and\ \bibinfo {author} {\bibfnamefont {A.~A.}\ \bibnamefont
  {{van Veggel}}},\ }\href {\doibase 10.1088/0264-9381/29/3/035003} {\bibfield
  {journal} {\bibinfo  {journal} {Classical and Quantum Gravity}\ }\textbf
  {\bibinfo {volume} {29}},\ \bibinfo {eid} {035003} (\bibinfo {year}
  {2012})}\BibitemShut {NoStop}%
\bibitem [{\citenamefont {{Robertson}}\ and\ \citenamefont
  {{Majorana}}(2019)}]{2019aigw.book..423R}%
  \BibitemOpen
  \bibfield  {author} {\bibinfo {author} {\bibfnamefont {N.~A.}\ \bibnamefont
  {{Robertson}}}\ and\ \bibinfo {author} {\bibfnamefont {E.}~\bibnamefont
  {{Majorana}}},\ }in\ \href {\doibase 10.1142/9789813146082\_0016} {\emph
  {\bibinfo {booktitle} {Advanced Interferometric Gravitational-Wave Detectors.
  Volume I: Essentials of Gravitational-Wave Detectors. Edited by Reitze D et
  al. Published by World Scientific Publishing Co. Pte. Ltd}}}\ (\bibinfo
  {year} {2019})\ pp.\ \bibinfo {pages} {423--457}\BibitemShut {NoStop}%
\bibitem [{\citenamefont {{Barsotti}}\ \emph {et~al.}(2010)\citenamefont
  {{Barsotti}}, \citenamefont {{Evans}},\ and\ \citenamefont
  {{Fritschel}}}]{2010CQGra..27h4026B}%
  \BibitemOpen
  \bibfield  {author} {\bibinfo {author} {\bibfnamefont {L.}~\bibnamefont
  {{Barsotti}}}, \bibinfo {author} {\bibfnamefont {M.}~\bibnamefont {{Evans}}},
  \ and\ \bibinfo {author} {\bibfnamefont {P.}~\bibnamefont {{Fritschel}}},\
  }\href {\doibase 10.1088/0264-9381/27/8/084026} {\bibfield  {journal}
  {\bibinfo  {journal} {Classical and Quantum Gravity}\ }\textbf {\bibinfo
  {volume} {27}},\ \bibinfo {eid} {084026} (\bibinfo {year}
  {2010})}\BibitemShut {NoStop}%
\bibitem [{\citenamefont {{Hua}}\ \emph {et~al.}(2004)\citenamefont {{Hua}},
  \citenamefont {{Adhikari}}, \citenamefont {{DeBra}}, \citenamefont
  {{Giaime}}, \citenamefont {{Hammond}}, \citenamefont {{Hardham}},
  \citenamefont {{Hennessy}}, \citenamefont {{How}}, \citenamefont {{Lantz}},
  \citenamefont {{Macinnis}}, \citenamefont {{Mittleman}}, \citenamefont
  {{Richman}}, \citenamefont {{Robertson}}, \citenamefont {{Rollins}},
  \citenamefont {{Shoemaker}},\ and\ \citenamefont
  {{Stebbins}}}]{2004SPIE.5500..194H}%
  \BibitemOpen
  \bibfield  {author} {\bibinfo {author} {\bibfnamefont {W.}~\bibnamefont
  {{Hua}}}, \bibinfo {author} {\bibfnamefont {R.}~\bibnamefont {{Adhikari}}},
  \bibinfo {author} {\bibfnamefont {D.~B.}\ \bibnamefont {{DeBra}}}, \bibinfo
  {author} {\bibfnamefont {J.~A.}\ \bibnamefont {{Giaime}}}, \bibinfo {author}
  {\bibfnamefont {G.~D.}\ \bibnamefont {{Hammond}}}, \bibinfo {author}
  {\bibfnamefont {C.}~\bibnamefont {{Hardham}}}, \bibinfo {author}
  {\bibfnamefont {M.}~\bibnamefont {{Hennessy}}}, \bibinfo {author}
  {\bibfnamefont {J.~P.}\ \bibnamefont {{How}}}, \bibinfo {author}
  {\bibfnamefont {B.~T.}\ \bibnamefont {{Lantz}}}, \bibinfo {author}
  {\bibfnamefont {M.}~\bibnamefont {{Macinnis}}}, \bibinfo {author}
  {\bibfnamefont {R.}~\bibnamefont {{Mittleman}}}, \bibinfo {author}
  {\bibfnamefont {S.}~\bibnamefont {{Richman}}}, \bibinfo {author}
  {\bibfnamefont {N.~A.}\ \bibnamefont {{Robertson}}}, \bibinfo {author}
  {\bibfnamefont {J.}~\bibnamefont {{Rollins}}}, \bibinfo {author}
  {\bibfnamefont {D.~H.}\ \bibnamefont {{Shoemaker}}}, \ and\ \bibinfo {author}
  {\bibfnamefont {R.~T.}\ \bibnamefont {{Stebbins}}},\ }in\ \href {\doibase
  10.1117/12.552518} {\emph {\bibinfo {booktitle} {Gravitational Wave and
  Particle Astrophysics Detectors}}},\ \bibinfo {series} {Society of
  Photo-Optical Instrumentation Engineers (SPIE) Conference Series}, Vol.\
  \bibinfo {volume} {5500},\ \bibinfo {editor} {edited by\ \bibinfo {editor}
  {\bibfnamefont {J.}~\bibnamefont {{Hough}}}\ and\ \bibinfo {editor}
  {\bibfnamefont {G.~H.}\ \bibnamefont {{Sanders}}}}\ (\bibinfo {year} {2004})\
  pp.\ \bibinfo {pages} {194--205}\BibitemShut {NoStop}%
\bibitem [{\citenamefont {{Abbott}}\ \emph {et~al.}(2004)\citenamefont
  {{Abbott}}, \citenamefont {{Adhikari}}, \citenamefont {{Allen}},
  \citenamefont {{Baglino}}, \citenamefont {{Campbell}}, \citenamefont
  {{Coyne}}, \citenamefont {{Daw}}, \citenamefont {{DeBra}}, \citenamefont
  {{Faludi}}, \citenamefont {{Fritschel}}, \citenamefont {{Ganguli}},
  \citenamefont {{Giaime}}, \citenamefont {{Hammond}}, \citenamefont
  {{Hardham}}, \citenamefont {{Harry}}, \citenamefont {{Hua}}, \citenamefont
  {{Jones}}, \citenamefont {{Kern}}, \citenamefont {{Lantz}}, \citenamefont
  {{Lilienkamp}}, \citenamefont {{Mailand}}, \citenamefont {{Mason}},
  \citenamefont {{Mittleman}}, \citenamefont {{Nayfeh}}, \citenamefont
  {{Ottaway}}, \citenamefont {{Phinney}}, \citenamefont {{Rankin}},
  \citenamefont {{Robertson}}, \citenamefont {{Scheffler}}, \citenamefont
  {{Shoemaker}}, \citenamefont {{Wen}}, \citenamefont {{Zucker}},\ and\
  \citenamefont {{Zuo}}}]{2004CQGra..21S.915A}%
  \BibitemOpen
  \bibfield  {author} {\bibinfo {author} {\bibfnamefont {R.}~\bibnamefont
  {{Abbott}}}, \bibinfo {author} {\bibfnamefont {R.}~\bibnamefont
  {{Adhikari}}}, \bibinfo {author} {\bibfnamefont {G.}~\bibnamefont {{Allen}}},
  \bibinfo {author} {\bibfnamefont {D.}~\bibnamefont {{Baglino}}}, \bibinfo
  {author} {\bibfnamefont {C.}~\bibnamefont {{Campbell}}}, \bibinfo {author}
  {\bibfnamefont {D.}~\bibnamefont {{Coyne}}}, \bibinfo {author} {\bibfnamefont
  {E.}~\bibnamefont {{Daw}}}, \bibinfo {author} {\bibfnamefont
  {D.}~\bibnamefont {{DeBra}}}, \bibinfo {author} {\bibfnamefont
  {J.}~\bibnamefont {{Faludi}}}, \bibinfo {author} {\bibfnamefont
  {P.}~\bibnamefont {{Fritschel}}}, \bibinfo {author} {\bibfnamefont
  {A.}~\bibnamefont {{Ganguli}}}, \bibinfo {author} {\bibfnamefont
  {J.}~\bibnamefont {{Giaime}}}, \bibinfo {author} {\bibfnamefont
  {M.}~\bibnamefont {{Hammond}}}, \bibinfo {author} {\bibfnamefont
  {C.}~\bibnamefont {{Hardham}}}, \bibinfo {author} {\bibfnamefont
  {G.}~\bibnamefont {{Harry}}}, \bibinfo {author} {\bibfnamefont
  {W.}~\bibnamefont {{Hua}}}, \bibinfo {author} {\bibfnamefont
  {L.}~\bibnamefont {{Jones}}}, \bibinfo {author} {\bibfnamefont
  {J.}~\bibnamefont {{Kern}}}, \bibinfo {author} {\bibfnamefont
  {B.}~\bibnamefont {{Lantz}}}, \bibinfo {author} {\bibfnamefont
  {K.}~\bibnamefont {{Lilienkamp}}}, \bibinfo {author} {\bibfnamefont
  {K.}~\bibnamefont {{Mailand}}}, \bibinfo {author} {\bibfnamefont
  {K.}~\bibnamefont {{Mason}}}, \bibinfo {author} {\bibfnamefont
  {R.}~\bibnamefont {{Mittleman}}}, \bibinfo {author} {\bibfnamefont
  {S.}~\bibnamefont {{Nayfeh}}}, \bibinfo {author} {\bibfnamefont
  {D.}~\bibnamefont {{Ottaway}}}, \bibinfo {author} {\bibfnamefont
  {J.}~\bibnamefont {{Phinney}}}, \bibinfo {author} {\bibfnamefont
  {W.}~\bibnamefont {{Rankin}}}, \bibinfo {author} {\bibfnamefont
  {N.}~\bibnamefont {{Robertson}}}, \bibinfo {author} {\bibfnamefont
  {R.}~\bibnamefont {{Scheffler}}}, \bibinfo {author} {\bibfnamefont {D.~H.}\
  \bibnamefont {{Shoemaker}}}, \bibinfo {author} {\bibfnamefont
  {S.}~\bibnamefont {{Wen}}}, \bibinfo {author} {\bibfnamefont
  {M.}~\bibnamefont {{Zucker}}}, \ and\ \bibinfo {author} {\bibfnamefont
  {L.}~\bibnamefont {{Zuo}}},\ }\href {\doibase 10.1088/0264-9381/21/5/081}
  {\bibfield  {journal} {\bibinfo  {journal} {Classical and Quantum Gravity}\
  }\textbf {\bibinfo {volume} {21}},\ \bibinfo {pages} {S915} (\bibinfo {year}
  {2004})}\BibitemShut {NoStop}%
\bibitem [{\citenamefont {{Schwartz}}(2020)}]{2020CQGra..37w5007S}%
  \BibitemOpen
  \bibfield  {author} {\bibinfo {author} {\bibfnamefont {E.~e.~a.}\
  \bibnamefont {{Schwartz}}},\ }\href {\doibase 10.1088/1361-6382/abbc8c}
  {\bibfield  {journal} {\bibinfo  {journal} {Classical and Quantum Gravity}\
  }\textbf {\bibinfo {volume} {37}},\ \bibinfo {eid} {235007} (\bibinfo {year}
  {2020})},\ \Eprint {http://arxiv.org/abs/2007.12847} {arXiv:2007.12847
  [physics.ins-det]} \BibitemShut {NoStop}%
\bibitem [{\citenamefont {{Walker}}(2017)}]{2017RScI...88l4501W}%
  \BibitemOpen
  \bibfield  {author} {\bibinfo {author} {\bibfnamefont {M.~e.~a.}\
  \bibnamefont {{Walker}}},\ }\href {\doibase 10.1063/1.5000264} {\bibfield
  {journal} {\bibinfo  {journal} {Review of Scientific Instruments}\ }\textbf
  {\bibinfo {volume} {88}},\ \bibinfo {eid} {124501} (\bibinfo {year}
  {2017})},\ \Eprint {http://arxiv.org/abs/1702.04701} {arXiv:1702.04701
  [astro-ph.IM]} \BibitemShut {NoStop}%
\bibitem [{\citenamefont {{Daveloza}}\ \emph {et~al.}(2012)\citenamefont
  {{Daveloza}}, \citenamefont {{Afrin Badhan}}, \citenamefont {{Diaz}},
  \citenamefont {{Kawabe}}, \citenamefont {{Konverski}}, \citenamefont
  {{Landry}},\ and\ \citenamefont {{Savage}}}]{2012JPhCS.363a2007D}%
  \BibitemOpen
  \bibfield  {author} {\bibinfo {author} {\bibfnamefont {H.~P.}\ \bibnamefont
  {{Daveloza}}}, \bibinfo {author} {\bibfnamefont {M.}~\bibnamefont {{Afrin
  Badhan}}}, \bibinfo {author} {\bibfnamefont {M.}~\bibnamefont {{Diaz}}},
  \bibinfo {author} {\bibfnamefont {K.}~\bibnamefont {{Kawabe}}}, \bibinfo
  {author} {\bibfnamefont {P.~N.}\ \bibnamefont {{Konverski}}}, \bibinfo
  {author} {\bibfnamefont {M.}~\bibnamefont {{Landry}}}, \ and\ \bibinfo
  {author} {\bibfnamefont {R.~L.}\ \bibnamefont {{Savage}}},\ }in\ \href
  {\doibase 10.1088/1742-6596/363/1/012007} {\emph {\bibinfo {booktitle}
  {Journal of Physics Conference Series}}},\ \bibinfo {series} {Journal of
  Physics Conference Series}, Vol.\ \bibinfo {volume} {363}\ (\bibinfo {year}
  {2012})\ p.\ \bibinfo {pages} {012007}\BibitemShut {NoStop}%
\bibitem [{\citenamefont {{Strain}}\ and\ \citenamefont
  {{Shapiro}}(2012)}]{2012RScI...83d4501S}%
  \BibitemOpen
  \bibfield  {author} {\bibinfo {author} {\bibfnamefont {K.~A.}\ \bibnamefont
  {{Strain}}}\ and\ \bibinfo {author} {\bibfnamefont {B.~N.}\ \bibnamefont
  {{Shapiro}}},\ }\href {\doibase 10.1063/1.4704459} {\bibfield  {journal}
  {\bibinfo  {journal} {Review of Scientific Instruments}\ }\textbf {\bibinfo
  {volume} {83}},\ \bibinfo {eid} {044501-044501-9} (\bibinfo {year}
  {2012})}\BibitemShut {NoStop}%
\bibitem [{\citenamefont {{Stochino}}\ \emph {et~al.}(2009)\citenamefont
  {{Stochino}}, \citenamefont {{Abbot}}, \citenamefont {{Aso}}, \citenamefont
  {{Barton}}, \citenamefont {{Bertolini}}, \citenamefont {{Boschi}},
  \citenamefont {{Coyne}}, \citenamefont {{DeSalvo}}, \citenamefont {{Galli}},
  \citenamefont {{Huang}}, \citenamefont {{Ivanov}}, \citenamefont {{Marka}},
  \citenamefont {{Ottaway}}, \citenamefont {{Sannibale}}, \citenamefont
  {{Vanni}}, \citenamefont {{Yamamoto}},\ and\ \citenamefont
  {{Yoshida}}}]{2009NIMPA.598..737S}%
  \BibitemOpen
  \bibfield  {author} {\bibinfo {author} {\bibfnamefont {A.}~\bibnamefont
  {{Stochino}}}, \bibinfo {author} {\bibfnamefont {B.}~\bibnamefont {{Abbot}}},
  \bibinfo {author} {\bibfnamefont {Y.}~\bibnamefont {{Aso}}}, \bibinfo
  {author} {\bibfnamefont {M.}~\bibnamefont {{Barton}}}, \bibinfo {author}
  {\bibfnamefont {A.}~\bibnamefont {{Bertolini}}}, \bibinfo {author}
  {\bibfnamefont {V.}~\bibnamefont {{Boschi}}}, \bibinfo {author}
  {\bibfnamefont {D.}~\bibnamefont {{Coyne}}}, \bibinfo {author} {\bibfnamefont
  {R.}~\bibnamefont {{DeSalvo}}}, \bibinfo {author} {\bibfnamefont
  {C.}~\bibnamefont {{Galli}}}, \bibinfo {author} {\bibfnamefont
  {Y.}~\bibnamefont {{Huang}}}, \bibinfo {author} {\bibfnamefont
  {A.}~\bibnamefont {{Ivanov}}}, \bibinfo {author} {\bibfnamefont
  {S.}~\bibnamefont {{Marka}}}, \bibinfo {author} {\bibfnamefont
  {D.}~\bibnamefont {{Ottaway}}}, \bibinfo {author} {\bibfnamefont
  {V.}~\bibnamefont {{Sannibale}}}, \bibinfo {author} {\bibfnamefont
  {C.}~\bibnamefont {{Vanni}}}, \bibinfo {author} {\bibfnamefont
  {H.}~\bibnamefont {{Yamamoto}}}, \ and\ \bibinfo {author} {\bibfnamefont
  {S.}~\bibnamefont {{Yoshida}}},\ }\href {\doibase 10.1016/j.nima.2008.10.023}
  {\bibfield  {journal} {\bibinfo  {journal} {Nuclear Instruments and Methods
  in Physics Research A}\ }\textbf {\bibinfo {volume} {598}},\ \bibinfo {pages}
  {737} (\bibinfo {year} {2009})}\BibitemShut {NoStop}%
\bibitem [{\citenamefont {{Canuel}}\ \emph {et~al.}(2013)\citenamefont
  {{Canuel}}, \citenamefont {{Genin}}, \citenamefont {{Vajente}},\ and\
  \citenamefont {{Marque}}}]{2013OExpr..2110546C}%
  \BibitemOpen
  \bibfield  {author} {\bibinfo {author} {\bibfnamefont {B.}~\bibnamefont
  {{Canuel}}}, \bibinfo {author} {\bibfnamefont {E.}~\bibnamefont {{Genin}}},
  \bibinfo {author} {\bibfnamefont {G.}~\bibnamefont {{Vajente}}}, \ and\
  \bibinfo {author} {\bibfnamefont {J.}~\bibnamefont {{Marque}}},\ }\href
  {\doibase 10.1364/OE.21.010546} {\bibfield  {journal} {\bibinfo  {journal}
  {Optics Express}\ }\textbf {\bibinfo {volume} {21}},\ \bibinfo {pages}
  {10546} (\bibinfo {year} {2013})}\BibitemShut {NoStop}%
\bibitem [{\citenamefont {{Ottaway}}\ \emph {et~al.}(2012)\citenamefont
  {{Ottaway}}, \citenamefont {{Fritschel}},\ and\ \citenamefont
  {{Waldman}}}]{2012OExpr..20.8329O}%
  \BibitemOpen
  \bibfield  {author} {\bibinfo {author} {\bibfnamefont {D.~J.}\ \bibnamefont
  {{Ottaway}}}, \bibinfo {author} {\bibfnamefont {P.}~\bibnamefont
  {{Fritschel}}}, \ and\ \bibinfo {author} {\bibfnamefont {S.~J.}\ \bibnamefont
  {{Waldman}}},\ }\href {\doibase 10.1364/OE.20.008329} {\bibfield  {journal}
  {\bibinfo  {journal} {Optics Express}\ }\textbf {\bibinfo {volume} {20}},\
  \bibinfo {pages} {8329} (\bibinfo {year} {2012})}\BibitemShut {NoStop}%
\end{thebibliography}%

\end{document}